\documentclass[12pt,preprint]{aastex}
\usepackage{epsfig}
\usepackage{natbib}
\usepackage{graphicx}
\usepackage{slashbox}
\usepackage{multirow}
\usepackage{lscape}
\usepackage{mathrsfs,amssymb}
\usepackage{amsmath}

\usepackage[caption=false]{subfig}

\newcommand	\beq	{\begin{equation}}	%{\begin{displaymath}}
\newcommand	\eeq	{\end{equation}}	%{\end{displaymath}}
\newcommand       \Angstrom     {\,{\rm \AA}}
\newcommand       \AU           {\,{\rm AU}}
\newcommand       \cm           {\,{\rm cm}}

\newcommand       \K            {\,{\rm K}}
\newcommand       \pc           {\,{\rm pc}}

\newcommand       \s            {\,{\rm s}}

\newcommand       \yr       {\,{\rm yr}}

\newcommand       \Myr      {\,{\rm Myr}}

\newcommand       \simlt        {\lesssim}
\newcommand       \simgt        {\gtrsim}

\newcommand       \um           {\mu{\rm m}}
\newcommand       \mum          {\,{\rm \mu m}}

\newcommand       \Teff         {T_{\rm eff}}
\newcommand       \msun         {\,{M_\odot}}
\newcommand       \Lsun         {\,{L_\odot}}

\newcommand       \Mstar        {M_\star}
\newcommand       \Lstar        {L_\star}

\newcommand       \simali       {\sim\,}

\newcommand{\etal}{\textrm{et al.\ }}
\newcommand{\ie}{\textrm{i.e., }}
\newcommand{\eg}{\textrm{e.g., }}

\newcommand{\Msun}      {\,{M_\odot}}
\newcommand{\Mearth}    {\,{M_\earth}}

\newcommand       \apahmin      {a^{\rm PAH}_{\rm min}}

\newcommand	  \ap                 {a_{\rm p}}
\newcommand	  \phiion            {\phi_{\rm ion}}
\newcommand    \brp                {\beta_{\rm RP}}

\renewcommand\i   {\hbox{$i$}}

\shorttitle{PAHs in Protoplanetary Disks}
\shortauthors{Seok \& Li}
\title{
Polycyclic Aromatic Hydrocarbon
in Protoplanetary Disks around Herbig Ae/Be and T Tauri Stars
}
\author{Ji Yeon Seok\altaffilmark{1,2} and Aigen Li\altaffilmark{1}}

\altaffiltext{1}{Department of Physics and Astronomy, University of Missouri, Columbia, MO 65211, USA, seokji@missouri.edu, lia@missouri.edu}
\altaffiltext{2}{Key Laboratory of Optical Astronomy, National Astronomical Observatories, Chinese Academy of Sciences, Beijing 100012, China, jiseok@bao.ac.cn}

\begin{document}

\begin{abstract}

A distinct set of broad emission features at 3.3, 6.2,
7.7, 8.6, 11.3, and 12.7$\mum$, is often detected in
protoplanetary disks (PPDs). These features are
commonly attributed to polycyclic aromatic hydrocarbons (PAHs). 
We model these emission features in
the infrared spectra of 69 PPDs
around 14 T Tauri and 55 Herbig Ae/Be stars in terms of
astronomical-PAHs. For each PPD, we derive
the size distribution and the charge state of PAHs.
We then examine the correlations of the PAH properties
(\ie sizes and ionization fractions) with the stellar properties
(\eg stellar effective temperature, luminosity, and mass).
We find that the characteristic size
of PAHs shows a tendency of correlating with 
the stellar effective temperature ($\Teff$)
and interpret this as the preferential 
photodissociation of small PAHs 
in systems with higher $\Teff$ of which 
the stellar photons are more energetic.  
In addition, the PAH size shows a moderate
correlation with the red-ward wavelength-shift
of the 7.7$\mum$ PAH feature that is commonly
observed in disks around cool stars.
The ionization fraction of PAHs 
does not seem to correlate with any stellar parameters.
This is because the charging of PAHs depends on
not only the stellar properties
(\eg $\Teff$, luminosity)
but also the spatial distribution of PAHs in the disks.
The mere negative correlation between the
PAH size and the stellar age suggests that continuous
replenishment of PAHs via the outgassing of cometary
bodies and/or the collisional grinding of
planetesimals and asteroids is
required to maintain the abundance of small PAHs
against complete destruction by photodissociation. 

\end{abstract}

\keywords{circumstellar matter --- infrared: stars
          --- planetary systems: protoplanetary disks}

\section{Introduction}\label{sec:intro}

A number of infrared (IR) observations of disks around
low- and intermediate-mass pre-main-sequence stars
clearly evince a distinct set of broad emission features at
3.3, 6.2, 7.7, 8.6, 11.2, and $12.7\mum$ 
(\eg Brooke \etal1993; Meeus \etal 2001;
Acke \& van den Ancker 2004;
Habart \etal2004a; Sloan \etal2005;
Geers \etal2006, 2007a, 2007b; 
Keller \etal2008; Acke \etal2010; Maaskant \etal2014).
These IR spectral features are commonly 
attributed to polycyclic aromatic hydrocarbons (PAHs;
L\'{e}ger \& Puget 1984, Allamandola \etal1985).
While $\la10\%$ of the disks around T Tauri stars 
(TTSs; $M_\star\la2\msun$) show PAH features in their IR spectra
(\eg Furlan \etal2006; Geers \etal2006),
for Herbig Ae/Be (HAeBe) stars, the higher-mass counterparts
of TTSs ($M_\star$$\simali2$--$8\msun$),
PAH features are more commonly detected from their disks
(\eg detection rate of $\simali$70\%, Acke \etal2010).
On the other hand, more evolved disks around main-sequence stars,
so-called debris disks around Vega-type stars, hardly show PAH features.
For instance, Chen \etal(2006) reported no detection of PAH emission
in the IR spectra of 59 debris disks obtained with the Infrared
Spectrograph (IRS) on board the {\it Spitzer Space Telescope}.
This suggests that the characteristics of PAH emission of these
protoplanetary disks (PPDs) are contingent on the stellar properties
and evolutionary phases of the disks and the
PAH features can be used to probe the physical and
chemical properties of PPDs.

The PAH emission features are often used as a
diagnostic tool of the physical conditions of PPDs
and a tracer of the structure of PPDs.
Their band ratios are sensitive to the charge state
of PAHs, which is controlled by electron density,
gas temperature, and starlight intensity (\eg
Bakes \& Tielens 1994; Weingartner \& Draine 2001).
Due to their stochastic heating nature (Draine \& Li 2001),
PAH emission features can be detected far away from
the central star (\eg the PAH features in the disk
around WL\,16, a HAeBe star, are spatially 
extended up to $\simali$440$\AU$ from the central star, 
Ressler \& Barsony 2003) and can be spatially more
extended than the emission of large grains in thermal
equilibrium with the stellar radiation. In addition,
ionized PAHs could trace protoplanetary gaps
(Maaskant \etal2014). 

PAHs profoundly influence the thermal budget and
chemistry of PPDs, where planet(s) would newly form. 
PAH molecules in the surface layer of the disk,
where PAHs are directly exposed to the stellar radiation,
efficiently get photoionized and provide photoelectrons
for heating the ambient gas (\eg Kamp \& Dullemond 2004).
PAHs have large surface areas for chemical reactions
(\eg Jonkheid \etal2004; Habart \etal 2004b).
In particular, PAHs can alter carbon chemistry of the disk
through transferring electrons from neutral and
negatively charged PAHs to ionized carbon atoms.
Also, PAHs have large far-ultraviolet (UV) opacity
so that the inner parts of the disk can be shielded 
from the stellar UV radiation. In this context,
examining the physical properties of PAHs in PPDs and 
their correlations with individual stellar properties
of the PPDs would be essential to understand
the diverse characteristics of PPDs.

%[lack of the systematic study and the significance of this work]
Profuse spectroscopic observations obtained with IR
space telescopes such as the
{\it Infrared Space Observatory} ($ISO$) and/or $Spitzer$
as well as ground-based telescopes such as the Very Large
Telescope (VLT) or the NASA Infrared Telescope
Facility (IRTF) have been utilized to conduct various studies
of PAH features in PPDs, in particular using a large number
of sample.
Acke \& van den Ancker (2004) presented $ISO$ spectra
of a sample of 46 HAeBe stars and detect PAH
features in 27 sources of their sample ($\simali$57\%).
Later, using $Spitzer$/IRS data, Geers \etal(2006) analyzed
the PAH features of 38 TTSs in comparison with a sample
of HAeBe stars and found that the 6.2 and 11.3$\mum$
PAH features are commonly detected while those at 7.7
and 8.6$\mum$ are relatively rare due to the strong
9.7$\mum$ silicate emission. As PAH features are often strong
in HAeBe stars, a number of previous studies focus
on the PAH features of HAeBe stars; Sloan \etal(2005)
found that the ionization fraction of PAHs is higher
for PPDs with hotter and brighter central stars.
Keller \etal(2008) used $Spitzer$/IRS spectra of 22 targets
including 18 HAeBe stars and reported spectral
variations among these stars, which is associated
with the effective temperature of the host stars. With a
larger sample of HAeBe stars (\ie 53 Herbig Ae stars), 
Acke \etal(2010) pointed out that the diversity of
PAH emission features shown in the Herbig Ae stars
is mainly attributed to chemical processing induced
by the stellar UV radiation field. In addition,
spatially resolved spectroscopy using ground-based
telescopes has revealed that the PAH features
in some PPDs around TTSs and HAeBe stars
are spatially extended over a few tens to hundred AU
(\eg van Boekel \etal2004; Habart \etal2006;
Geers \etal2007b; Maaskant \etal2014), and
the observed PAH extent does not always coincide
with that of the dust continuum (\eg Geers \etal2007b;
Maaskant \etal2014).

While the characteristics of PAH emission features and
their band ratios have been extensively studied for
a number of PPDs in the literature, the physical properties
of PAH molecules in the PPDs are still mostly unknown,
which requires detailed modeling of PAH spectra. 
Habart \etal(2004a) performed a disk modeling using
a radiation transfer code assuming either fully ionized or
neutral PAHs with a single PAH size (\ie $N_{\rm C}=40$
or 100). They compared the model results with observations
of HAeBe stars from $ISO$ and ground-based telescopes
but limited it to the band strengths (\ie not aim to
reproduce the PAH features of individual sources).
Later, Maaskant \etal(2014) implemented
a model to calculate the charge state of PAHs in a
radiative transfer code and
applied it to the PAH spectra of four PPDs. 
Since they intended to verify the effect of ionization,
a single PAH size is adopted. However, the
typical size of PAHs in the disk might vary substantially
from one object to another and
could affect the characteristics of PAH spectral features. 

In previous papers, we
performed comprehensive modelings of the PAH features
for three specific sources (HD141569A: Li \& Lunine 2003b;
HD 34700: Seok \& Li 2015; 
HD 169142: Seok \& Li 2016) adopting the astro-PAH
model of Li \& Draine (2001b) and Draine \& Li (2007)
and derived the properties of PAHs in the disks.
Since, only for limited sources,
their PAH properties have been well-constrained so far,
an analysis with a statistically relevant sample is further required
to probe the possible physical association
between PAHs in PPDs and their ambient conditions.
In this paper, we present the PAH spectra of 69 PPDs
collated from the literature and compile the fundamental
stellar properties such as the effective temperature, 
stellar mass, and age to interpret the PAH properties
including the size and charge state of PAHs.   
By coherently modeling the PAH spectra of a large number of
sources including both TTSs and HAeBe stars 
and statistically examining their PAH properties,
we are able to characterize the physical and chemical
properties of PAHs in the PPDs and reveal their
relations with the stellar properties and disk environments. 

The paper is organized as follows. The sample
selection and observational data are summarized
in Section \ref{sec:data}, and the astro-PAH model adopted
here and the model-fitting procedure are
described in Section \ref{sec:model}. We present
the model results and statistics in Section \ref{sec:res}.
In Section \ref{sec:disc}, we discuss the
correlations between the PAH properties
and stellar parameters, the destruction of PAHs in PPDs,
the aliphatic fraction of PAH molecules, and
the effects of different methods of extracting
the PAH spectra. Finally, we summarize
the main results in Section \ref{sec:summ}.

\section{Data}\label{sec:data}

%To carry out a statistical study of PAHs related to stellar
%properties, 
\subsection{Sample Selection}
We search for PPDs reported to show PAH emission in
the literature, and 69 disks are found,
consisting of 14 TTSs (Furlan \etal2006; Geers \etal2006, 2007b;
Keller \etal2008; Sch{\"u}tz \etal2009; Mer{\'{\i}}n \etal2010) 
and 55 HAeBe stars (Acke \& van den Ancker 2004;
Sloan \etal2005, 2007;
Keller \etal2008;
Boersma \etal2009;
Sch{\"u}tz \etal2009;
Acke \etal2010;
K{\'o}sp{\'a}l \etal2012;
Maaskant \etal2013),
which allow us to cover a wide range of stellar parameters.
To the best of our knowledge, this is the largest sample of
PPDs showing PAH features in their IR spectra, which are to
be modeled taking precise PAH chemistry into account.
The final sample with their stellar parameters is listed 
in Table \ref{tab:obj}.

\subsection{Data Set}\label{sub:dataset}

For the 69 sources, we collate both near-IR (NIR; $\la4\mum$)
and mid-IR (MIR; $\ga5\mum$) spectra available in the literature 
to analyze the major PAH features at 3.3--$11.3\mum$. 
This is important because a simultaneous modeling of
as many PAH bands as possible can more accurately
constrain the properties of PAHs in a disk
although it is the 6.2, 7.7, and $8.6\mum$ features 
that mainly control the best-fit model parameters.
For the MIR data, the $Spitzer$/IRS spectra
(SL module: $R\approx60$--127 at 5.2--$14.5\mum$) 
of all except 11 sources are available in the literature
(\eg Furlan \etal2006;
Geers \etal2006;
Kessler-Silacci \etal2006;
Keller \etal2008;
Acke \etal2010; 
Mer{\'{\i}}n \etal2010;
K{\'o}sp{\'a}l \etal2012;
Maaskant \etal2013).
Among those with $Spitzer$/IRS data, the
IRS spectra of 8 sources do not sufficiently cover the
wavelength range of the SL module for viable model calculation.
For four of them (HD 100546, HD 163296, HD 179218,
and Wray 15-1484), we take fully processed $ISO$/SWS
spectra (2.38--$45.2\mum$, $R\simali$1500--2000) from
the $ISO$/SWS atlas (Sloan \etal2003).\footnote{%
 http://isc.astro.cornell.edu/$\simali$sloan/library/swsatlas/atlas.html}
For SR 21N, we adopt the $N$-band spectrum (7.7--$12.5\mum$,
$R\simali350$) obtained with the VLT Imager and Spectrometer
for the mid-IR (VISIR) installed at the VLT from Geers \etal(2007b).
For the remaining three sources (HD 98922, HD 259431, and T Cha),
no other supplementary data are available in the literature.
While more than two PAH bands are present in the IRS spectrum
of HD 98922, only the $11.3\mum$ feature is covered by the spectra
of HD 259431 and T Cha. We exclude the latter two sources
for model calculations. 
For those without $Spitzer$/IRS data,
we take the $ISO$/ISOPHOT-SL spectra (5.8--$11.6\mum$,
$R\approx100$) of K{\'o}sp{\'a}l \etal(2012)\footnote{%
 http://vizier.cfa.harvard.edu/viz-bin/VizieR?-source=J/ApJS/201/11}
for three sources (IRAS 03260+3111, VX Cas, and WL 16)
the $N$-band spectrum (8--$13\mum$) obtained with
the ESO TIMMI2 camera at La Silla Observatory
for PDS 144N (Sch{\"u}tz \etal2009),
and the $ISO$/SWS spectra of eight sources (BD+40$\degr$4124,
HD 200775, IRAS 06084-0611, LkH$\alpha$ 224, MWC 297,
MWC 865, MWC 1080,
and TY CrA) from the $ISO$/SWS atlas (Sloan \etal2003).

Unlike the MIR spectra, the NIR spectra
are not available for all sources. 
For the 27 sources observed by $ISO$/SWS,
we adopt the SWS spectra from the $ISO$/SWS atlas.   
For IRAS 03260+3111, VX Cas, WL 16,
%together with three additional sources (
HD 97300, RR Tau, and WW Vul,
we take the ISOPHOT-SS (2.5--$4.9\mum$, $R\approx100$)
data from K{\'o}sp{\'a}l \etal(2012).
For AK Sco and BF Ori (Acke \& van den Ancker 2006),
Oph IRS48 (Geers \etal2007a), HD 98922, HD 101412,
HD 141569, SR21N, T Cha, and VV Ser (Geers \etal2007b),
we collect the $L$-band (2.8--$4.2\mum$) spectra
obtained with ISAAC, the Infrared Spectrometer And Array Camera,
installed at the VLT. For HD 34700, we use 
the $L$-band ($\simali$1.9--$4.2\mum$) spectrum
obtained with the medium-resolution spectrograph SpeX at the
IRTF (Smith \etal2004).
  
%matching absolute scale
When we use two different observational sets
for the NIR and MIR spectra of one object, we need to
match their flux levels. If there is an overlapping wavelength
range of the two spectra (\ie $ISO$/SWS or ISOPHOT and 
$Spitzer$/IRS data for the NIR and MIR spectra, respectively),
we derive a scaling factor using the overlapping part and
scale the NIR spectrum with respect to the MIR spectrum.\footnote{%
We scaled the NIR spectra (\ie $ISO$ data) with
respect to the MIR spectra (\ie $Spitzer$ data) because
the $Spitzer$ spectra are mostly consistent
with the IR photometry whereas the $ISO$ spectra
of some objects are rather noisy and/or have large
uncertainties in their absolute flux calibration.}
If no overlapping part is available, for those with
an NIR spectrum from a ground-based telescope and an MIR
spectrum from $Spitzer$, we scale both spectra with respect
to the photometric measurements taken from VizieR.\footnote{%
 http://vizier.u-strasbg.fr/vizier}
 
In summary, we accumulate the IR spectroscopic data of
69 disks from the literature and data archives,
composed of 43 NIR and 69 MIR spectra.
Table \ref{tab:obj} presents a summary of the data set
compiled in this work. 

%%%%%%%% Table 1: Source list %%%%%%%%
%% LaTeX deluxetable generator for the AASTeX package.
%% Written by Greg Schwarz (5/1/2001).

%% Table generated: Mon Mar 28 12:09:53 2016

%% Remove the two lines and the last line if you want
%% want to incorporate this table into another LaTex document.
%\documentclass{aastex}
%\begin{document}

%% The values (usually only l,r and c) in the last part of
%% \begin{deluxetable}{} command tell LaTeX how many columns
%% there are and how to align them.
\begin{deluxetable}{lrrrrccccc}
\scriptsize
\center
%% Keep a portrait orientation

%% Over-ride the default font size
%% Use Default (12pt)
\tablewidth{500pt}
%% Use \tablewidth{?pt} to over-ride the default table width.
%% If you are unhappy with the default look at the end of the
%% *.log file to see what the default was set at before adjusting
%% this value.

%% This is the title of the table.
\tablecaption{\label{tab:obj}
Sample of Protoplanetary Disks with PAH Emission}
%\scriptsize
\footnotesize
%% This command over-rides LaTeX's natural table count
%% and replaces it with this number.  LaTeX will increment 
%% all other tables after this table based on this number
%\tablenum{1}

%% The \tablehead gives provides the column headers.  It
%% is currently set up so that the column labels are on the
%% top line and the units surrounded by ()s are in the 
%% bottom line.  You may add more header information by writing
%% another line between these lines. For each column that requries
%% extra information be sure to include a \colhead{text} command
%% and remember to end any extra lines with \\ and include the 
%% correct number of &s.
\tablehead{\colhead{Object} & \colhead{$T_{\rm eff}$} & \colhead{$L_{\star}$} &
\colhead{$M_{\star}$} & \colhead{Age} & \colhead{$d$} & \colhead{Ref.} &
\multicolumn{2}{c}{Data\tablenotemark{a}}& \colhead{9.7$\mum$\tablenotemark{b}}  \\ 
\cline{8-9} \colhead{} & \colhead{(K)} & \colhead{($L_{\sun}$)} &
\colhead{($M_{\sun}$)} & \colhead{(Myr)} & \colhead{(pc)} & \colhead{}&
\colhead{NIR}& \colhead{MIR} & \colhead{} } 

%% All data must appear between the \startdata and \enddata commands
\startdata
AB Aur & 9800 & 57.5 & 2.5 & 3.7 & 139 & (1) & SWS & A10 & em \\
AK Sco & 6500 & 8.9 & 1.66 & 9.3 & 103 & (1) & A06 & A10 & em \\
BD+40$\degr$4124 & 22000 & 5900 & $\ga5.99$ & $\la0.01$ & 980 & (2), (3), (4) & SWS & SWS & non \\
BF Ori & 8750 & 56 & 2.58 & 3.15 & 375 & (1) & A06 &  K12 & em \\
DoAr 21 & 5080 & 7.5 & 2.40 & 0.4 & 120 & (5), (6), (7) & SWS &  M10 & abs \\
EC 82 & 4060 & 3.21 & 0.75 & 1.5 & 415 & (8), (9) & \dots & K06 & em \\
HD 31648 & 8200 & 15.1 & 1.93 & 7.8 & 137 & (1) & SWS & A10 & em\\
HD 34282 & 8625 & 13.5 & 1.59 & 6.4 & 191 & (1) & SWS & A10 & non \\
HD 34700 & 6000 & 20.4 & 1.20 & 10 & 260 & (10) & SpX & K12 & non \\
HD 35187 & 8900 & 14.1 & 1.93 & 10.7 & 114 & (1) & SWS & A10 & em \\
HD 36112 & 7800 & 66 & 2.90 & 2.1 & 279 & (1) & \dots & A10 &em  \\
HD 36917 & 10000 & 245.5 & 3.98 & 0.72 & 375 & (1) & \dots & A10 & em \\
HD 37357 & 9250 & 52.5 & 2.48 & 3.7 & 375 & (1) & \dots & A10 & em \\
HD 37411 & 9100 & 34.4 & 1.90 & 9 & 510 & (11), (12), (13) &\dots  & A10 & non \\
HD 37806 & 11000 & 282 & 3.94 & 0.88 & 375 & (1) & \dots & A10 & em \\
HD 38120 & 11000 & 41.7 & 2.49 & 5.1 & 375 & (1) & \dots & A10 & em \\
HD 58647 & 10500 & 912 & 6.0 & 1 & 543 & (14) & \dots & A10 & em \\
HD 72106 & 11000 & 21.9 & 2.40 & 9 & 289 & (1) & \dots & A10 & em \\
HD 85567 & 20900 & 14791 & 12.0 & \dots & 1500 & (15) & \dots & A10 & em \\
HD 95881 & 9000 & 7.6 & 1.7 & $\ga3.16$ &118 & (16) & SWS & A10 & em \\
HD 97048 & 10000 & 44 & 2.53 & 4.61 &150 & (3) & SWS & A10 & non \\
HD 97300 & 10700 & 37 & 2.50 & $\ga3$ &188 & (17), (18) & PHT & K12 & non \\
HD 98922 & 10500 & 5888 & $\ga4.95$ & $\la0.01$ & 1150 & (1), (4) & G07 & A10 & em \\
HD 100453 & 7600 & 10 & 1.8 & 15 & 114 & (14) & SWS & A10 & non \\
HD 100546 & 10500 & 32 & 2.4 & $\ga10$ &103 & (16) & SWS & SWS & em\\
HD 101412 & 8600 & 83 & 3.0 & 1.2 & 600 & (19) & G07 &  A10 & em\\
HD 135344B & 6750 & 14.5 & 1.9 & 6.6 & 142 & (1) & SWS & A10 & non \\
HD 139614 & 7600 & 12.6 & 1.76 & 8.8 & 142 & (1) & SWS & A10 & em \\
HD 141569 & 9800 & 30.9 & 2.33 & 5.7 & 116 & (1) & G07 & K12& non \\
HD 142527 & 6360 & 23.58 & 2.3 & 2 & 140 & (6), (20) & SWS & A10 & em \\
HD 142666 & 7900 & 27.5 & 2.15 & 5 & 145 & (1) & SWS & A10 & em \\
HD 144432 & 7500 & 19.1 & 1.95 & 6.4 & 145 & (1) & SWS & K12 & em \\
HD 145718 & 8100 & 19.5 & 1.93 & 7.4 & 145 & (1) & \dots & K08 & em \\
HD 163296 & 9200 & 33 & 2.23 & 5.1 & 119 & (1) & SWS & SWS & em\\
HD 169142 & 8250 & 8.55 & 1.69 & 6 & 145 & (21) & SWS & A10 & non \\
HD 179218 & 9640 & 182 & 3.66 & 1.08 & 254 & (1) & SWS & SWS & em\\
HD 200775 & 18600 & 8912.5& 10.7 &  0.016 & 429 & (1) &  SWS & SWS & em \\
HD 244604 & 8200 & 55 & 2.66 & 2.79 & 375 & (1) & \dots & A10 & em \\
HD 250550 & 11000 & 138 & 3.1 & 1.42 & 700 & (15), (13) & \dots & A10 & em \\
HD 259431 & 14000 & 2239 & 7.1 & 0.059 & 660 & (1) & \dots & A10 & non \\
HD 281789 & 9520 & 10\tablenotemark{c} & \dots & \dots & 350 & (22) & \dots & K08 & em \\
IC348 LRL110 & 3778 & 0.22 & 0.78 & \dots & 250 & (23) & \dots & M10 & em  \\
IC348 LRL190 & 3306 & 0.09 & 0.35 & \dots & 250 & (23) & \dots & M10 & em\\
IRAS 03260+3111 & 13000 & $\simali$180\tablenotemark{d} & \dots & \dots & 290 & (24), (25) & PHT & PHT & non \\
IRAS 06084-0611 & 18600\tablenotemark{e} & 10\tablenotemark{c} & \dots & \dots & 1050 & (26) & SWS & SWS & non \\
J032903.9+305630 & 5630 & 0.05 & \dots & \dots & 250 & (23) & \dots & M10 & abs \\
J182858.1+001724 & 5830 & 4.76 & 1.40 & \dots & 260 & (23) & \dots & M10 & non \\
J182907.0+003838 & 4060 & 0.6 & 1.04 & \dots & 260 & (23) & \dots & M10 & em\\
LkH$\alpha$ 224 & 7850 & 115 & 6.05 & 0.312 & 980 & (2), (3) & SWS & SWS& non \\
LkH$\alpha$ 330 & 5800 & 11 & 2.50 & 3 & 250 & (5), (6), (27) & \dots & G06 & em \\
MWC 297 & 24000 & 10230 & 9.0 & 1 & 250 & (2), (3) & SWS & SWS & abs \\
MWC 865 & 14000 & 8710 & 9.0 & 1 & 400 & (28) & SWS & SWS & non \\
MWC 1080 & 30000 & 180000 & 10.0 & 1 & 2200 & (2), (3) & SWS & SWS & non \\
Oph IRS48 & 10000 & 14.3 & 2.25 & 15 & 120 & (11), (9) & G07 & M13 &non \\
PDS 144N & 8750 & 10\tablenotemark{c} & \dots & \dots & 1000 & (29) & \dots & TMM & non \\
RR Tau & 8460 & 2 & 3.57 & 1.34 & 160 & (2), (3) & PHT & A10 & non \\
RXJ1615.3-3255 & 4590 & 0.85 & 1.28 & \dots & 120 & (23) &\dots & M10& em \\
SR 21N & 5830 & 6.5 & 1.70 & 1 & 125 & (6), (30) & G07 & VSR & non \\
SU Aur & 5860 & 11 & 1.88 & 6.3 & 146 & (5), (31) & \dots & K08 & em\\
T Cha & 5250 & 1.34 & 1.10 & 7 &108 & (6), (32) & G07 & G06 & non \\
TY CrA & 12000 & 98 & 3.16 & 3 & 140 & (33) & SWS & SWS & non \\
UX Tau & 5520 & 3.5 & 1.50 & 3.0 &140 & (34), (35) & \dots & F06 & em \\
V590 Mon & 13000 & 295 & 4.71 & 2.8 & 800 & (36) & \dots & A10 &  em\\
V892 Tau & 8000 & 21 & 2.06 & 6.11 &140 & (3) & SWS & K08 & em\\
VV Ser & 14000 & 324 & 4.0 & 0.64 & 260 & (1) & G07 & A10 & em\\
VX Cas & 9500 & 60 & 2.55 & 3.4 & 620 & (1) & PHT & PHT & em \\
WL 16 & 9000 & 250 & 4.0 & 1 & 125 & (37) & PHT & PHT & non \\
Wray 15-1484 & 30000 & 977 & \dots & \dots & 750 & (2) & SWS & SWS & non \\
WW Vul & 9000 & 170 & 3.70 & 0.9 & 700 & (1) & PHT & K12 & em \\
\enddata

%% Include any \tablenotetext{key}{text}, \tablerefs{ref list},
%% or \tablecomments{text} between the \enddata and 
%% \end{deluxetable} commands
\tablenotetext{a}{The sources of the NIR ($\la4\mum$) and MIR
($\ga5\mum$) spectroscopic data used in this work.
A06: VLT/ISAAC data from Acke \& van den Ancker (2006),
A10: $Spitzer$/IRS data from Acke \etal(2010),
%c2d: $Spitzer$/IRS data from the c2d data archive
%(https://irsa.ipac.caltech.edu/data/SPITZER/C2D/, Evans \etal2003),
F06: $Spitzer$/IRS data from Furlan \etal(2006),
G06: $Spitzer$/IRS data from Geers \etal(2006),
G07: VLT/ISAAC data from Geers \etal(2007a, 2007b),
K06: $Spitzer$/IRS data from Kessler-Silacci \etal(2006),
K08: $Spitzer$/IRS data from Keller \etal(2008),
K12: $Spitzer$/IRS data from K{\'o}sp{\'a}l \etal(2012),
M10: $Spitzer$/IRS data from Mer{\'{\i}}n \etal(2010),
M13: $Spitzer$/IRS data from Maaskant \etal(2013),
PHT: $ISO$/ISOPHOT-S data from K{\'o}sp{\'a}l \etal(2012),
SpX: IRTF/SpeX data from Smith \etal(2004),
SWS: Fully processed $ISO$/SWS atlas 
(http://isc.astro.cornell.edu/$\sim$sloan/library/swsatlas/atlas.html, Sloan \etal2003),
TMM: ESO/TIMMI2 data from Sch{\"u}tz \etal(2009), and
VSR: VLT/VISIR data from Geers \etal(2007b).
}
\tablenotetext{b}{Note for the presence or absence of
the $9.7\mum$ silicate spectral feature in the disk.
If present, we label it ``em'' if it is detected in emission
or ``abs'' if in absorption. The symbol ``non'' indicates
the disk does not exhibit the 9.7$\mum$ silicate feature
either in emission or in absorption. 
}
\tablenotetext{c}{Stellar luminosity is not available in the 
literature. A luminosity of 10$\Lsun$ is arbitrarily given 
for modeling, which does not affect the model results.
}
\tablenotetext{d}{$\Lstar$ of IRAS 03260+3111 is roughly
estimated from 
$L_{\rm tot}-L_{\rm bol}$, where $L_{\rm tot}=318\Lsun$
and $L_{\rm bol}=138\Lsun$ are taken from Harvey
\etal(1984) and Connelley \etal(2008), respectively.
}
\tablenotetext{e}{$\Teff$ of IRAS 06084-0611 is not
well constrained in the literature. We adopt the
typical $\Teff$ (18,600 K) for its spectral type (B3 for IRAS 06084
VLA4; Boersma \etal2009). Unless the spectral type
significantly differs, the model results would not be
altered much.
}
\tablecomments{``\dots'' indicates that the stellar
parameters or the NIR data are not available in the literature.
}

\tablerefs{(1) Alecian \etal(2013),
(2) Acke \& van den Ancker (2004),
(3) Alonso-Albi \etal(2009),
(4) Manoj \etal(2006),
(5) Menu \etal(2015),
(6) van der Marel \etal(2016),
(7) Jensen \etal(2009),
(8) Rigliaco \etal(2015),
(9) Sturm \etal(2013),
(10) Seok \& Li (2015), 
(11) Maaskant \etal(2014),
(12) Juh{\'a}sz \etal(2010),
(13) Fairlamb \etal(2015),
(14) Mari{\~n}as \etal(2011),
(15) Verhoeff \etal(2012),
(16) van Boekel \etal(2005),
(17) Siebenmorgen \etal(2000),
(18) Hubrig \etal(2009),
(19) Folsom \etal(2012),
(20) Fukagawa \etal(2010),
(21) Seok \& Li (2016), 
(22) Keller \etal(2008),
(23) Mer{\'{\i}}n \etal(2010),
(24) Harvey \etal(1984),
(25) K{\'o}sp{\'a}l et al.(2012),
(26) Boersma \etal(2009),
(27) Brown \etal(2008),
(28) Borges Fernandes \etal(2007),
(29) Sch{\"u}tz \etal(2009),
(30) Prato \etal(2003),
(31) Jeffers \etal(2013),
(32) Hu{\'e}lamo \etal(2011),
(33) Casey \etal(1998),
(34) Pinilla \etal(2014),
(35) Magazzu \etal(1991),
(36) Liu \etal(2011),
(37) Ressler \& Barsony (2003)
}
\end{deluxetable}
%\end{document}
%%%%%%%% Table 1: Source list %%%%%%%%

\subsection{Extraction of the PAH Emission Spectrum}\label{sub:extspec}

Since we aim to mainly model the PAH emission features,
for each disk, we need to extract the PAH features from the original
data by subtracting the underlying continuum.
The data from Acke \etal(2010) 
are already continuum-subtracted,
with the dust thermal continuum 
emission fitted with a spline function.%
Similarly, Keller \etal(2008) extracted the PAH features
by fitting a spline with two sets of anchor points for the spectra
with or without silicate emission. Following these two previous
studies, we take a similar approach to the rest of the data
that we have collected to obtain continuum-subtracted spectra
(hereafter residual spectra). %, and a brief description is as following. 

Acke \etal(2010) split the IRS spectrum into two wavelength
ranges (5--7 and 7--$14\mum$) and treated them differently.
The former range, containing the $6.2\mum$ PAH feature,
has a relatively smooth continuum underneath the feature,
so it is easy to reproduce
the continuum with a spline through a few anchor points. 
We use a cubic spline through the anchor points at 5.35, 5.45,
5.58, 6.66, 7.06, 7.40, 7.55, and $7.70\mum$, adopted from
Acke \etal(2010) for the 5--$7\mum$ interval, with a slight shift
to avoid peculiar spectral features appearing in individual spectra.
The latter range, however, is not as simple as the former,
especially for the sources whose spectra are dominated
by the strong $9.7\mum$ silicate
emission (or absorption) feature.
Following Keller \etal(2008), we use two sets of anchor points
for the spectra dominated by PAH bands (PAH-dominated sources)
and those dominated by the $9.7\mum$ silicate emission or
absorption band. For PAH-dominated sources, we adopt
anchor points at 5.55, 5.80, 6.7, 7.0, 9.15, 9.45, 9.7,
10.2, 10.7, 11.8, 12.2, 13.05, 13.25, 13.8, 14.0, and $14.8\mum$.
Like the 5--$7\mum$ interval, we slightly adjust the anchor points,
but for most cases, the variations are marginal. 
For the silicate emission (or absorption) sources,
we use anchor points at 7.35, 8.95, and $9.35\mum$
instead of those at 9.15 and $9.45\mum$
while keeping other points the same as those for
PAH-dominated sources. As pointed out by Acke \etal(2010),
spline fits result in considerable extra residuals 
(especially between 7--$9\mum$) for those with strong
silicate emission and weak PAH features. Also, some
objects (\eg BF Ori, EC 82, HD 144432, and WW Vul)
show a crystalline silicate feature at 11.2--$11.3\mum$,
which overlaps with the $11.2\mum$ PAH feature. We subtract
this feature by fitting the blended $11.3\mum$ feature with
two Drude profiles.

A continuum underlying the PAH emission feature
at $3.3 \mum$ does not show a strong curvature
for most cases, which is dominated by starlight with a
possible contribution from a hot dust component.
Since a minor feature at $3.43\mum$,
generally attributed to the C--H vibrational modes in
aliphatic hydrocarbons (\eg Chiar \etal2000; 
Pendleton \& Allamandola 2002),
has a broad emission plateau blended with the $3.3\mum$
PAH feature, we use the short- and
long-wavelength sides of the 3.3 and $3.43\mum$ features
to fit the continuum using a linear function. 
We consider a wavelength range of 3.05--3.15 and
3.6--$3.8\mum$, respectively, for the short- and
long-wavelength sides. If HI recombination
lines such as Pf$\gamma$ at $3.741\mum$ are strong,
we exclude them from the fit. 
For a few sources (Oph IRS48, SR 21N, WL16, and V892 Tau),
we use a second or fourth degree polynomial function
as their NIR spectra seem to have a recognizable curvature. 

There is one caveat for continuum-subtraction using a spline:
it essentially forces the baselines between the PAH features in the
residual spectrum to be zero, which might oversubtract the continuum.
Since PAH emission features, which can be described
by a combination of Drude profiles, have a considerably wide
wing, even a spectrum consisting of PAH emission features only
does not have a zero-level baseline.  
To investigate this effect of subtraction of a spline fit,
we use the PAHFIT decomposition tool (Smith \etal2007),
which fits simultaneously the PAH features, dust and stellar continuum,
and atomic and molecular emission lines, to fit the spectra
of selected objects in our sample. Since PAHFIT uses blended
Drude profiles to fit the PAH features, it results in non-zero baselines
between the features. Although the residual spectra
from the spline fit and PAHFIT result in some differences in the spectra,
we find these differences do not alter our main result significantly
(see Section \ref{sub:cntsub}). 
Thus, we keep the spline fit to extract the residual spectra for our analysis,
and detailed comparisons will be discussed in Section \ref{sub:cntsub}.
 
\begin{figure}
\epsscale{1.0}
\plotone{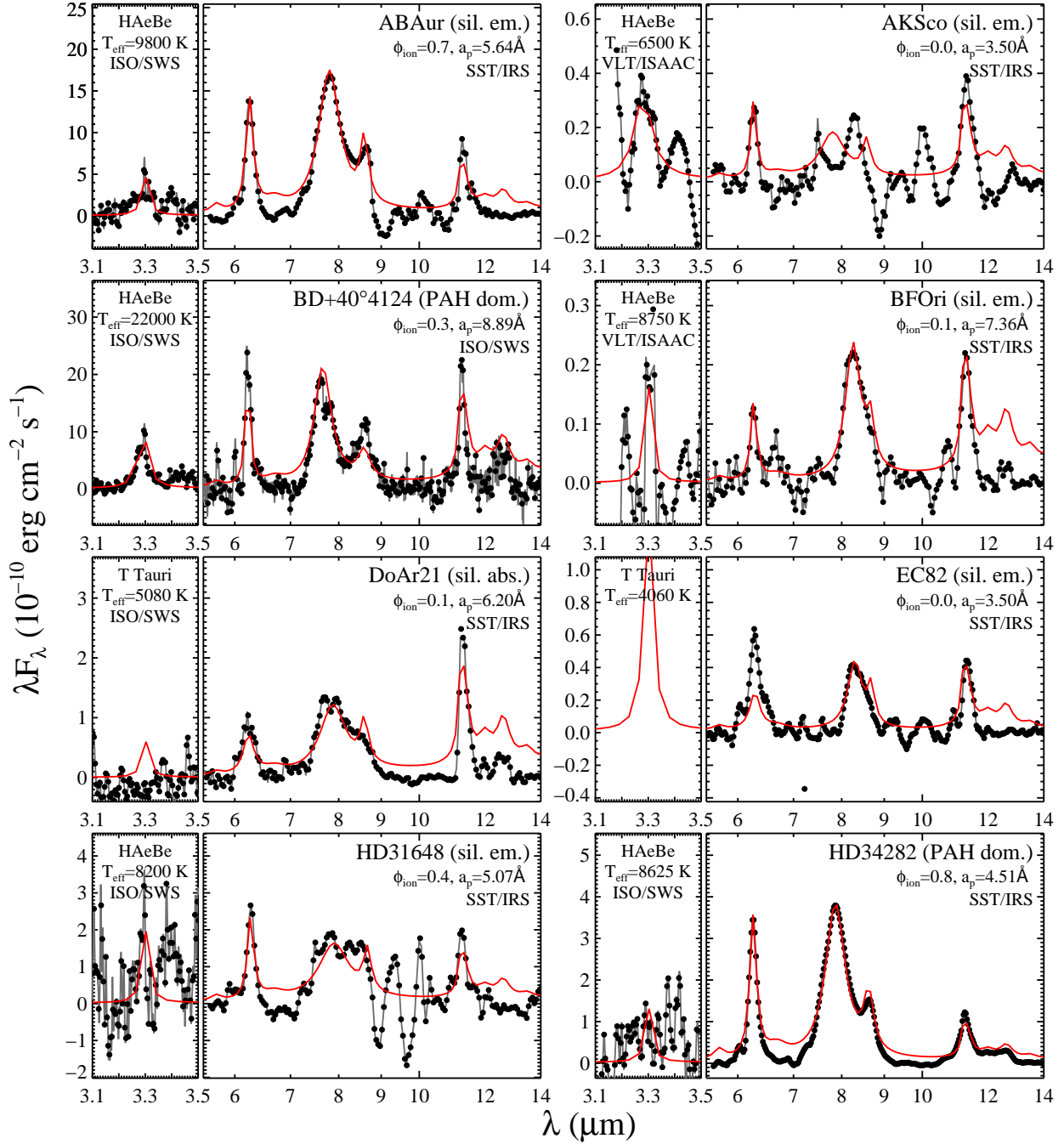}
\caption{\footnotesize\label{fig:spec}
         Comparison of the best-fit model spectra (red lines)
         with the residual spectra (\ie continuum-subtracted spectra;
         grey lines) of AB Aur, AK Sco, BD+$40\degr$ 4124,
         BF Ori, DoAr 21, EC 82, HD 31648, and HD 34282.
         The interpolated data points actually used for model
          calculations are overlaid with black dots. The best-fit model
          parameters ($\phi_{\rm ion}$ and $a_{\rm p}$) are
          given in the upper-right side of each panel. 
         The label ``PAH dom.'' specifies that the IR spectrum of
         the source is dominated by PAH features, while the
         label ``sil. em.'' or ``sil. abs.'' denotes that
         the source emits or absorbs strongly at the 9.7$\mum$
         silicate feature, respectively. For each source, we also
         specify the nature of the central star (\ie HAeBe
         or T Tauri), the stellar effective temperature ($\Teff$),
         and the instruments used to obtain the observed spectrum
         (\eg $ISO$/SWS, $Spitzer$/IRS [shortened as SST/IRS],
         VLT/ISAAC, etc.). 
%         Instruments that the NIR and MIR spectra are
%          obtained with are marked. 
          }
\end{figure}

\begin{figure}
%\ContinuedFloat
\plotone{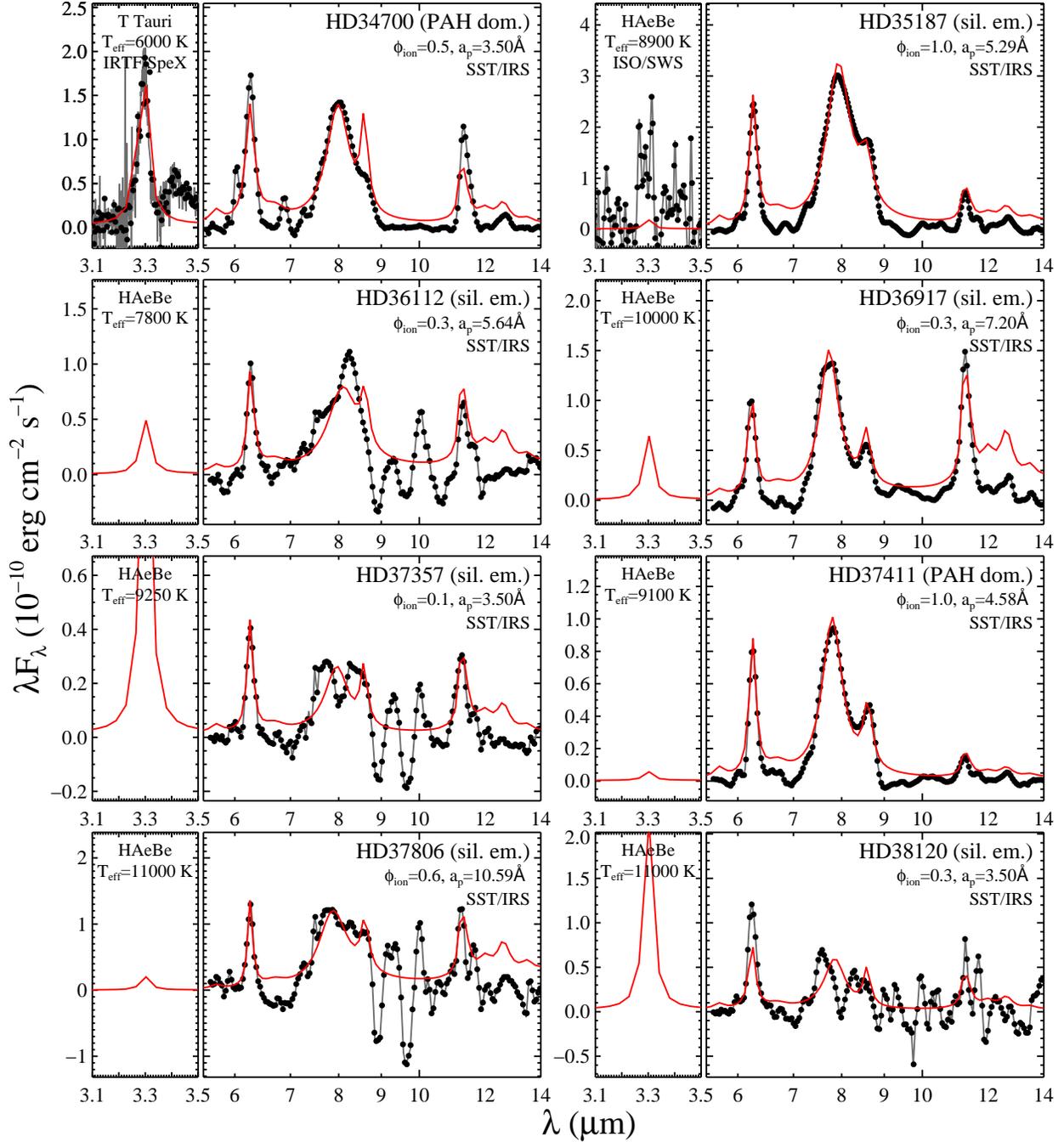}
\caption{\label{fig:spec2}
         Same as Figure \ref{fig:spec} but for HD 34700, HD 35187,
         HD 36112, HD 36917, HD 37357, HD 37411, HD 37806,
         and HD 38120. 
          }
\end{figure}

\begin{figure}
%\ContinuedFloat
\plotone{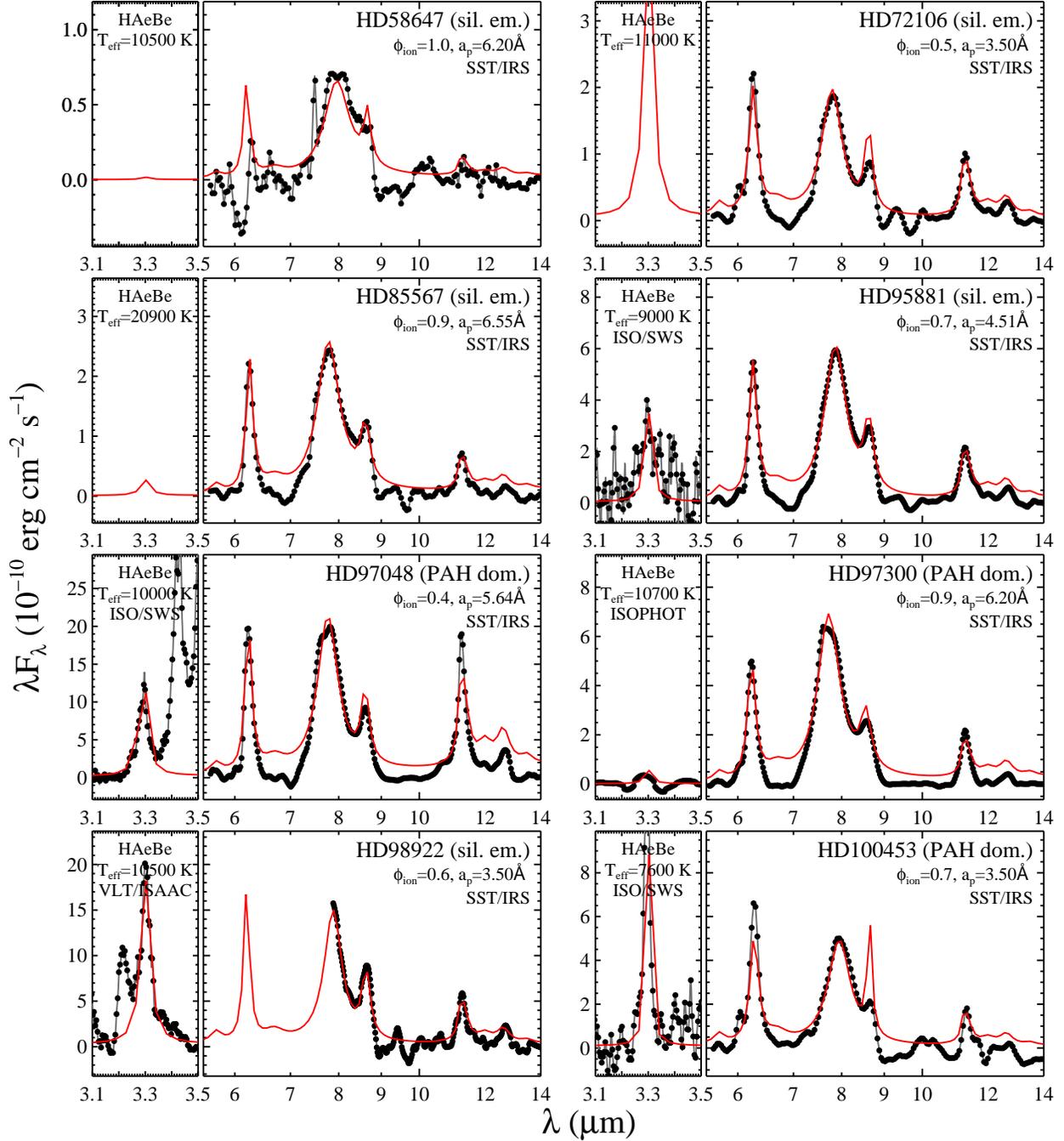}
\caption{\label{fig:spec3}
         Same as Figure \ref{fig:spec} but for HD 58647, HD 72106,
         HD 85567, HD 95881, HD 97048, HD 97300, HD 98922,
         and HD 100453.
          }
\end{figure}

\begin{figure}
%\ContinuedFloat
\plotone{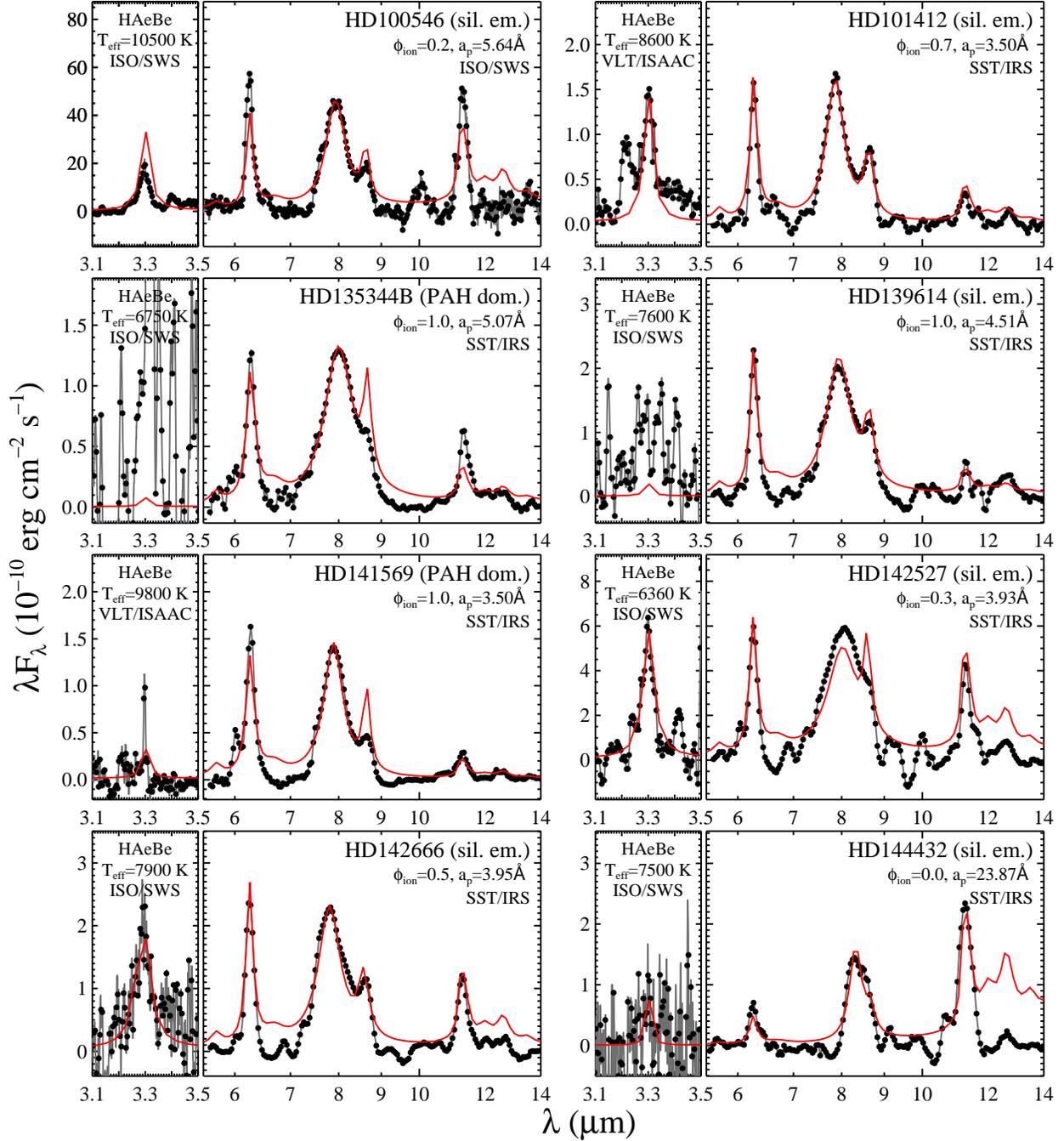}
\caption{\label{fig:spec4}
         Same as Figure \ref{fig:spec} but for HD 100546, 
         HD 101412, HD 135344B, HD 139614, HD 141569,
         HD 142527, HD 142666, and HD 144432.
          }
\end{figure}

\begin{figure}
%\ContinuedFloat
\plotone{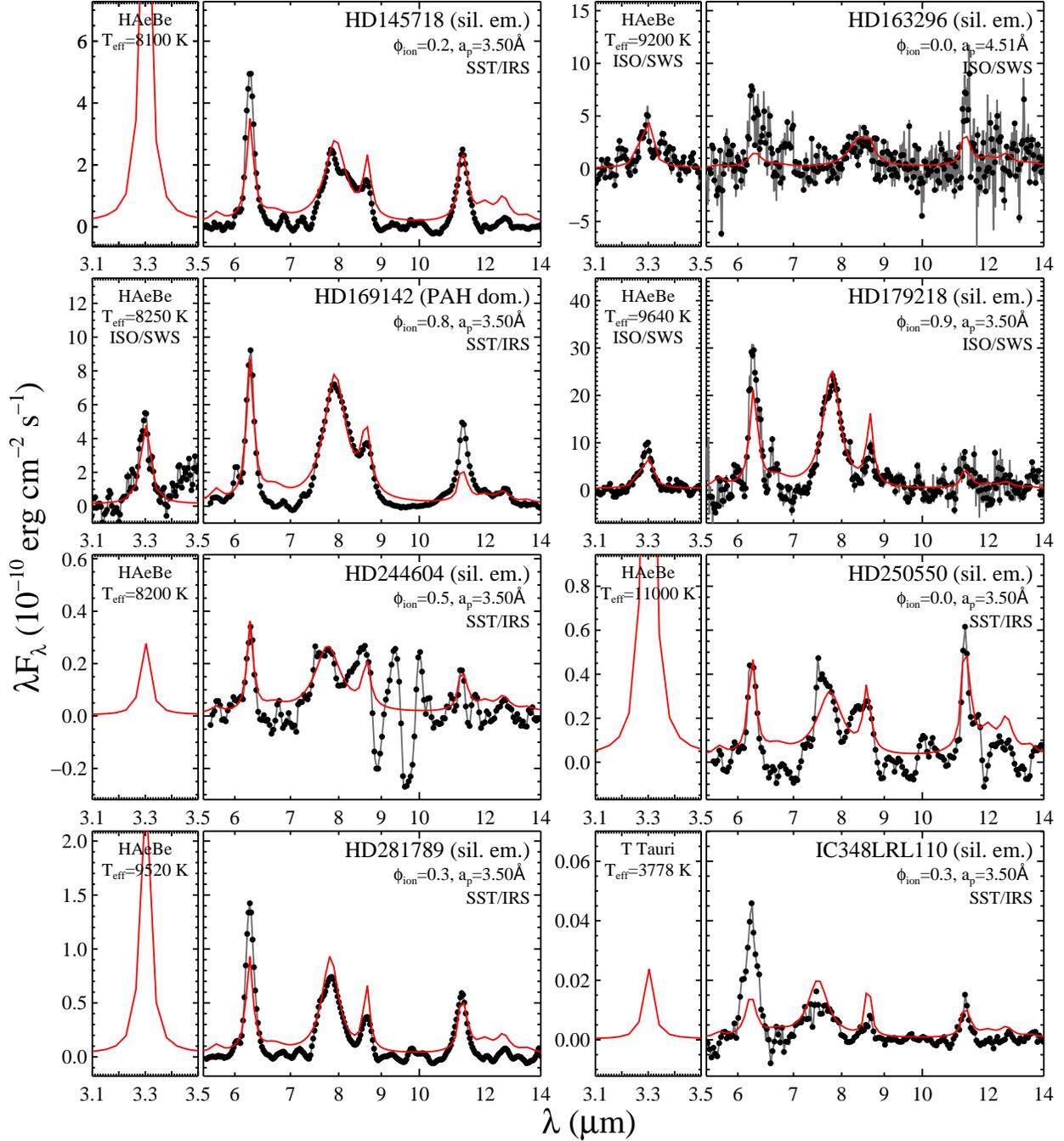}
\caption{\label{fig:spec5}
         Same as Figure \ref{fig:spec} but for HD 145718, 
         HD 163296, HD 169142, HD 179218, HD 244604,
         HD 250550, HD 281789, and IC348 LRL110.          }
\end{figure}

\begin{figure}
%\ContinuedFloat
\plotone{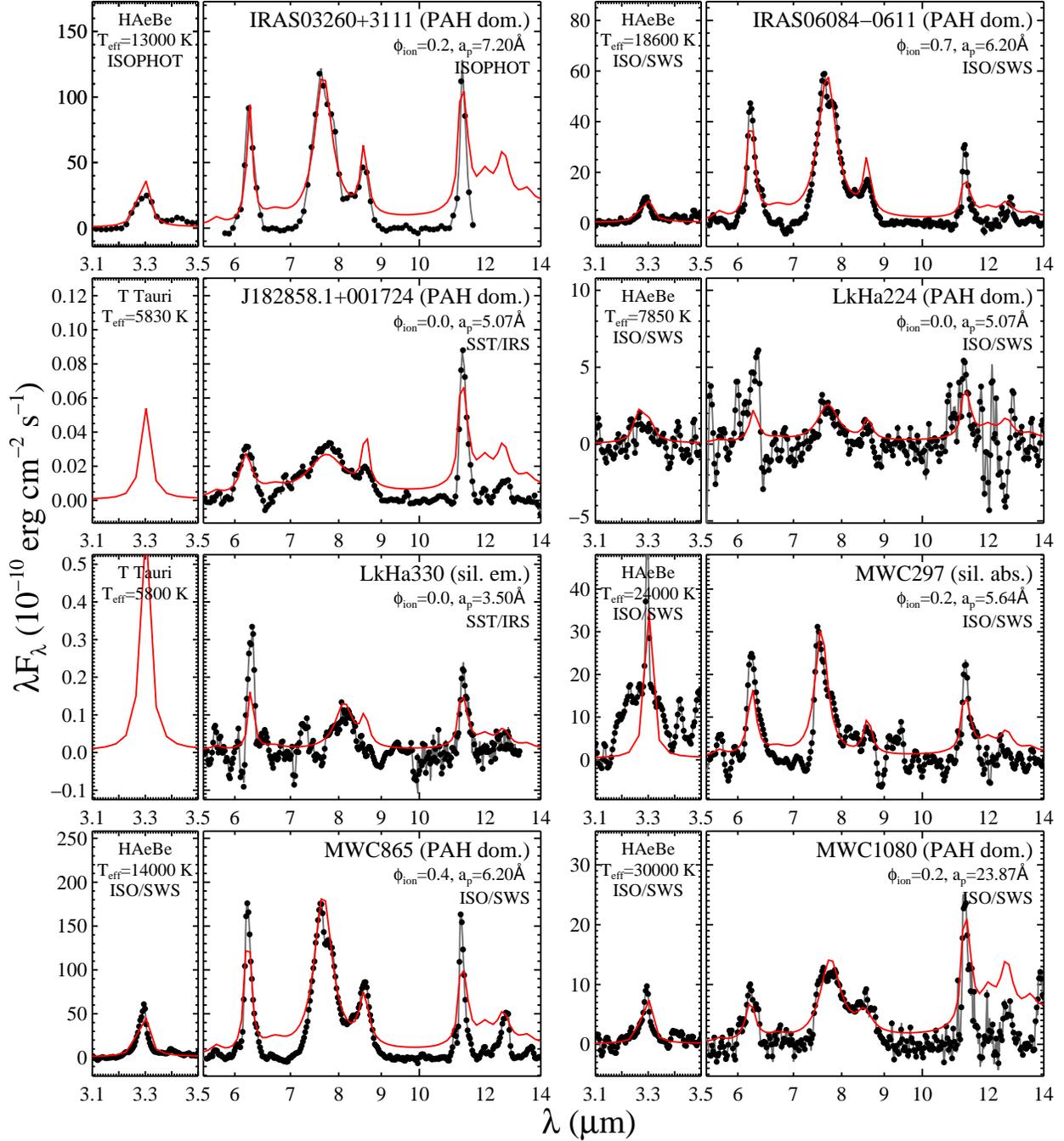}
\caption{\label{fig:spec6}
         Same as Figure \ref{fig:spec} but for IRAS 03260+3111,
         IRAS 06084-0611, J182858.1+001724, LkH$\alpha$ 224,
         LkH$\alpha$ 330, MWC 297, MWC 865, and MWC 1080.
          }
\end{figure}

\begin{figure}
%\ContinuedFloat
\plotone{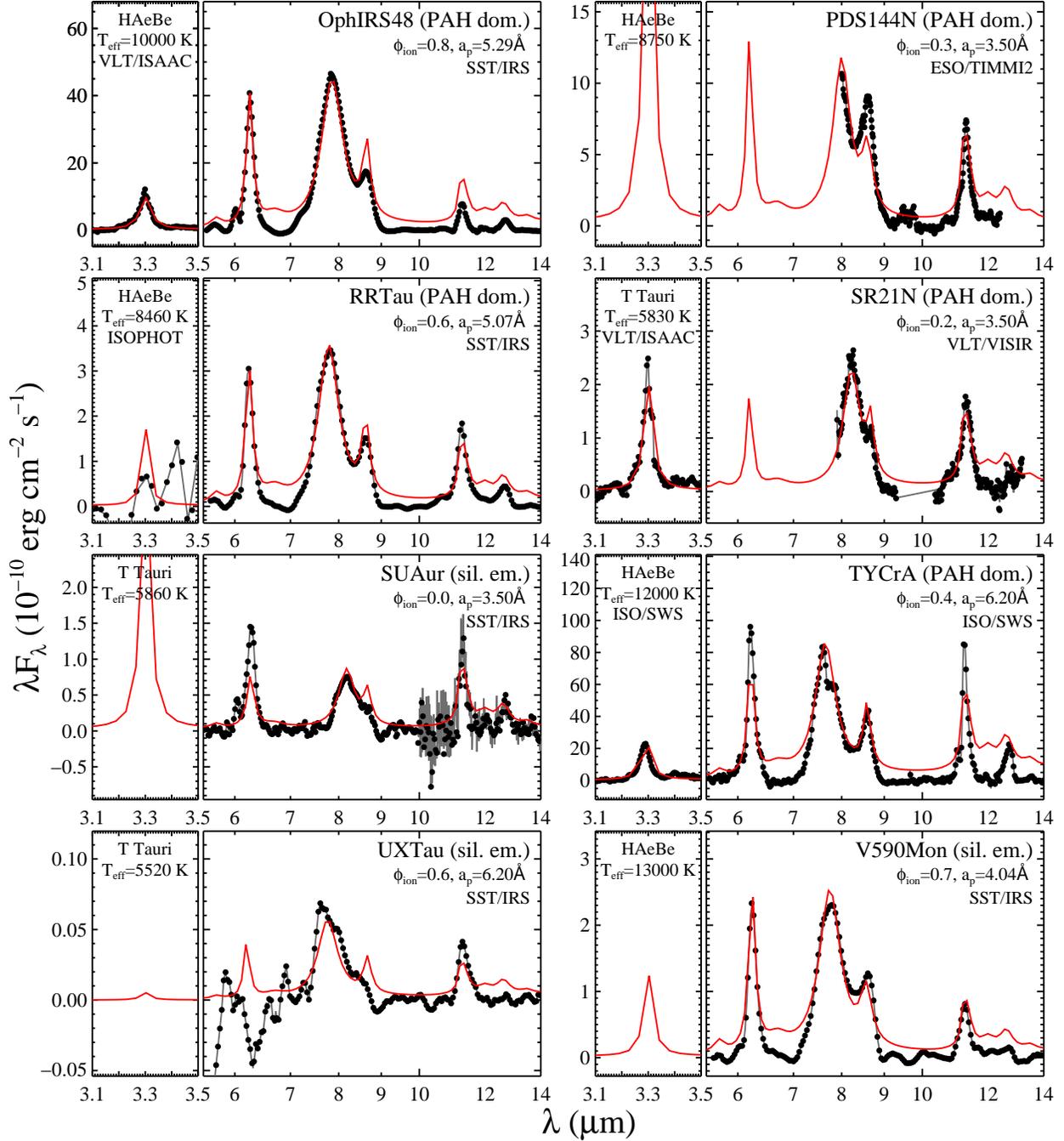}
\caption{\label{fig:spec7}
         Same as Figure \ref{fig:spec} but for Oph IRS48,
         PDS 144N, RR Tau, SR 21N, SU Aur, TY CrA,
         UX Tau, and V590 Mon.
          }
\end{figure}

\begin{figure}
%\ContinuedFloat
%\plotone{model_best8}
\epsscale{0.6}
\includegraphics[trim={0 5.3cm 0 0},clip,width=16.5cm]{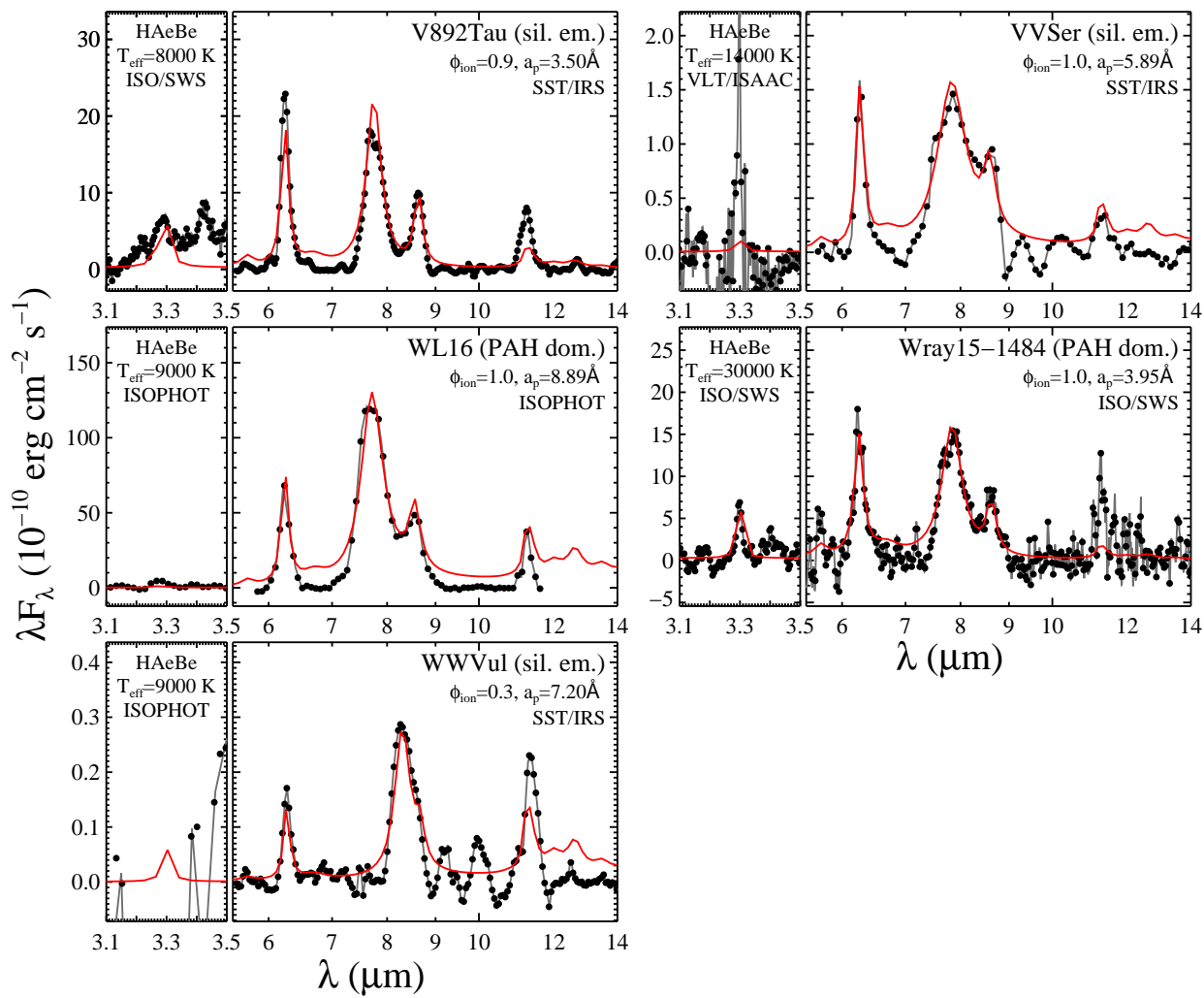}
\caption{\label{fig:spec8}
          Same as Figure \ref{fig:spec} but for V892 Tau, VV Ser,
          WL 16, Wray 15-1484, and WW Vul.
          }
 \end{figure}

\begin{figure}
\epsscale{1}
\plotone{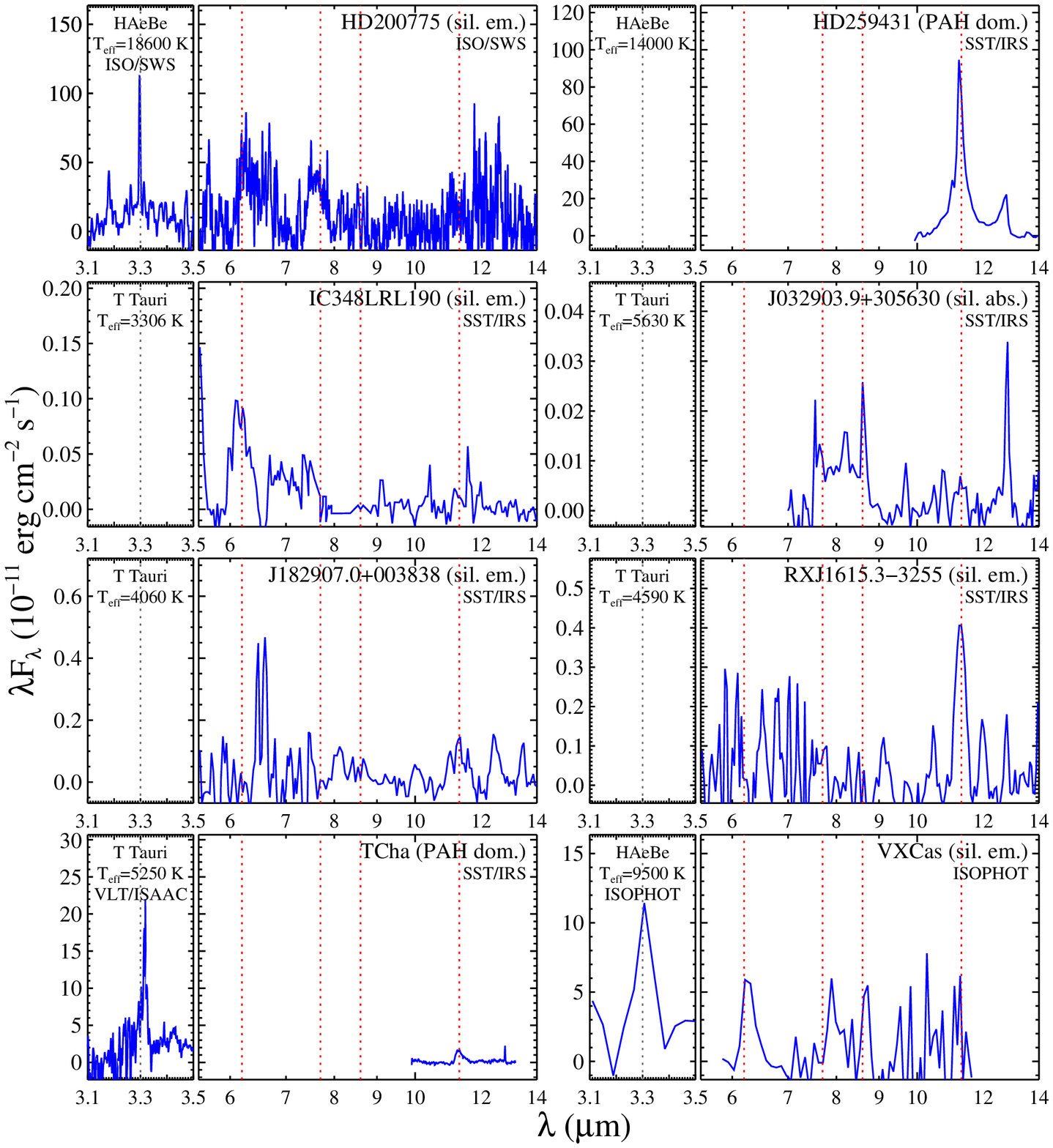}
\caption{\footnotesize
         \label{fig:nomod}
        Residual spectra of the eight sources (blue lines)
        that are excluded for PAH modeling
        due to either a limited wavelength coverage
        or an insufficient S/N (see Section \ref{sub:extspec}).
        As a reference, the peak wavelengths of the major PAH features
        (\ie 3.3, 6.2, 7.7, 8.6, and $11.2\mum$) are marked (dotted lines).
          }
 \end{figure}

The residual spectra of the 69 sources are 
shown in Figures \ref{fig:spec}--\ref{fig:nomod}. The eight
sources in Figure \ref{fig:nomod} are those excluded for further
model calculations. The original spectra of these sources 
either have the limited wavelength coverage
(\eg HD 259431 and T Cha, see Section \ref{sub:dataset}) 
or low signal-to-noise ratios (S/N),
so their residual spectra are inadequate
to perform model calculations.  

%%%%%%%%%%%%% Figure 1-8: %%%%%%%%%%%%%%

%%%%%%%%%%%%% Figure 1-8: %%%%%%%%%%%%%%

%%%%%%%%%%%%% Figure 9 %%%%%%%%%%%%
%%%%%%%%%%%%% Figure 9 %%%%%%%%%%%%

\section{Model}\label{sec:model}

\begin{figure}[tbp]
\epsscale{0.8}
\plotone{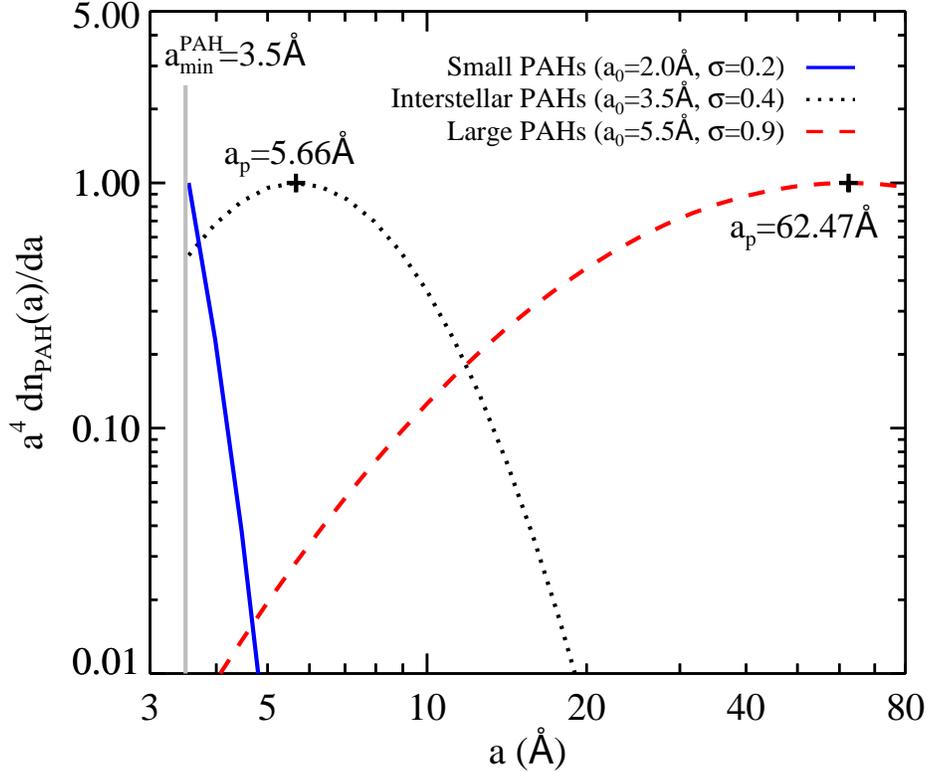}
\caption{\footnotesize
         \label{fig:dnda}
          Log-normal size distribution ($dn_{\rm PAH}/da$) of PAHs,
          which is characterized by $a_0$ and $\sigma$ (Equation
          \ref{eq:dnda}). We plot $a^4 dn_{\rm PAH}/da$
          (normalized to its peak value) to show
          the mass distribution per logarithmic PAH radius, which
          peaks at $a_{\rm p}=a_0$exp$(3\sigma^2)$. For illustration,
          we show three different distributions: small PAHs
          ($a_0=2.0\Angstrom$ and $\sigma=0.2$), PAHs in
          the diffuse ISM ($a_0=3.5\Angstrom$ and $\sigma=0.4$;
          Li \& Draine 2001b), and large PAHs ($a_0=5.5\Angstrom$
          and $\sigma=0.9$). We mark the peak sizes for the
          diffuse ISM PAH mass distribution ($a_{\rm p}\approx5.66\Angstrom$) 
          and the distribution of the large PAHs ($\ap\approx62.47\Angstrom$). 
          For the distribution of the small PAHs,
          however, $a_{\rm p}$ is not calculated because the 
          derived $a_{\rm p}$ ($\approx2.25\Angstrom$) 
          is smaller than $\apahmin$ ($=3.5\Angstrom$), and we set
          $a_{\rm p}=\apahmin$ in such a case.
          }
\end{figure}

To derive the PAH properties from the residual spectra,
we adopt the astro-PAH model of Li \& Draine (2001b)
and Draine \& Li (2007). We assume 
%that the size of PAHs ($a$) follows 
a log-normal distribution function ($dn_{\rm PAH}/da$)
for the PAH size distribution,
\beq\label{eq:dnda}
dn_{\rm PAH}/da = 
\frac{1}{\sqrt{\pi/2}\,\sigma
        \left\{ 1 - {\rm erf}\left[
        \ln\left(a^{\rm PAH}_{\rm min}/a_0\right)/\sqrt{2}\sigma
        \right] \right\}}     
        \frac{1}{a} \exp \left\{ - \frac{1}{2} 
       \left[ \frac{\ln (a/a_{0})}{\sigma} \right]^2 \right\},
       ~~{\rm for}\  a > \apahmin ~,
\eeq
where $a_0$ and $\sigma$ are, respectively, the peak and the
width of the log-normal distribution, and $\apahmin$ is
the lower cutoff of the PAH size. We set
$\apahmin\equiv3.5\Angstrom$, equivalent to a PAH
molecule containing $\simali20$ carbon atoms (\ie
$N_{\rm C}\approx20$), which is the minimum size of
PAHs required to survive in the diffuse interstellar medium
(ISM; see Li \& Draine 2001b). Figure \ref{fig:dnda} shows
the PAH size distribution expressed by multiplying $a^4$
to show the mass distribution of PAHs.
For this distribution, $a^4dn/da$ peaks at 
$a_{\rm p}=a_0$exp(${3\sigma^2}$) (Li \& Draine 2001a).
Since $a^4dn/da$ is truncated at $\apahmin=3.5\Angstrom$,
we set $a_{\rm p}=\apahmin=3.5\Angstrom$ 
when $a_0\exp(3\sigma^2)<\apahmin$.
%which the peak of the mass distribution is actually located at. 

%%%%%%%%%%%%% Figure 10 %%%%%%%%%%%%%%
%%%%%%%%%%%%% Figure 10 %%%%%%%%%%%%%%

We adopt the PAH absorption
cross-sections that are expressed as a series of Drude profiles
(Li \& Draine 2001b; Draine \& Li 2007). It has been noticed
that the feature profiles of PAH emission vary from one object to
another (\eg Peeters \etal2002; Sloan \etal2007), and the PAH
emission features seen in PPDs often show deviations
from those of the diffuse ISM (\eg Keller \etal2008).
For each source, the peak wavelengths and widths of some of
the Drude profiles are slightly modified to be consistent
with the observed spectrum.
%taking into account the peak wavelengths and widths
%of PAH features measured from the residual spectrum. 
In this way,
the observed residual spectra can be fitted more closely
(\eg see Seok \& Li 2015, 2016). Note that the integrated area
of each Drude profile is kept unchanged so that the model
calculations are coherent through the entire sample. 
We distinguish the absorption cross-sections of neutral PAHs
from those of charged PAHs, but neither cations from anions
nor multiply charged PAHs are considered separately.
The absorption cross sections of PAH cations and anions
are rather close (see Li \& Draine 2001b) since the IR
properties of PAH anions closely resemble those of
PAH cations except for the very strong 3.3$\mum$
C-H stretch enhancement in the anion 
(\eg Szczepanski \etal1995; Langhoff 1996; Hudgins \etal2000).

%Charge STATe
The charging of PAHs is determined by two competing
mechanisms, photoionization and electron
recombination (Bake \& Tielens 1994; Weingartner \&
Draine 2001). We quantify the charge state of PAHs in our
sample disks by using the ionization fraction ($\phi_{\rm ion}$),
which is the probability of finding a PAH molecule in a 
nonzero charge state (Li \& Lunine 2003). 
The ionization fraction $\phi_{\rm ion}$
is determined by the PAH size ($a$), the UV starlight intensity
($U$), the electron density ($n_{\rm e}$), and the gas
temperature ($T_{\rm gas}$). We define $U$ as
the intensity of the stellar radiation between $912\Angstrom$
and 1$\mum$ with respect to that of the local interstellar
radiation field (ISRF) of Mathis \etal(1983; MMP),
\beq
 U(r)=\frac{(R_*/2r)^2\int^{1\mum}_{912\Angstrom}F^*_{\lambda}d\lambda}
                 {\int^{1\mum}_{912\Angstrom}cu^{\rm ISRF}_\lambda d\lambda},
\eeq
where $R_*$ is the stellar radius, $F^*_\lambda$ is the flux
per unit wavelength (in units of ergs $\s^{-1}\cm^{-2}\mum^{-1}$)
at the top of the central star's atmosphere, which is 
approximated by the Kurucz model atmospheric spectrum
(Kurucz 1979), $c$ is the speed
of the light, and $u^{\rm ISRF}_{\lambda}$ is the energy
density of the MMP interstellar radiation field. To calculate
$\phi_{\rm ion}$ precisely, it is required to take into account
the spatial (radial) variations of $U$ and $n_{\rm e}$, which
are not always available in our sample. Thus, we
assume a constant $\phi_{\rm ion}$ for all PAH sizes, which
can be a representative charge state of PAHs in the disk.

Because of their small heat contents 
which are often smaller than the energy
of a single stellar UV photon, PAHs will
not acquire an equilibrium temperature, and
instead, they will undergo transient heating
and experience temperature fluctuations (see Li 2004). 
Because of this, we calculate the temperature 
probability distribution functions of PAHs of 
a given size $a$ and a given charge state 
(i.e., neutral or ionized) exposed to starlight of 
a given intensity $U$ and a given spectral shape.
The spectral shape of the UV/optical starlight
is approximated by the model atmospheric spectrum 
of Kurucz (1979), characterized by $T_{\rm eff}$, 
the effective temperature of the star. 
For this purpose, we employ the ``thermal-discrete'' 
method which treats both heating (absorption of UV/optical 
stellar photons) and cooling (emission of IR photons) 
as discrete transitions, using a thermal approximation 
for the downward transition probabilities (Draine \& Li 2001). 
Then, we calculate the IR emissivity
for a mixture of neutral and ionized PAHs 
with a log-normal size distribution.
Since PAHs have the single-photon heating nature,
the spectral appearance of PAH emission is
independent of the starlight intensity $U$, so the absolute
emissivity level simply scales with $U$ (Draine \& Li 2001;
Li \& Draine 2001b). Therefore, the total mass of PAHs
($M_{\rm PAH}$) required to account for the observed
PAH emission bands in the disk is inversely proportional
to $U$ (see Li \& Lunine 2003).

In summary, there are three key free parameters in modeling
the residual spectra: $a_0$ and $\sigma$ for the PAH size
distribution and $\phi_{\rm ion}$ for the charge state.
We vary $a_0$ from 2.0 to 5.5 $\Angstrom$ with a step of
$0.5\Angstrom$, $\sigma$ from 0.2 to 0.9 with a step of
0.1, and $\phi_{\rm ion}$ from 0 to 1 with a step of 0.1.
For a given set of the PAH parameters, the total mass of PAHs
($M_{\rm PAH}$) is determined by the absolute level of the
observed PAH flux density.
For each source, 704 ($=8\times8\times11$) models are
calculated according to the stellar parameters of each source
(\eg $\Lstar$ and $\Teff$ listed in Table \ref{tab:obj}),
and we search for the best-fit model that reproduces
the residual spectrum most closely. 
 
\section{Results}\label{sec:res}

Among the 704 model calculations of each source, we select the
best-fit model based on the least-$\chi^2/$dof value. 
The best-fit models of the 61 sources are
shown with their residual spectra in Figures
\ref{fig:spec}--\ref{fig:spec8}, and the PAH parameters
($a_0$, $\sigma$, and $\phi_{\rm ion}$) of the best-fit models
are listed in Table \ref{tab:para}.
We also derive the peak of the PAH mass distribution
(per logarithmic PAH radius), 
$a_{\rm p}$, from $a_0$ and $\sigma$ (see Figure
\ref{fig:dnda} and Section \ref{sec:model}). The peak
size $a_{\rm p}$ is beneficial to
describe and compare the overall PAH size distributions 
since $a_0$ or $\sigma$ cannot solely represent the
log-normal size distribution. Instead, the peak size is
determined by the combination of these two parameters. 

In addition to the best-fit model, we select
the top 70 models of each source 
(\ie $\simali10\%$ of the total 704 model calculations), 
which in fact reproduce the residual spectra
reasonably well, and calculate the median value
(\eg $\langle\ap\rangle$ and $\langle\phiion\rangle$) 
and the median absolute deviation (MAD; 
\eg $\Delta\ap$ and $\Delta\phiion$) 
for each parameter (see Table \ref{tab:para}). The MAD is defined as 
$\Delta X=\langle |X_i-\langle X_i\rangle |\rangle$,
where $X_i$ is a univariate data set. The MAD is 
less sensitive to the outliers in a data set than the 
standard deviation, so it is considered as a robust
estimator of the spread in the top 70 models for each parameter.
We adopt the MAD of the top 70 models as
uncertainties for each parameter.
By taking the median and MAD for each parameter,
we evaluate the robustness of the PAH parameters and
look for the correlation of PAH properties with stellar
parameters. %(see Figure \ref{fig:corr_teff}--\ref{fig:corr_age}).

Finally, we examine the reliability of the model-fit parameters
by taking into account both $\chi^2/$dof and visual inspection.
The reason why we perform
the visual inspection supplementally is that,
for some sources their $\chi^2/$dof values 
may be low (which means that the best-fit model 
successfully reproduces the residual spectra),
however, the residual spectra themselves,
in particular of those sources 
with weak PAH features superimposed on 
the strong 9.7$\mum$ silicate emission feature, 
may not be very reliable, probably due to the imperfect 
continuum-subtraction (\eg Acke \etal2010).
In these cases, the derivation of the PAH features 
at $\simali$7--9$\mum$ from the observed spectra
is most likely to be affected, which is critical for 
constraining the model parameters. 
As a result, the best-fit parameters we derive 
might deviate from the true properties of 
the PAH molecules in the disks.
Therefore, we shall distinguish these sources
from those whose residual spectra are highly 
reliable for further analysis.
We mainly use two criteria for 
determining the reliabilities of the model fits:
(i) $\chi^2/{\rm dof}\la3.0$, and 
(ii) the absence of any appreciable
extra residual in the observed spectrum 
except the PAH features 
after subtracting the 9.7$\mum$ 
silicate feature  and the dust thermal 
emission continuum from the observed spectrum.
It turns out that, for most of the sources which are
classified as with low reliabilities, 
their $\chi^2/$dof values are  greater than 3.0; 
but for some sources, even though they have 
low $\chi^2/$dof values (\eg HD 37806, HD 58647),
they are still classified as with low reliabilities 
due to the presence of extra residual,
in addition to the PAH features,
in the observed spectra after subtracting
the dust thermal continuum emission
and the 9.7$\mum$ silicate feature.
We note that there are also some exceptions 
due to an incomplete wavelength coverage 
of the data sets (e.g., HD\,98922 and PDS\,144N 
are considered as having a low reliability 
since their data are only available 
at $\lambda\ga8\um$,
in spite of their low $\chi^2/$dof values 
and the absence of any extra residual 
in the their subtracted spectra).
The reliability of each source is listed 
in Table \ref{tab:para}. 

We classify the sample targets into two groups: 
PAH-dominated spectra (``PAH dom.'') and those with
a strong silicate emission (``sil. em.'') or absorption
(``sil. abs'') feature at 9.7$\mum$.
When the PAH parameters are compared with the stellar
parameters, we examine if the two groups show a different
trend and calculate the correlation coefficients both separately
and combined (see Figure \ref{fig:corr_teff}--\ref{fig:corr_peak}).
Note that we only consider those sources with reliable, high 
S/N residual spectra when we examine the correlations. 

Using the best-fit model, we examine the total mass
of the PAHs ($M_{\rm PAH}$) in PPDs. As $M_{\rm PAH}$
depends on the starlight intensity ($U$), which is a
function of distance, $r$, from the central star
(\ie $M_{\rm PAH}\propto1/U$ and $U\propto1/r^2$),
we cannot constrain $M_{\rm PAH}$ exactly unless
the spatial distribution of PAHs in the disk is properly taken
into account. Instead, we derive $M_{\rm PAH}$ at $10\AU$
($M^{10\AU}_{\rm PAH}$) assuming that all PAHs
are located at $10\AU$ from the central star.
Then, if the spatial distribution of PAHs is revealed,
we can easily calculate $M_{\rm PAH}$ by scaling with $r$
(\ie $M_{\rm PAH}=M^{10\AU}_{\rm PAH}\left[r/10\AU\right]^2$).
$M^{10\AU}_{\rm PAH}$ of each disk is listed in
Table \ref{tab:para}.

%%%%%%%% Table 2: Model Parameters  %%%%%%%%%
%% The values (usually only l,r and c) in the last part of
%% \begin{deluxetable}{} command tell LaTeX how many columns
%% there are and how to align them.
\begin{deluxetable}{lccccccccccc}
\center
%% Keep a portrait orientation
\rotate
%% Over-ride the default font size
%% Use Default (12pt)
\tablewidth{590pt}
%% Use \tablewidth{?pt} to over-ride the default table width.
%% If you are unhappy with the default look at the end of the
%% *.log file to see what the default was set at before adjusting
%% this value.

%% This is the title of the table.
\tablecaption{\label{tab:para}
PAH Parameters of 61 PPDs as Derived from 
the Astro-PAH Model Calculations 
}
\footnotesize

%% This command over-rides LaTeX's natural table count
%% and replaces it with this number.  LaTeX will increment 
%% all other tables after this table based on this number
%\tablenum{1}

%% The \tablehead gives provides the column headers.  It
%% is currently set up so that the column labels are on the
%% top line and the units surrounded by ()s are in the 
%% bottom line.  You may add more header information by writing
%% another line between these lines. For each column that requries
%% extra information be sure to include a \colhead{text} command
%% and remember to end any extra lines with \\ and include the 
%% correct number of &s.
\tablehead{\colhead{Object} & \multicolumn{5}{c}{Best-fit} & \colhead{} &
 \multicolumn{4}{c}{Median of Top 70 Models\tablenotemark{a}} & 
 \colhead{Reliability\tablenotemark{b}} \\ 
\cline{2-6} \cline{8-11}\colhead{} & \colhead{$a_0$} & \colhead{$\sigma$} &
\colhead{$a_{\rm p}$} & \colhead{$\phi_{\rm ion}$} & 
\colhead{$M^{10\AU}_{\rm PAH}$\tablenotemark{c}} & \colhead{} & 
\colhead{$\langle a_0\rangle$} & \colhead{$\langle \sigma\rangle$} & 
\colhead{$\langle a_{\rm p}\rangle$} & \colhead{$\langle \phi_{\rm ion}\rangle$} &
\colhead{}\\
\colhead{} & \colhead{$(\Angstrom)$} & \colhead{} & \colhead{$(\Angstrom)$} &
\colhead{} & \colhead{$(10^{-6}\Mearth)$} & \colhead{} & \colhead{$(\Angstrom)$} &
\colhead{} & \colhead{$(\Angstrom)$} & \colhead{} & \colhead{} 
} 

%% All data must appear between the \startdata and \enddata commands
\startdata
AB Aur & 5.0 & 0.2 & 5.64 & 0.7 & 1.96 & &4.0$\pm$1.0 & 0.4$\pm$0.1 & 5.64$\pm$0.56 & 0.8$\pm$0.1 & h \\
AK Sco & 2.5 & 0.3 & 3.50 & 0.0 & 0.56 & &3.0$\pm$0.5 & 0.3$\pm$0.1 & 3.95$\pm$0.45 & 0.1$\pm$0.1 & l \\
BD+40$\degr$4124 & 5.5 & 0.4 & 8.89 & 0.3 & 1.01 & & 4.0$\pm$1.0 & 0.6$\pm$0.1 & 10.59$\pm$3.18 & 0.4$\pm$0.1 & h \\
BF Ori & 2.5 & 0.6 & 7.36 & 0.1 & 0.44&& 4.0$\pm$1.0 & 0.5$\pm$0.1 & 8.08$\pm$1.73 & 0.1$\pm$0.1 & l \\
DoAr21 & 5.5 & 0.2 & 6.20 & 0.1 & 12.4 && 4.0$\pm$1.0 & 0.6$\pm$0.1 & 10.31$\pm$2.94 & 0.3$\pm$0.1 & l \\
EC82 & 2.0 & 0.2 & 3.50 & 0.0 & 540 && 2.5$\pm$0.5 & 0.2$\pm$0.0 & 3.50$\pm$0.00 & 0.2$\pm$0.2 & l \\
HD 31648 & 4.5 & 0.2 & 5.07 & 0.4 & 1.56 && 3.5$\pm$1.0 & 0.3$\pm$0.1 & 5.07$\pm$0.82 & 0.5$\pm$0.1 & l \\
HD 34282 & 4.0 & 0.2 & 4.51 & 0.8 & 3.86 && 3.5$\pm$1.0 & 0.3$\pm$0.1 & 4.58$\pm$0.65 & 0.8$\pm$0.1 &h\\
HD 34700 & 2.0 & 0.2 & 3.50 & 0.5 & 17.7 && 2.5$\pm$0.5 & 0.3$\pm$0.1 & 3.93$\pm$0.43 & 0.2$\pm$0.1 &h\\
HD 35187 & 2.5 & 0.5 & 5.29 & 1.0 & 1.12 && 3.5$\pm$1.0 & 0.4$\pm$0.1 & 5.89$\pm$0.66 & 0.9$\pm$0.1 &h\\
HD 36112 & 5.0 & 0.2 & 5.64 & 0.3 & 0.96 && 4.0$\pm$1.0 & 0.3$\pm$0.1 & 5.29$\pm$0.60 & 0.3$\pm$0.1 & l \\
HD 36917 & 5.5 & 0.3 & 7.20 & 0.3 & 0.62 && 4.0$\pm$1.0 & 0.5$\pm$0.1 & 7.41$\pm$1.42 & 0.3$\pm$0.1 &h\\
HD 37357 & 2.0 & 0.2 & 3.50 & 0.1 & 0.69 && 3.0$\pm$0.5 & 0.3$\pm$0.1 & 3.93$\pm$0.43 & 0.2$\pm$0.1 & l\\
HD 37411 & 3.5 & 0.3 & 4.58 & 1.0 & 2.27 && 3.5$\pm$1.0 & 0.3$\pm$0.1 & 4.58$\pm$0.65 & 0.9$\pm$0.1 &h\\
HD 37806 & 5.0 & 0.5 & 10.59 & 0.6 & 0.41 && 3.5$\pm$1.0 & 0.7$\pm$0.1 & 13.64$\pm$3.34 & 0.7$\pm$0.1 & l\\
HD 38120 & 2.0 & 0.2 & 3.50 & 0.3 & 0.96 && 2.5$\pm$0.5 & 0.2$\pm$0.0 & 3.50$\pm$0.00 & 0.4$\pm$0.2 & l\\
HD 58647 & 5.5 & 0.2 & 6.20 & 1.0 & 0.052 && 4.0$\pm$1.0 & 0.4$\pm$0.1 & 6.55$\pm$1.26 & 0.9$\pm$0.1 & l\\
HD 72106 & 2.0 & 0.2 & 3.50 & 0.5 & 2.87 && 2.5$\pm$0.5 & 0.2$\pm$0.0 & 3.50$\pm$0.00 & 0.6$\pm$0.2 &h\\
HD 85567 & 5.0 & 0.3 & 6.55 & 0.9 & 0.054 & &4.0$\pm$1.0 & 0.4$\pm$0.1 & 5.89$\pm$0.60 & 0.9$\pm$0.1 &h\\
HD 95881 & 4.0 & 0.2 & 4.51 & 0.7 &4.32 & &3.0$\pm$1.0 & 0.3$\pm$0.1 & 4.23$\pm$0.73 & 0.8$\pm$0.1 &h\\
HD 97048 & 5.0 & 0.2 & 5.64 & 0.4 & 6.42 && 4.0$\pm$1.0 & 0.4$\pm$0.1 & 5.89$\pm$0.65 & 0.5$\pm$0.1 &h\\
HD 97300 & 5.5 & 0.2 & 6.20 & 0.9 & 1.60 && 4.0$\pm$1.0 & 0.4$\pm$0.1 & 5.89$\pm$0.65 & 0.9$\pm$0.1 &h\\
HD 98922 & 2.5 & 0.2 & 3.50 & 0.6 & 0.99 && 3.0$\pm$0.5 & 0.3$\pm$0.1 & 3.95$\pm$0.45 & 0.5$\pm$0.1 & l\\
HD 100453 & 2.0 & 0.2 & 3.50 & 0.7 & 4.72 && 3.0$\pm$0.5 & 0.3$\pm$0.1 & 4.04$\pm$0.54 & 0.4$\pm$0.2 &h\\
HD 100546 & 5.0 & 0.2 & 5.64 & 0.2 & 5.37 && 3.5$\pm$1.0 & 0.3$\pm$0.1 & 5.07$\pm$0.82 & 0.5$\pm$0.2 &h\\
HD 101412 & 2.0 & 0.2 & 3.50 & 0.7 & 3.13 && 3.0$\pm$0.5 & 0.3$\pm$0.1 & 3.95$\pm$0.45 & 0.5$\pm$0.1 &h\\
HD 135344B & 4.5 & 0.2 & 5.07 & 1.0 & 1.55 && 3.5$\pm$1.0 & 0.3$\pm$0.1 & 5.07$\pm$1.03 & 0.9$\pm$0.1 &h\\
HD 139614 & 4.0 & 0.2 & 4.51 & 1.0 & 1.83 && 3.0$\pm$1.0 & 0.3$\pm$0.1 & 4.23$\pm$0.73 & 0.8$\pm$0.1 &h\\
HD 141569 & 2.0 & 0.2 & 3.50 & 1.0 & 0.19 && 3.0$\pm$1.0 & 0.3$\pm$0.1 & 4.51$\pm$0.73 & 0.9$\pm$0.1 &h\\
HD 142527 & 3.0 & 0.3 & 3.93 & 0.3 &7.51 && 3.0$\pm$1.0 & 0.3$\pm$0.1 & 4.23$\pm$0.73 & 0.3$\pm$0.1 &h\\
HD 142666 & 3.5 & 0.2 & 3.95 & 0.5 & 2.08 && 3.0$\pm$0.5 & 0.3$\pm$0.1 & 4.04$\pm$0.54 & 0.5$\pm$0.1 &h\\
HD 144432 & 3.5 & 0.8 & 23.87 & 0.0 & 6.50 && 4.5$\pm$1.0 & 0.7$\pm$0.1 & 19.57$\pm$8.70 & 0.0$\pm$0.0 & l \\
HD 145718 & 2.0 & 0.2 & 3.50 & 0.2 & 5.17 && 2.5$\pm$0.5 & 0.2$\pm$0.0 & 3.50$\pm$0.00 & 0.4$\pm$0.2 & l \\
HD 163296 & 4.0 & 0.2 & 4.51 & 0.0 & 0.80 && 3.0$\pm$0.5 & 0.3$\pm$0.1 & 3.93$\pm$0.43 & 0.2$\pm$0.2 & l \\
HD 169142 & 2.5 & 0.2 & 3.50 & 0.8 & 9.06 && 3.0$\pm$0.5 & 0.3$\pm$0.1 & 4.58$\pm$1.05 & 0.5$\pm$0.2 &h\\
HD 179218 & 2.0 & 0.2 & 3.50 & 0.9 & 2.58 && 3.0$\pm$0.5 & 0.3$\pm$0.1 & 3.95$\pm$0.45 & 0.7$\pm$0.1 &h\\
HD 244604 & 2.0 & 0.4 & 3.50 & 0.5 & 0.44 & &3.0$\pm$0.5 & 0.3$\pm$0.1 & 4.04$\pm$0.54 & 0.5$\pm$0.1 & l \\
HD 250550 & 2.0 & 0.2 & 3.50 & 0.0 & 0.98 && 3.0$\pm$0.5 & 0.3$\pm$0.1 & 4.23$\pm$0.73 & 0.2$\pm$0.1 & l\\
HD 281789 & 2.0 & 0.2 & 3.50 & 0.3 & 5.60 && 2.5$\pm$0.5 & 0.2$\pm$0.0 & 3.50$\pm$0.00 & 0.4$\pm$0.2 & l\\
IC348LRL110 & 2.0 & 0.2 & 3.50 & 0.3 & 82.7 && 2.5$\pm$0.5 & 0.2$\pm$0.0 & 3.50$\pm$0.00 & 0.6$\pm$0.2 &l \\
IRAS 03260+3111 & 5.5 & 0.3 & 7.20 & 0.2 & 23.0 &&4.0$\pm$1.0 & 0.5$\pm$0.1 & 8.89$\pm$1.98 & 0.3$\pm$0.1 &h\\
IRAS 06084-0611 & 5.5 & 0.2 & 6.20 & 0.7 & 7.91& &3.5$\pm$1.0 & 0.4$\pm$0.1 & 6.35$\pm$1.11 & 0.8$\pm$0.1 &h\\
J182858.1+001724 & 4.5 & 0.2 & 5.07 & 0.0& 2.20 && 3.5$\pm$1.0 & 0.4$\pm$0.1 & 5.89$\pm$1.30 & 0.1$\pm$0.1 &h\\
LkH$\alpha$224 & 4.5 & 0.2 & 5.07 & 0.0 & 25.2 && 3.0$\pm$1.0 & 0.3$\pm$0.1 & 4.23$\pm$0.73 & 0.2$\pm$0.1 & l\\
LkH$\alpha$330 & 2.0 & 0.2 & 3.50 & 0.0& 3.34 && 2.5$\pm$0.5 & 0.2$\pm$0.0 & 3.50$\pm$0.00 & 0.2$\pm$0.2 & l\\
MWC 297 & 5.0 & 0.2 & 5.64 & 0.2 & 0.023 && 3.5$\pm$1.0 & 0.3$\pm$0.1 & 5.24$\pm$1.11 & 0.3$\pm$0.2 & l\\
MWC 865 & 5.5 & 0.2 & 6.20 & 0.4 & 5.61 && 3.5$\pm$1.0 & 0.4$\pm$0.1 & 6.46$\pm$1.62 & 0.5$\pm$0.1 &h\\
MWC 1080 & 3.5 & 0.8 & 23.87 & 0.2 & 0.40 && 4.0$\pm$1.0 & 0.7$\pm$0.1 & 19.57$\pm$5.93 & 0.2$\pm$0.1 &h\\
Oph IRS48 & 2.5 & 0.5 & 5.29 & 0.8 & 24.3&& 3.5$\pm$1.0 & 0.5$\pm$0.1 & 7.27$\pm$1.62 & 0.7$\pm$0.1 &h\\
PDS144N & 2.0 & 0.2 & 3.50 & 0.3 & 737 && 3.0$\pm$1.0 & 0.3$\pm$0.1 & 3.93$\pm$0.43 & 0.4$\pm$0.2 & l\\
RR Tau & 4.5 & 0.2 & 5.07 & 0.6 & 19.7 && 3.5$\pm$1.0 & 0.3$\pm$0.1 & 4.85$\pm$0.61 & 0.7$\pm$0.1 &h\\
SR21N & 2.0 & 0.4 & 3.50 & 0.2 & 10.5 && 3.0$\pm$0.5 & 0.3$\pm$0.1 & 3.95$\pm$0.45 & 0.2$\pm$0.1 &h\\
SU Aur & 2.0 & 0.2 & 3.50 & 0.0 & 6.78 && 2.5$\pm$0.5 & 0.2$\pm$0.0 & 3.50$\pm$0.00 & 0.3$\pm$0.2 & l\\
TY CrA & 5.5 & 0.2 & 6.20 & 0.4 & 6.94 && 3.5$\pm$1.0 & 0.4$\pm$0.1 & 6.46$\pm$1.62 & 0.5$\pm$0.2 &h\\
UX Tau & 5.5 & 0.2 & 6.20 & 0.6 & 0.50 && 4.5$\pm$1.0 & 0.5$\pm$0.1 & 8.08$\pm$1.45 & 0.7$\pm$0.1 & l\\
V590 Mon & 2.5 & 0.4 & 4.04 & 0.7 & 1.24 && 3.5$\pm$1.0 & 0.3$\pm$0.1 & 4.58$\pm$0.64 & 0.8$\pm$0.1 &h\\
V892 Tau & 2.0 & 0.2 & 3.50 & 0.9 & 12.2 && 3.0$\pm$0.5 & 0.3$\pm$0.1 & 3.95$\pm$0.45 & 0.7$\pm$0.2 &h\\
VV Ser & 2.0 & 0.6 & 5.89 & 1.0 & 0.08 && 4.0$\pm$1.0 & 0.4$\pm$0.1 & 6.35$\pm$0.85 & 0.9$\pm$0.1 &h\\
WL 16 & 5.5 & 0.4 & 8.89 & 1.0 & 3.45 && 4.0$\pm$1.0 & 0.6$\pm$0.1 & 9.53$\pm$2.16 & 0.9$\pm$0.1 &h\\
Wray15-1484 & 3.5 & 0.2 & 3.95 & 1.0 & 1.26 && 3.5$\pm$1.0 & 0.5$\pm$0.2 & 8.08$\pm$2.84 & 0.8$\pm$0.1 &h\\
WW Vul & 5.5 & 0.3 & 7.20 & 0.3 & 0.27 && 4.0$\pm$1.0 & 0.5$\pm$0.1 & 8.89$\pm$1.68 & 0.4$\pm$0.1 & l\\
\enddata

%% Include any \tablenotetext{key}{text}, \tablerefs{ref list},
%% or \tablecomments{text} between the \enddata and 
%% \end{deluxetable} commands
\tablenotetext{a}{The uncertainties for each parameter are calculated from the
median absolute deviation
(MAD: $\Delta X\equiv\langle|X_i-\langle X_i\rangle|\rangle$,
where $X_i$ is a univariate data set).
The MAD is an estimator for each parameter of the spread 
in the top 70 models that
closely fit the observed PAH spectrum
(see Section \ref{sec:res}).
}
\tablenotetext{b}{Reliability of the model-fit parameters (see Section \ref{sec:res}).
If the model-fit is highly reliable, we label it with ``h''.
If the model-fit is less reliable, we label it ``l''. Note that
only those with high reliability are used to investigate
the correlations of the PAH properties}
with stellar parameters (see Section \ref{sub:corr} and
Figures \ref{fig:corr_teff}--\ref{fig:corr_peak}).  
\tablenotetext{c}{Total mass of PAHs under the assumption
that all PAHs are located at $r=10\AU$ from the central star.
Total mass of PAHs at a given distance $r$ can be derived
from $M_{\rm PAH}(r)=M^{10\AU}_{\rm PAH}(r/10\AU)^2$.}
%% No \tablecomments indicated

%% No \tablerefs indicated

\end{deluxetable}

%%%%%%%% Table 2: Model Parameters %%%%%%%%%

%%%%%%%%%%%%%%%%% Results %%%%%%%%%%%%%%%%%
\section{Discussion}\label{sec:disc}

\subsection{Correlating PAH Properties with Stellar Parameters}\label{sub:corr}

\subsubsection{Stellar Effective Temperature}

\begin{figure}[tbp]
\epsscale{1.}
\plotone{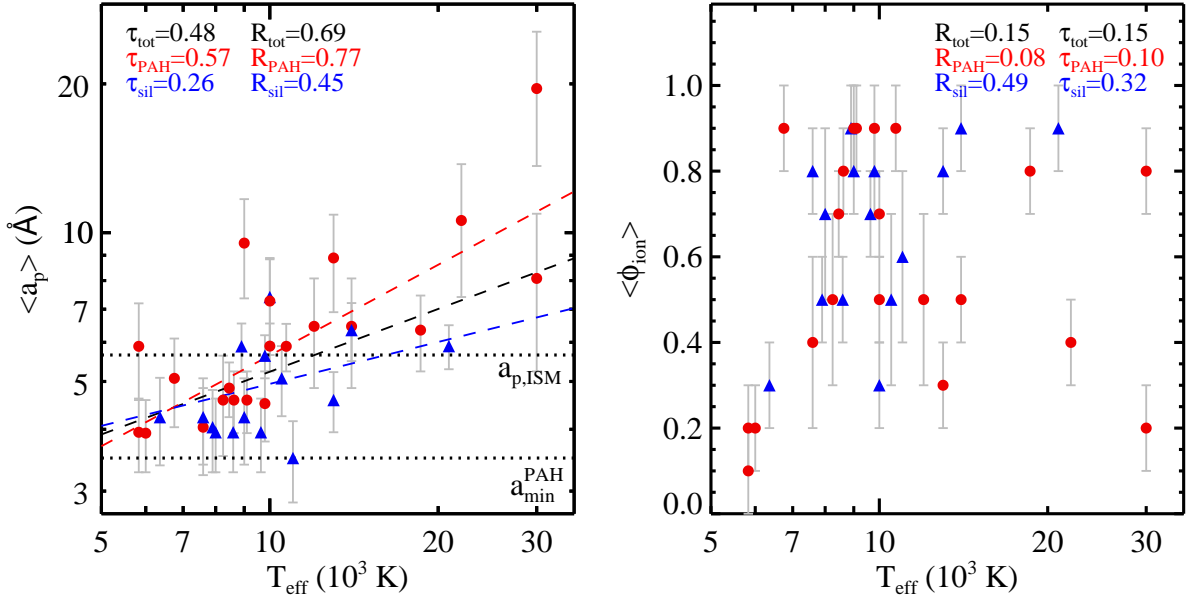}
\caption{ \label{fig:corr_teff}
         Comparison between the PAH parameters
         ($a_{\rm p}$ and $\phi_{\rm ion}$ in the left and right panels, 
         respectively) and the stellar effective temperatures ($\Teff$). 
         $\langle\ap\rangle$ and $\langle\phiion\rangle$ are
         median values in the top 70 models of each source,
         and uncertainties are the MAD (\ie $\Delta\ap$
         and $\Delta\phiion$, see Section \ref{sec:res} and
         Table \ref{tab:para}).
         Circle and triangle symbols denote sources with
         PAH emission-dominated and silicate emission
         (or absorption) dominated spectra, respectively.
         Note that we only consider those sources with the
         high reliability of the model-fit parameters (see
         Section \ref{sec:res} and Table \ref{tab:para}).
         In the left panel, $a_{\rm p}$ for the diffuse ISM
         (\ie $a_{\rm p, ISM}=5.66\Angstrom$)
         and $\apahmin$ ($3.5\Angstrom$,
         the minimum size of $a_{\rm p}$, see Section \ref{sec:model})
         are denoted by dotted lines as a reference. 
         The correlations between $\ap$ and $\Teff$ are
         fitted with a linear function in logarithmic scale 
         (red, blue, and black dashed lines for the PAH-dominated
         sources, the silicate-dominated sources, and combined
         data sets, respectively).
         Correlation coefficients for the PAH-dominated spectra
         ($R_{\rm PAH}$) and those with the strong 9.7$\mum$ silicate
         emission (or absorption) feature ($R_{\rm sil}$)
         are measured separately, and a total correlation
         coefficient is also given ($R_{\rm tot}$).
         Hereafter, symbol designation is applied in the same way
         for Figures \ref{fig:corr_LM}--\ref{fig:corr_peak}.
         }
\end{figure}

The effective temperature of the central star ($\Teff$) is one
of the key environmental parameters that control
the physical and chemical properties of PAHs in a disk. 
In Figure \ref{fig:corr_teff}, the peak size of the PAH
size distribution ($\ap$) and the ionization fraction
($\phiion$) are compared with $\Teff$. With a correlation
coefficient of $R_{\rm PAH}\approx0.77$, the PAH
peak size ($\ap$) derived for those PAH-dominated sources
seems to have a tendency of positively correlating
with $\Teff$. Such a trend is relatively weak 
($R_{\rm sil}\approx0.45$) for those emitting or absorbing 
strongly the 9.7$\mum$ silicate feature. 
Combining both data sets,
we obtain a correlation coefficient of $R_{\rm tot}\approx0.69$.
We also derive the Kendall rank correlation coefficient
($\tau$) and the significance of its deviation from zero ($p$)
for a non-parametric hypothesis test. In Kendall $\tau$ test,
$p$ is in the interval between 0.0 and 1.0, 
and a smaller value indicates a more significant correlation. 
The Kendall test for the correlation between $\ap$ and $\Teff$
gives $\tau\approx0.57$, 0.26, and 0.48, 
and $p\approx2.81\times10^{-4}$, 0.25, 
and $3.26\times10^{-5}$ for the PAH-dominated sources, 
the silicate-dominated sources, 
and the combined data sets, respectively. 
The low $p$ value for the PAH-dominated sources 
(and the combined data sets) indicates that 
the correlation between $\ap$ and $\Teff$
is meaningful.%

We perform a simple linear-fit in logarithmic scale, which
gives $\langle\ap\rangle=0.02^{+0.04}_{-0.01}\times(\Teff/\K)^{0.6\pm0.1}$,
$0.37^{+0.75}_{-0.25}\times(\Teff/\K)^{0.3\pm0.1}$, and
$0.11^{+0.12}_{-0.06}\times(\Teff/\K)^{0.4\pm0.1}$
for the PAH-dominated sources, the silicate-dominated
sources, and the combined data sets, respectively. 
This trend can be readily explained: 
in those sources with a high $\Teff$,
the stellar photons are more energetic,
and therefore, the photodissociation of 
small PAHs is more likely to occur. In particular,
small PAHs are more vulnerable than large ones
because their internal vibrational modes 
are not large enough to quickly re-distribute 
the energy transferred from the absorbed stellar photons 
(see Section \ref{sub:dest} for details). 
Thus, in disks around stars with a higher $\Teff$,
small PAHs are preferentially photodissociated.

Unlike $\ap$, $\phiion$ does not show
any significant correlation with $\Teff$ (and other
stellar parameters such as $\Lstar/\Mstar$ or stellar age,
which are not shown here). 
From the Kendall test, we obtain $p\approx0.54$, 0.10,
and 0.19 for the PAH-dominated
sources, the silicate-dominated sources, and the combined data sets,
respectively, which is much larger than those for $\ap$.
We note that $\phiion$ depends on
$U\sqrt{T_{\rm gas}}/n_{\rm e}$ (see Section \ref{sec:model})
while $U$ is proportional to $r^{-2}$, 
so $\phiion$ depends on the distance of PAHs 
from the central star.    
Then, this lack of correlation
between $\phiion$ and $\Teff$ could be attributed 
to the diversity of the spatial distribution of PAHs 
from one disk to another. 
In addition, $n_{\rm e}$ within a disk can
span a wide range from a few to $\ga10^{5-6} \cm^{-3}$
depending on the disk geometry, so the spatial distribution
of PAHs should be taken into account to 
properly interpret $\phiion$. 
Note that there is also an environmental aspect
to be considered. For example, in the case of MWC 1080,
a member of a small star forming region,
the diffuse nebulosity surrounding it is known
to be the dominant origin of the PAH emission rather than
its own disk (\eg Li \etal2014). 
Then, the PAH molecules in the nebula 
could be neutral despite of 
the rather high $\Teff$ (=\,30,000\,K)
probably due to the highly diluted stellar radiation 
(\ie low $U$) compared with that of a typical PPD. 

%%%%%%%%% Figure 11  %%%%%%%%
%%%%%%%%% Figure 11%%%%%%%%

\subsubsection{Luminosity-to-Mass Ratio}

\begin{figure}[tbp]
\epsscale{0.8}
\plotone{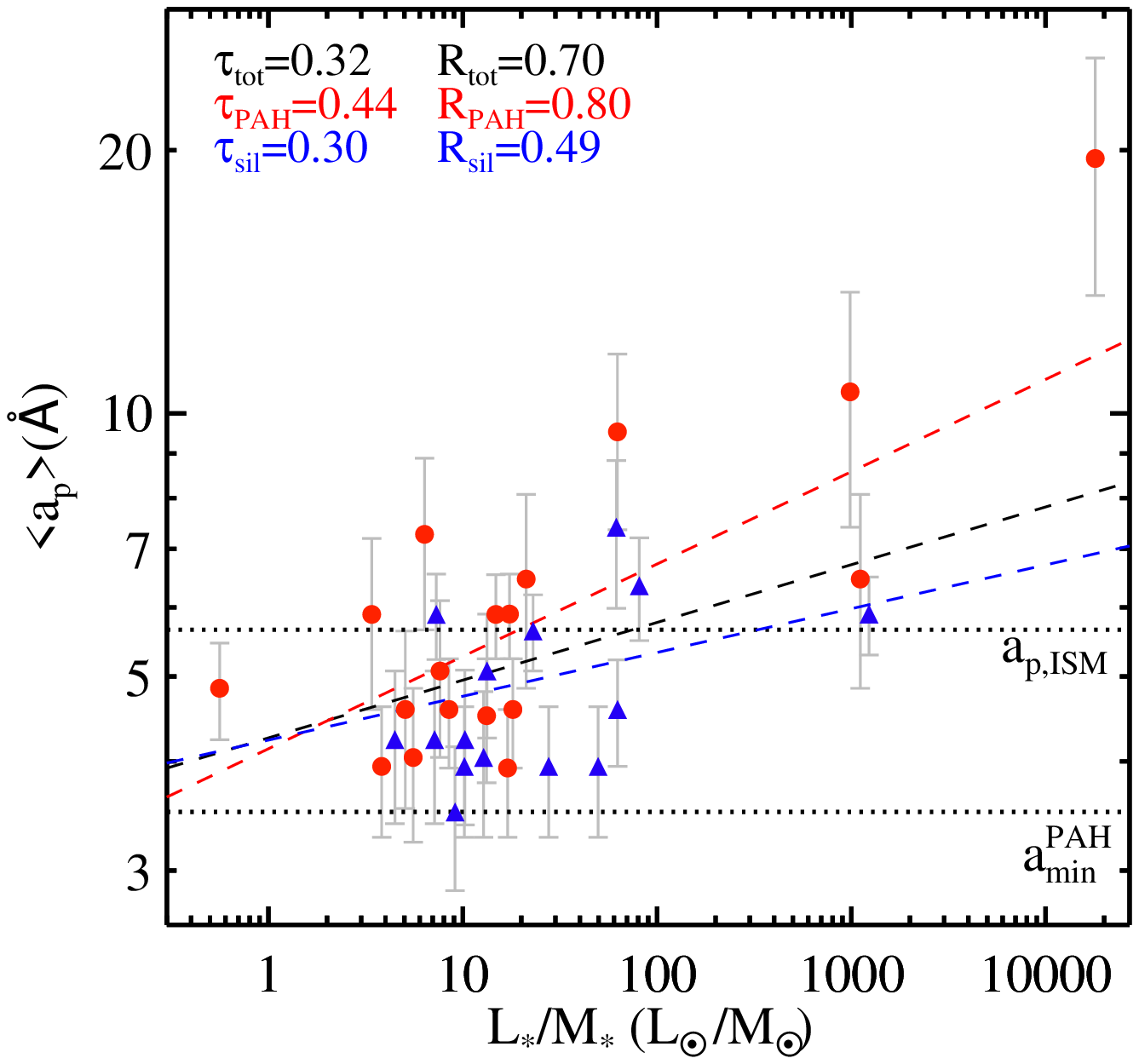}
\caption{\label{fig:corr_LM}
          Comparison between $\ap$ and $\Lstar/\Mstar$.  
         }
\end{figure}

In Figure \ref{fig:corr_LM}, we compare $\ap$ with 
$\Lstar/\Mstar$, the ratio of stellar luminosity to stellar mass.
The correlation between $\ap$ and $\Lstar/\Mstar$ appears
as strong as that between $\ap$ and $\Teff$ (for the
PAH-dominated spectra, $R_{\rm PAH}\approx0.80$).
From the Kendall test, we derive $\tau\approx0.44$, 0.30,
and 0.32 and $p\approx0.01$, 0.11, and 0.01 for the
PAH-dominated sources, the silicate-dominated sources, 
and the combined data sets, respectively. 
Like $\Teff$, we perform a simple linear-fit in logarithmic scale, which
gives $\langle\ap\rangle=4.1^{+0.3}_{-0.3}\times(\Lstar/\Mstar)^{0.11\pm0.02}$,
$4.2^{+0.4}_{-0.4}\times(\Lstar/\Mstar)^{0.05\pm0.02}$, and
$4.3^{+0.2}_{-0.2}\times(\Lstar/\Mstar)^{0.07\pm0.02}$
for the PAH-dominated sources, the silicate-dominated
sources, and the combined data sets, respectively. 

%%%%%%%%% Figure 12%%%%%%%%
%%%%%%%%% Figure 12 %%%%%%%%

The luminosity-to-mass ratio can be a diagnostics of the relative
strength of the radiative force against the gravitational force
($\brp$, see more in Section \ref{sub:dest}). 
The radiation pressure blows materials away from the
disk so that the size distribution of PAHs can be
rearranged. However, as most of the sources in our sample
are expected to possess a gas-rich disk, the gas drag
would dominate the motion of PAHs rather than the radiation
pressure. Then, the correlation we see in Figure \ref{fig:corr_LM}
merely reflects the dependence of $\Lstar$ on $\Teff$
(\ie $\Lstar\propto\Teff^4$). To investigate any influences
of the radiation pressure, it is required to distinguish
gas-rich and gas-depleted disks (see Section \ref{sub:dest}).
 
\subsubsection{Stellar Age}

\begin{figure}[tbp]
\epsscale{0.8}
\plotone{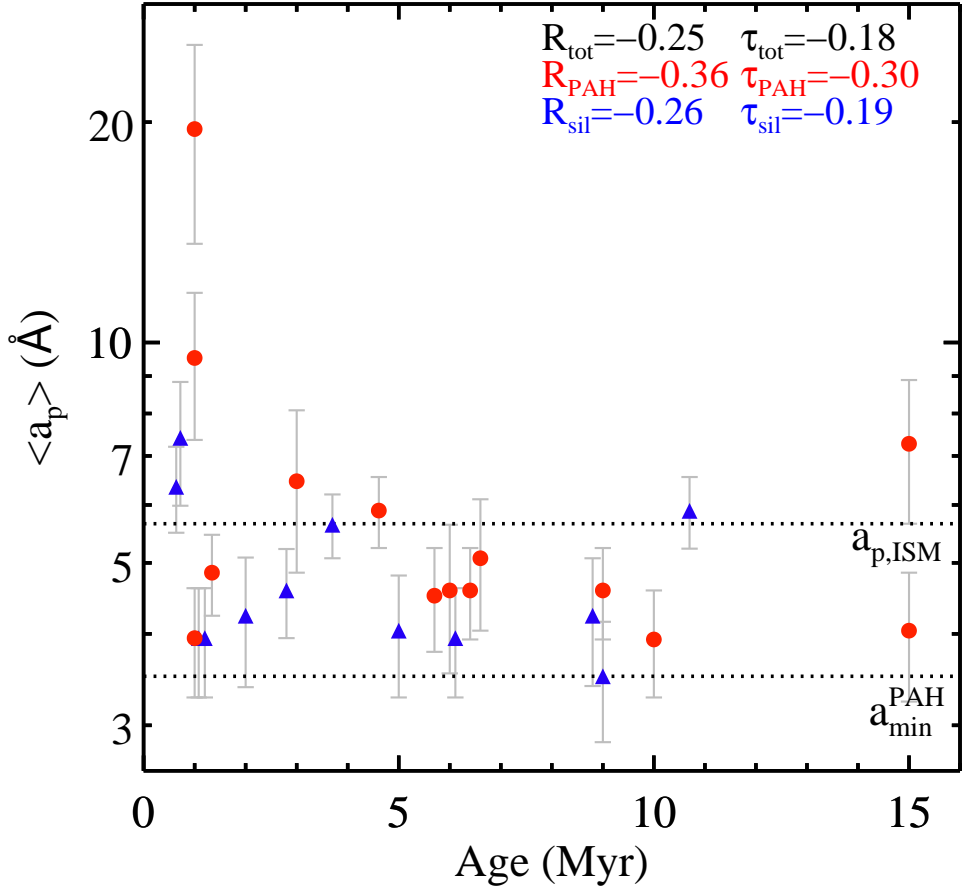}
\caption{ \label{fig:corr_age}
          Comparison between $\ap$ and stellar age.  
         }
\end{figure}

We examine the correlation between $\ap$ and the
stellar age. As shown in Figure \ref{fig:corr_age},
$\ap$ does not seem to exhibit any correlation
with the stellar age, but the overall comparison might
indicate a decreasing trend of $\ap$ with the stellar age
since all correlation coefficients (\ie both $R$ and $\tau$) 
are negative (with Kendall's $\tau\approx-0.30$ 
and $p=0.14$ for the PAH dominated-spectra).
If this is true, it would be 
interesting because small PAHs
are more quickly destroyed by photodissociation
than large PAHs
(see Section \ref{sub:dest} and Figure \ref{fig:tdes}).
To maintain the abundance of small PAHs, it is
required that small PAHs are rapidly replenished 
unless the photodissociation is not effective due to
very mild environment such as low $\Teff$ and
low $\Lstar$ (\eg IC348 LRL110: $\Teff=3778$ K
and $\Lstar=0.22\Lsun$). If fragmentation
is a primary mechanism of the 
replenishment, the existence of small PAH molecules
in aged disks could be explained.  

%%%%%%%%% Figure 13%%%%%%%%
%%%%%%%%% Figure 13%%%%%%%%

Note that the estimation of stellar age often has
large uncertainties and is model-dependent. Also,
for some sources, their ages have not been estimated
mostly due to the poorly constrained distance to the sources. 
For this reason, fewer sources are used to find a
correlation compared with other cases. 
Once $Gaia$, a space observatory of ESA, dedicated to
astrometry releases the data,\footnote{%
 The first data release of $Gaia$ is scheduled on 14
 September 2016, which will include the positions and
 G magnitude for about one billion stars using observations
 taken between 25 July 2014 and 16 September 2015.
 See more at http://sci.esa.int/gaia/.
}
we expect that the accuracy of age estimation would be much
improved, and the correlation with the stellar age can be revisited.

\subsection{The $7.7\mum$ feature}\label{sub:7p7}

\begin{figure}[tbp]
\epsscale{0.8}
\plotone{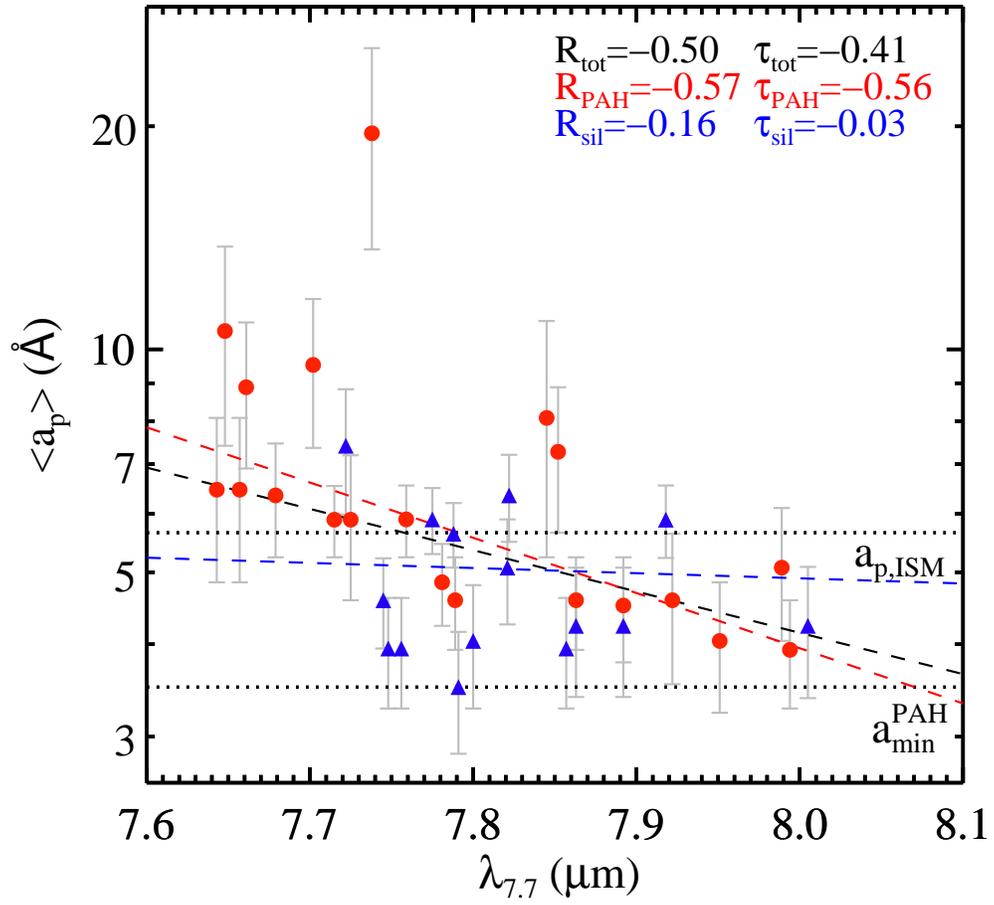}
\caption{\label{fig:corr_peak}
         Comparison between $\ap$ and the peak wavelength of
         the $7.7\mum$ PAH feature ($\lambda_{7.7}$).  
         }
\end{figure}

The PAH feature at $7.7\mum$ attributed to the C--C
stretching mode has been known
to show large variation in its emission profile,
and its shift from $7.7\mum$ is not fully understood
(\eg see Peeters \etal2002; van Diedenhoven \etal2004;
Sloan \etal2005, 2007).
Sloan \etal(2005) suggest that the $7.7\mum$ shift might
be influenced by not the charge state but the size of PAHs. 
In Figure \ref{fig:corr_peak}, we compare $\ap$ and the
peak wavelength of the $7.7\mum$ feature
($\lambda_{7.7}$), which shows
a trend between $\ap$ and the peak wavelength:
as PAH molecules become smaller, $\lambda_{7.7}$
tends to shift toward longer wavelengths. We perform a 
simple linear-fit in logarithmic scale of $\langle\ap\rangle$, which
gives log$\langle\ap\rangle={6.6^{+1.3}_{-1.3}-0.75^{+0.17}_{-0.17}\times
(\lambda_{7.7}/\um)}$, ${1.2^{+2.0}_{-2.0}-0.07^{+0.25}_{-0.25}\times
(\lambda_{7.7}/\um)}$, and
${5.1^{+1.1}_{-1.1}-0.56^{+0.14}_{-0.14}\times(\lambda_{7.7}/\um)}$
for the PAH-dominated sources, the silicate-dominated,
and combined data sets, respectively. 

%%%%%%%%% Figure 14 %%%%%%%%
%%%%%%%%% Figure 14 %%%%%%%%

\begin{figure}[tbp]
\epsscale{0.8}
\plotone{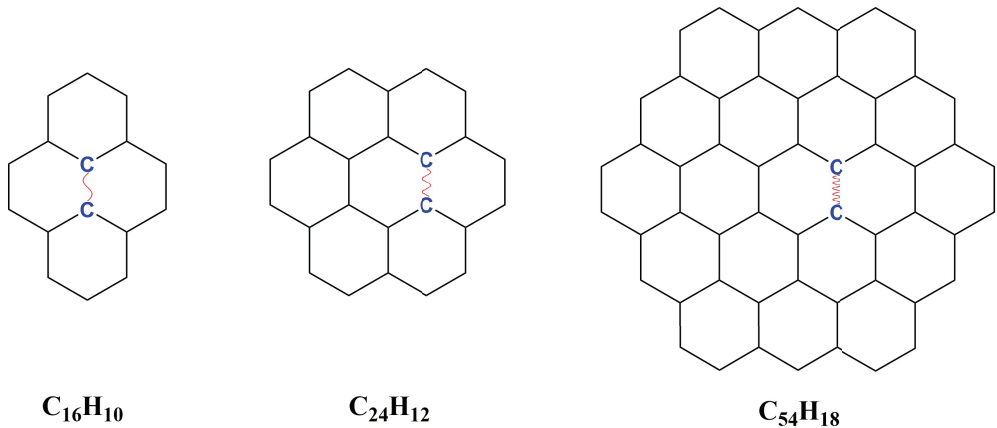}
\caption{\label{fig:3pah}
         Illustration of PAH molecules with three different sizes:
         pyrene ($\rm C_{16}H_{10}$), coronene ($\rm C_{24}H_{12}$),
         and circumcoronene ($\rm C_{54}H_{18}$). In each
         PAH molecule, two carbon atoms connected by a spring
         delineate the C--C stretching mode at $7.7\mum$ resembling
         a classical harmonic oscillator. Different stiffness of springs
         represents the variation of the spring constant ($k$) where
         the wavelength of the oscillation is defined as
         $\lambda_{\rm C-C}=2\pi c\sqrt{m/k}$ ($c$: the speed of light,
         $m\approx6m_{\rm H})$, see text). The stiffness of $k$
         increases from smaller PAHs to larger PAHs. 
          }
\end{figure}

This trend can be elucidated by postulating
that the C--C stretching mode acts like a classical harmonic
oscillator (see Figure \ref{fig:3pah}). If two carbon atoms
oscillate, the wavelength of the oscillation is given as
$\lambda_{\rm C-C}=2\pi c\sqrt{m/k}$, where $c$ is the speed
of light, $m$ is the reduced mass ($m\approx6m_{\rm H}$,
where $m_{\rm H}$ is the mass of the hydrogen nuclei) of
the two carbon atoms, and $k$ is the spring constant,
which measures the strength of the C--C stretch in this case. 
Considering that the C--C stretch
of a small PAH molecule can oscillate more easily
than that residing in a large PAH molecule
(\ie smaller $k$ for smaller PAHs than for larger PAHs),
$\lambda_{\rm C-C}$ would be red-shifted for smaller
PAHs with respect to larger PAHs. Although this approach
might be oversimplified, however, it gives an idea how
the size of PAHs can affect the 7.7$\mum$ shift physically.
We note that some quantum-chemical computations 
based on the hybrid density functional theoretical method 
(B3LYP) have reported that larger PAH molecules tend to shift 
the peak wavelengths of the 6.2 and 7.7$\mum$ features 
to longer wavelengths (\eg see Bauschlicher \etal2010,
Ricca \etal2012). However, it is well known that the vibrational
frequencies computed from density functional theory 
are not accurate and should be scaled in order to agree 
with the experimentally-measured frequencies. 
The commonly-adopted frequency scaling factors 
(\eg  Andersson \etal2005, Merrick \etal2007, Borowski 2012)
were derived from small molecules
($\simlt$\,20 C atoms).
It is not clear whether they are valid for large molecules
($\simgt$\,100 C atoms) of interest here.

%%%%%%%%% Figure 15 %%%%%%%%
%%%%%%%%% Figure 15 %%%%%%%%

It has been pointed out that the 7.7$\mum$ shift is related
to the effective temperature of the central star
(\eg see Sloan \etal2007; Tielens 2008). Sloan \etal(2007) examined
the PAH features seen in a wide range of stellar types including T Tauri,
HAeBe, and post-AGB stars and found that cooler stars
show more red-shifted $7.7\mum$ features (see their Figure 5).
This dependency is consistent with the correlations of $\ap$
with $\Teff$ and $\lambda_{7.7}$ 
(see Figure \ref{fig:corr_teff} and \ref{fig:corr_peak}).
Small PAHs, which tend to emit at a larger $\lambda_{7.7}$,
seem to be dominant in disks around cool stars, so the $7.7\mum$
shift could be commonly observed around cool stars.

\subsection{Destruction of PAHs in PPDs}\label{sub:dest}

%%%%%%%%% Figure 16 %%%%%%%%
\begin{figure}[tbp]
\epsscale{0.8}
\plotone{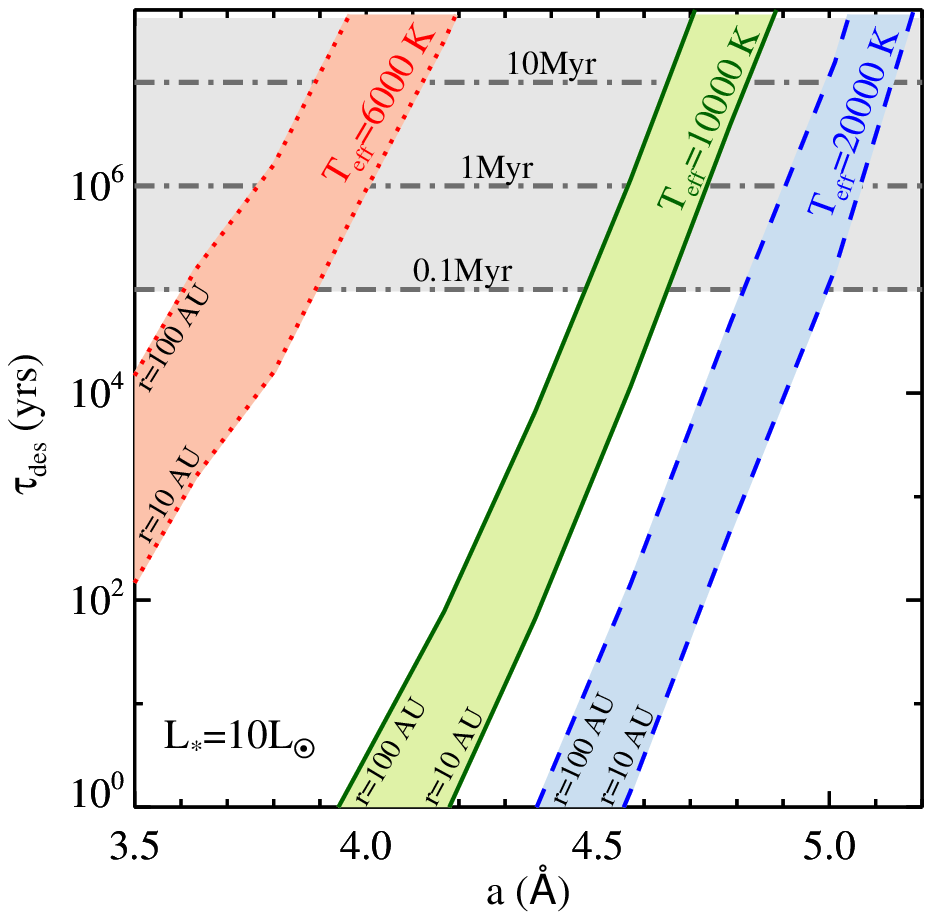}
\caption{\label{fig:tdes}
         Photodestruction timescales ($\tau_{\rm des}$) of PAHs
         as a function of size for $\Teff=0.6, 1,$ and $2\times10^4$ K
         at a given luminosity ($\Lstar=10\Lsun$).
         For each $\Teff$, $\tau_{\rm des}$ at $r=10$ and 100\,AU are
         calculated. A typical range of stellar ages of PPDs (\ie 0.1--50
         Myr) is designated by the shaded region with representative
         values of 0.1, 1, and 10 Myr (dash-dotted lines). 
         Note that $\tau_{\rm des}$ is proportional to $r^2$ and
         $\Lstar^{-1}$ (\ie $\tau_{\rm des}\propto r^2/\Lstar$). 
         }
\end{figure}         
%%%%%%%%% Figure 16 %%%%%%%%

Photodissociation is one of the primary mechanisms for the
destruction of PAHs in PPDs. When PAHs, in particular small PAHs,
excited by an energetic photon, cannot redistribute the absorbed energy
from the photon via internal vibrational modes, they may lose
a hydrogen atom, a hydrogen molecule, or an acetylene
molecule (C$_2$H$_2$) and could be eventually photodissociated. 
We follow the photophysics of Li \& Lunine (2003,
see their Appendix A for details) to derive
the photodestruction rate ($k_{\rm des})$ 
approximated by the ejection rate 
of an acetylene molecule. 
The destruction timescale 
($\tau_{\rm des}\equiv1/k_{\rm des}$)
depends on $\Teff$, $\Lstar$, and the distance
of PAHs from the central star ($r$). 
For a given $\Lstar$ ($=10\Lsun$), we calculate
$\tau_{\rm des}$ for $\Teff=0.6,$ 1, and $2\times10^4\K$
at a distance of $r=10$ and 100\,AU (Figure \ref{fig:tdes}).
We note that, when we derive $\tau_{\rm des}$,
we do not consider the competition between photodestruction
and photoionization. Intuitively, one would expect a longer
$\tau_{\rm des}$ if such a competition is taken into account
because the photoionization process offers an additional
channel for losing the absorbed stellar UV photons and
will somewhat quench the competing photodestruction
process. These UV photons lost through photoionization
could otherwise potentially dissociate the molecule 
by kicking off a H atom, a H$_2$ molecule, 
and/or a C$_2$H$_2$ molecule. Due to lack of
experimental data, however, it is difficult to quantitatively
investigate the branching ratio between photoionization
and photodestruction. Recently, Zhen et al.\ (2015, 2016)
experimentally studied this competition by exposing
PAH cations to synchrotron radiation.\footnote{%
   Zhen et al.\ (2015 ) considered coronene (C$_{24}$H$_{12}$$^+$), 
   ovalene (C$_{32}$H$_{14}$$^+$) and hexa-peri-hexabenzocoronene 
   (C$_{42}$H$_{18}$$^+$) cations irradiated by synchrotron radiation 
   in the range of 8--40\,eV, whereas Zhen et al.\ (2016) studied 
   a set of eight PAH cations ranging in sizes from 14 to 24 carbon atoms
   exposed to vacuum UV photons in the 7--20\,eV range.
   }
They found that, at photon energies below $\simali$13.6\,eV,
photodestruction dominates over photoionization for
small PAH cations. This supports our treatment
which ignores the competition between photodestruction
and photoionization for PAHs in PPDs.
Note that most of these PPDs
lack photons of $>$10\,eV.

In spite of the variation of $\tau_{\rm des}$ with $\Teff$,
it is clear that in general small PAHs are readily
destroyed during the typical lifetime of PPDs. The lifetime of 
most of the sources in our sample ranges from 1 to 10 Myr,
then PAHs smaller than $\approx4\Angstrom$ are expected
to be destroyed by photodissociation. For hotter sources
(\ie $\Teff\ga10,000\K$), even PAHs with $a\approx4.5\Angstrom$
are destroyed within $\simali$1$\Myr$.
This can naturally explain the correlation between $\ap$
and $\Teff$ shown in Figure \ref{fig:corr_teff}. Those 
with high $\Teff$ (\ie $\Teff\ga20,000\K$) are usually
more luminous (\ie $\Lstar\ga10^3\Lsun$) so that PAHs
($\la5\Angstrom$) can be rapidly destroyed.  
Meanwhile, for those with $\Teff\la5000\K$, small
PAHs ($\la4\Angstrom$) can survive against the
photodissociation. 
In this context, the large $\ap$ of DoAr 21
($\langle\ap\rangle=10.31\pm2.94\Angstrom$) is surprising because
it is a cool star ($\Teff=5080\K$) with a young age ($\simali$0.4$\Myr$).
Although our model parameters are considered to be less
reliable due to the uncertain subtraction of the
silicate absorption feature (see Figure \ref{fig:pahfit} and 
more in Section \ref{sub:cntsub}), 
our results based on the current data indicate the absence
of small PAHs in its disk. If the large $\ap$ of DoAr 21 is
real, it could be attributed to its unique physical conditions.
DoAr 21 is known to be a strong X-ray emitter ($T\simali$$10^8\K$),
which produces sufficient (far-)UV radiation to excite PAHs
located at $\ga100\AU$ from the central star
(\eg Jensen \etal2009).\footnote{%
 Jensen \etal(2009) suggest that the origin of the observed
 emission around DoAr 21 including PAH features, 
 dust continuum, and H$_2$ emission could be a small-scale
 photodissociation region (PDR) rather than a PPD
 in the final stage of disk clearing, considering its youth
 ($\la1$ Myr) and the extreme radiation environment
 (\ie the brightest X-ray source in the $\rho$ Oph cloud core A
 with large X-ray flares). Unlike PPDs, a continuous
 replenishment via collisions of planetesimals or
 outgassing of cometary
 bodies would not occur in PDRs, 
so the lack of small PAHs in DoAr~21 can be plausible. }
High-resolution MIR imaging obtained with the T-ReCS camera
on Gemini South reveals that spatially resolved PAH emission
feature at $11.3\mum$ shows little or no emission within $\simali$$100\AU$
and bright emission extended over $\simali$$200\AU$ (Jensen \etal2009).
This supports the efficient destruction of PAHs near the star
through photodissociation, and the large $\ap$ would be a corollary
to this.

\begin{figure}[tbp]
\epsscale{1.1}
\plottwo{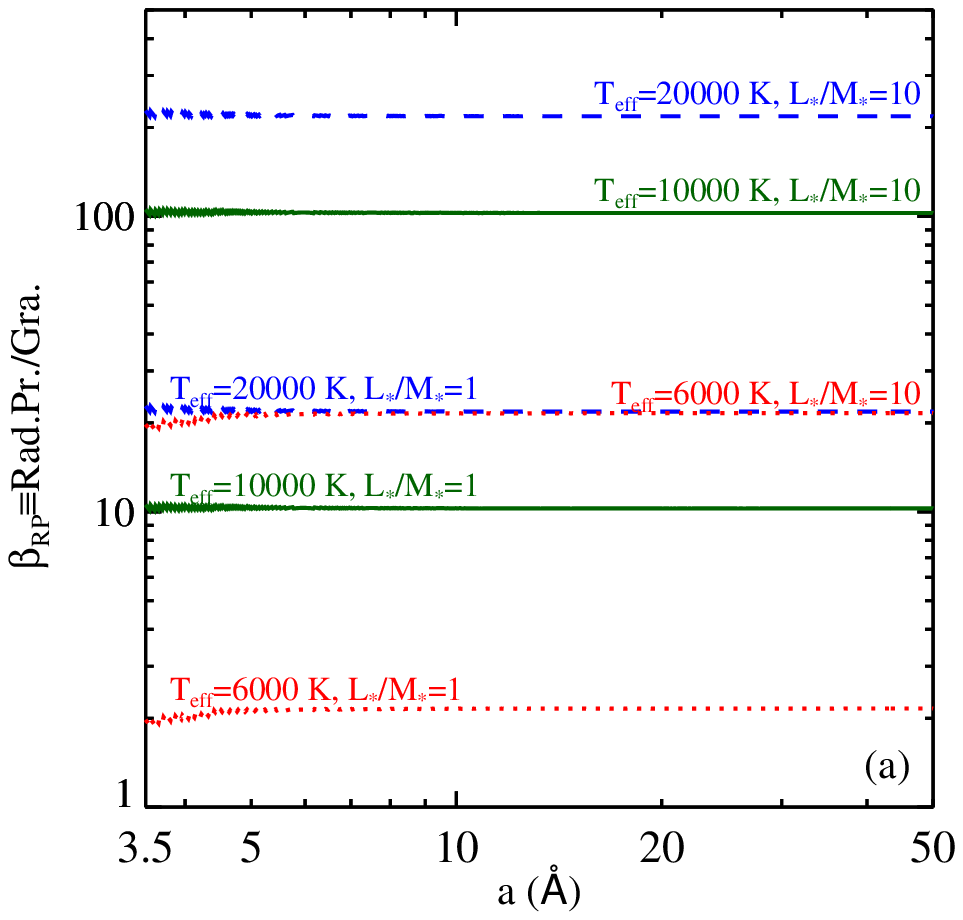}{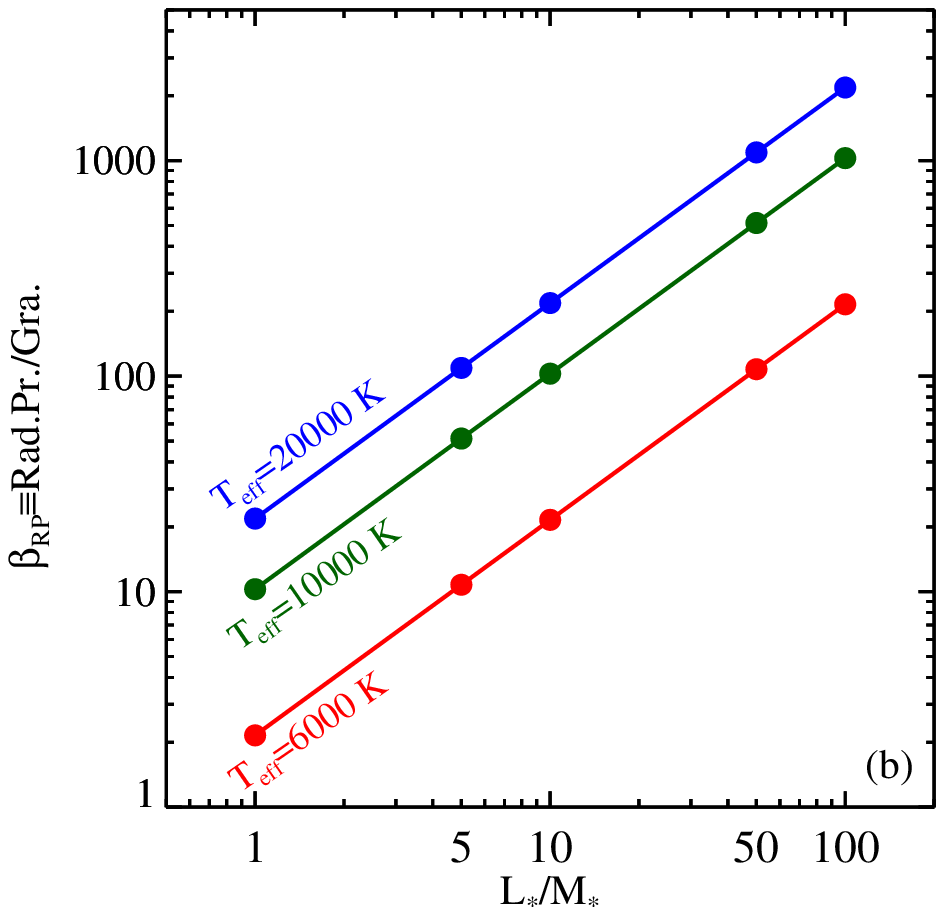}
\caption{\label{fig:brp}
         Ratio of the radiative force to the gravitational force 
         ($\beta_{\rm RP}$) at various $\Teff$ (0.6, 1, and
         $2\times10^4$ K). $\beta_{\rm RP}$ is nearly
         independent of PAH size (panel (a)), which makes
         $\beta_{\rm RP}$ directly proportional to $\Lstar/\Mstar$
         (panel (b)). $\beta_{\rm RP}$ is always larger than
         unity unless both $\Teff$ and $\Lstar/\Mstar$ are low.
          }
\end{figure}

During the evolution of PPDs, a disk experiences
a considerable amount of gas-loss and eventually becomes a
gas-poor disk (see also Mann \etal2006).
In such a condition, PAHs can be removed
from a PPD by radiative pressure (RP) and Poynting-Robertson 
(PR) drag in addition to photodissociation. As the 
RP usually overwhelms the PR drag for
small particles (\eg $\la 10\mum$, see Seok \& Li 2015), 
the RP can play a main role for the removal of PAHs. 
Following Equation (19) of Li \& Lunine (2003), we derive 
the ratio of the radiative force to the gravitational force
($\brp$) for different $\Teff$ (=0.6, 1, and $2\times10^4\K$).
As PAHs fall in the Rayleigh limit (see Li 2009), 
$\brp$ is not sensitive to their size but is directly 
proportional to $\Lstar/\Mstar$ (see Figure \ref{fig:brp}).
For those with $\Teff\ga 6,000 \K$, $\brp$ is expected
to be larger than unity, which implies that PAHs
would be continuously expelled outward from the disk.
The timescale of the RP ($\tau_{\rm RP}$) expelling
can be estimated by adopting the local dynamical 
timescale ($\tau_{\rm dyn}$, see Li \& Lunine 2003;
Seok \& Li 2015). We calculate $\tau_{\rm dyn}=2\pi/\Omega(r)$,
where $\Omega(r)\equiv(G\Mstar/r^3)^{1/2}$ is the Keplerian
frequency. Then, $\tau_{\rm RP}$ is derived from
\beq
\tau_{\rm RP}(r)\approx31.6\times(\Mstar/\Msun)^{-1/2}(r/10\AU)^{3/2} \yr.
\eeq
In general, $\tau_{\rm RP}$ is even shorter than
$\tau_{\rm des}$ except for very small PAHs
(see Figure \ref{fig:tdes}):
for example, HD 34700 ($\Mstar\approx1.20\Msun$)
has $\tau_{\rm RP}\approx110\yr$ at $r=25\AU$, and 
$\tau_{\rm des}\approx 240\yr$ and 
$\tau_{\rm des}\approx1.9\times10^6\yr$
for PAH size of 3.5 and 4.0$\Angstrom$ 
at $r=25\AU$, respectively 
(see Figure 6 in Seok \& Li 2015).
Thus, the RP expulsion would be the primary mechanism for
the removal of PAHs from the gas-depleted disk.

%%%%%%%%% Figure 17 %%%%%%%%
%%%%%%%%% Figure 17 %%%%%%%%

For gas-rich disks, on the other hand, the effect of gas drag
must be taken into account, which might
dominate the motion of PAHs in the disks. For example,
the pre-transitional disk around HD 169142 could have
a gas-drag timescale of only a few decades
(see Seok \& Li 2016), which is much shorter
than its RP timescale. Although a full treatment of the gas
drag using a two-dimensional disk structure is required
to properly estimate the gas-drag timescale in PPDs
(\eg see Chiang \& Youdin 2001 for a review), it is
clear that the rapid removal of PAHs would
take place in PPDs via photodissociation, RP expulsion,
and gas drag.
To maintain PAHs against the efficient removal of PAHs,
continuous replenishment in the disk would be required.  
A primary mechanism of the replenishment still remains unknown.
A continuous supply of PAHs via the outgassing of cometary
bodies and collisions of planetesimals and asteroids
seems to be the most reasonable way
(\eg see Seok \& Li 2015, 2016).
Mann \etal(2007) also discussed the origin and survival
of nanoparticles in the inner solar system.

\subsection{Aliphatics versus Aromatics}\label{sub:ali}

While PAH molecules produce conspicuous emission features
in the IR spectra of PPDs, relatively weak but distinguishable minor
features sometimes appear at $\simali$3.4, 6.9, and $7.3\mum$
(\eg Sloan \etal2005; Acke \etal2010).
They are usually attributed to the C--H vibrational modes in
aliphatic hydrocarbons (\eg Chiar \etal2000; Pendleton 
\& Allamandola 2002; Yang \etal2016a, 2016b) although
the $3.43\mum$ feature can also originate from 
anharmonicity and/or superhydrogenation of the aromatic
C--H stretch (Barker \etal1987; Bernstein \etal1996).
Among the 63 MIR spectra that have the spectral data at 
$\lambda\la6.9\mum$, 21 sources show the aliphatic feature
at $6.9\mum$, which is indicative of aliphatic CH sidegroups
of PAH molecules in those systems. 

For the 21 PPDs, we estimate the aliphatic fraction
of the PAH molecules from the measured intensities of
the $6.9\mum$ aliphatic feature and $7.7\mum$ aromatic
feature ($I_{6.9}$ and $I_{7.7}$, respectively)
following Li \& Draine (2012). 
Let $A_{6.9}$ and $A_{7.7}$ be the band strengths 
of the aliphatic C--H deformation band 
and the aromatic C--C stretching band, respectively.
We adopt $A_{6.9}\approx2.3\times10^{-18}\cm$ per CH$_2$ 
or CH$_3$ group and $A_{7.7}=5.4\times10^{-18}\cm$ 
per C atom for charged aromatic molecules (Li \& Draine 2012). 
Let $B_{\lambda}(T)$ be the Planck function 
at wavelength $\lambda$ and temperature $T$.
If the molecule emits at temperature $T$, the intensity
ratio of the 6.9$\mum$ aliphatic C--H band to the aromatic
C--C band is approximately $I_{6.9}/I_{7.7} \approx
\left[A_{6.9} B_{6.9}(T) N_{\rm C,\,aliph}\right]/\left[A_{7.7} B_{7.7}(T) N_{\rm C,\,arom}\right]$.
For a molecule to emit appreciably the PAH bands
at 3.3--11.3$\mum$, the thermal temperature needs to
be in range of $330 \simlt T \simlt 1000\K$. For such a
temperature range we obtain $B_{6.9}/B_{7.7}\approx0.9\pm0.2$
(Li \& Draine 2012).
Then, the ratio of the number of C atoms in aliphatic sidegroups
to that in aromatic benzene rings is derived from
$N_{\rm C,\,aliph}/N_{\rm C,\,arom}\approx
\left(I_{6.9}/I_{7.7}\right)\times\left(A_{7.7}/A_{6.9}\right)\times
\left(B_{7.7}/B_{6.9}\right)\approx 0.9\times  \left(I_{6.9}/I_{7.7}\right)\times\left(A_{7.7}/A_{6.9}\right) $.

%%%%%%%%% Figure 18%%%%%%%%
\begin{figure}[tbp]
\epsscale{0.8}
\plotone{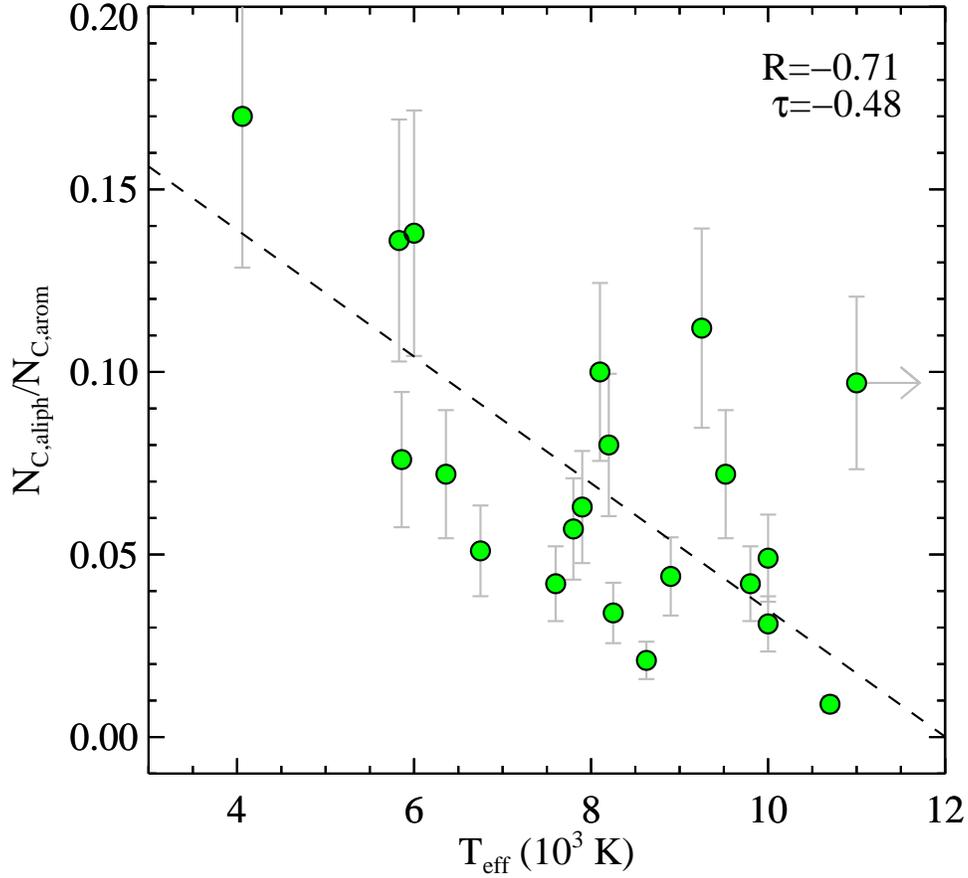}
\caption{\label{fig:ali}
         Ratios of the number of C atoms in aliphatic sidegroups
         to that in aromatic benzene rings
         ($N_{\rm C,\,aliph}/N_{\rm C,\,arom}$) derived from the
         6.9$\mum$ aliphatic C--H deformation band and the
         7.7$\mum$ aromatic C--C stretch band in correlation with $\Teff$.
         The correlation is linearly
         fitted ($N_{\rm C,\,aliph}/N_{\rm C,\,arom}=0.21{\pm0.03}
         -0.017{\pm0.004}\times(\Teff/10^3\K)$, dashed line), 
         and the correlation coefficient is measured as $R=-0.71$.
         For clarity, Wray 15-1484 is plotted with an arrow, which originally
         falls at $\Teff=30,000\K$. Note that we exclude Wray 15-1484 for
         the correlation analysis (see Section \ref{sub:ali}).        
         }
\end{figure}         
%%%%%%%%% Figure 18 %%%%%%%%

Figure \ref{fig:ali} shows the derived aliphatic fractions
($N_{\rm C,\,aliph}/N_{\rm C,\,arom}$) in comparison
with $\Teff$. It is clearly shown that
$N_{\rm C,\,aliph}/N_{\rm C,\,arom}$ decreases as $\Teff$
increases. This correlation has been pointed out previously
(\eg Acke \etal2010), and it is physically expected
since the aliphatic component is more highly susceptible
to local physical conditions than the aromatic rings
(\eg Joblin \etal1996). 
The aliphatic component would be easily destroyed 
in harsh environment (\eg $\Teff\ga10^4\K$), so 
$N_{\rm C,\,aliph}/N_{\rm C,\,arom}$ shows an
anti-correlation with $\Teff$, and even 
the aliphatic features are not present for those
with high $\Teff$.

In this context, the presence of aliphatic features in Wray 15-1484
($\Teff=30,000\K$) remains exceptional ($N_{\rm C,\,aliph}
/N_{\rm C,\,arom}\approx0.1$, see Figure \ref{fig:ali}). The nature
of Wray 15-1484 is still controversial. A protoplanetary nebula
or a pre-main-sequence star is a most likely scenario
(\eg Lachaume \etal2007), and the NIR polarization supports
an optically thin disk around a HAeBe star (Pereyra \etal2009).
Lachaume \etal(2007) resolved the innermost part of the
circumstellar matter at $N$-band (8--$13\mum$) with the 
Mid-Infrared Interferometric instrument (MIDI)
on board the Very Large Telescope Interferometer (VLTI)
and pointed out a large gap ($\simali30\AU$ in radius) with
a possible presence of an envelope or wind in addition to
the disk. If an additional envelope exists, the detected
aliphatic features may be ascribed to the envelope where
the environment is mild enough for the aliphatic component
to survive.

\subsection{Effect of Continuum-Subtraction}\label{sub:cntsub}

As mentioned in Section \ref{sub:extspec},
it is inevitable to force the baselines between
the PAH features in the residual spectrum to reach
a zero value when we use a spline fit to subtract
the continuum from the original spectrum. 
Hence, there would be a chance to oversubtract
the continuum, so we examine the effect
of the continuum-subtraction by comparing
between a spline fit and PAHFIT for two exemplary
sources (DoAr 21 and Oph IRS48).
Since PAHFIT can handle major and minor PAH
features, various dust features including
the silicate absorption bands at 9.7 and 18$\mum$ 
%but the 9.7$\mum$ silicate emission
(Smith \etal2007), we select DoAr 21 and Oph
IRS48 as a representative of those with the 9.7$\mum$
silicate absorption feature and those with the
PAH-dominated spectra, respectively.

%%%%%%%%% Figure 19%%%%%%%%
\begin{figure}[tbp]
\epsscale{1}
\plotone{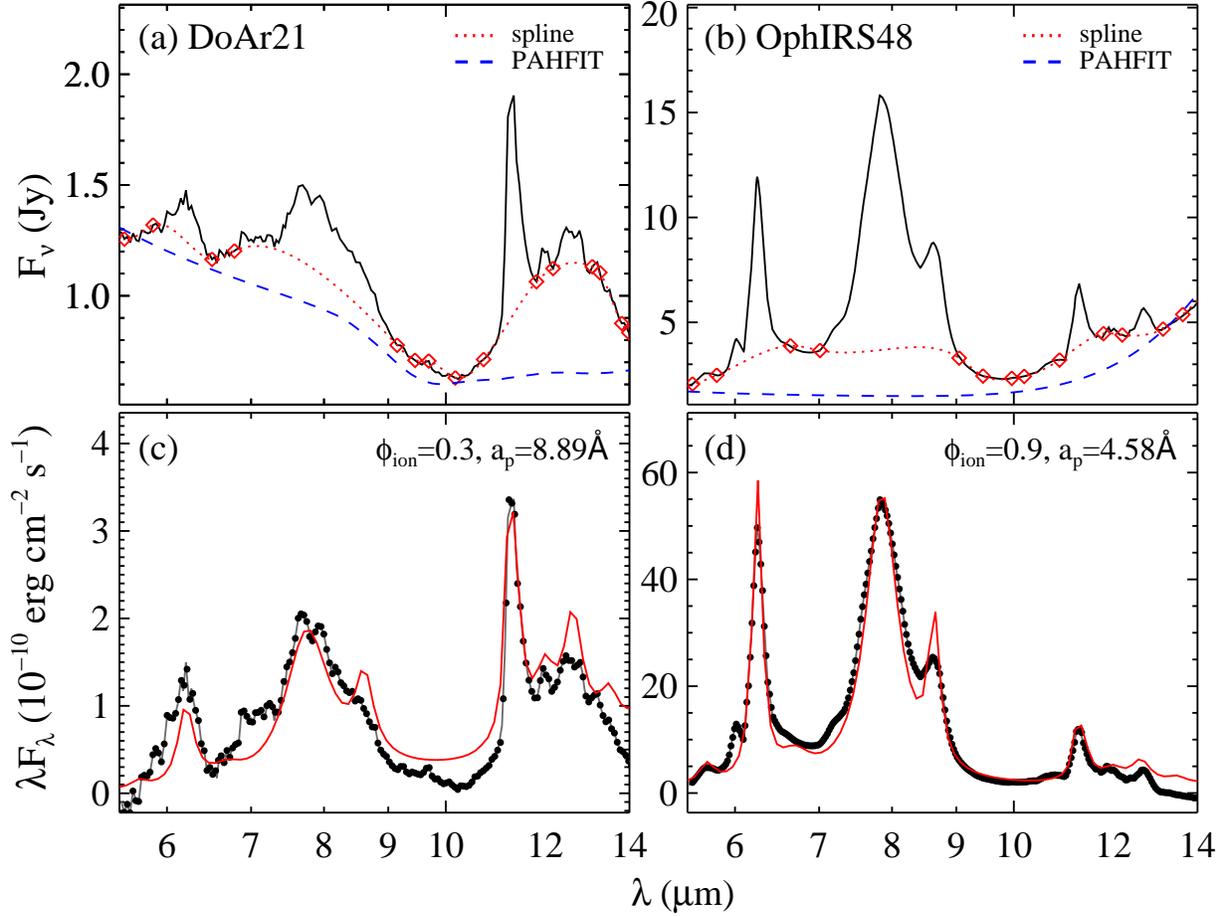}
\caption{\label{fig:pahfit}
        (a)--(b) Original spectra (solid lines) of two exemplary 
        sources (DoAr 21 and Oph IRS48) with a comparison
        between a spline fit (dotted lines) and PAHFIT
        (dashed lines) for continuum-subtraction.
        Anchor points used for a spline fit are
        marked with diamond symbols. In panels (c) and (d), 
        residual spectra of the two sources 
        using PAHFIT are shown with their best-fit model
        (red lines). Symbol
        designation is the same as Figure \ref{fig:spec}.
         }
\end{figure}         
%%%%%%%%% Figure 19 %%%%%%%%

The original spectra of the two sources before
the continuum-subtraction are shown in Figures
\ref{fig:pahfit}(a) and (b). Two continua derived from a cubic
spline and PAHFIT are overlaid (dotted and dashed
lines, respectively), which clearly
shows the difference between the two methods.
Note that the continuum of PAHFIT is the combination
of the total dust continuum and the starlight.
Since broad underlying plateaus commonly
attributed to large PAH clusters (\ie 100--1000
carbon atoms; Tielens 2008) are considered
as a continuum in a spline fit while they
are individually fitted as a PAH feature in PAHFIT,
differences at 6--9$\mum$ and
11--14$\mum$ between the two continua,
in particular for DoAr 21, are substantial. 
Even for wavelength ranges without any
significant features such as the 9--11$\mum$ range,
the absolute level of the PAHFIT continuum is
lower than that of the spline continuum, so
the baselines of the residual spectra are usually 
greater than zero (see Figure \ref{fig:pahfit}(c) and (d)).
In comparison with the residual spectra from the
spline fit (DoAr 21 and Oph IRS48 shown in Figures
\ref{fig:spec} and \ref{fig:spec7}, respectively),
the peak intensities of the PAHFIT residual spectra are
greater by a factor of $\simali$1.2--1.4.

We model the residual spectra from PAHFIT, and
the best-fit models are shown in Figure \ref{fig:pahfit}(c)
and (d). For both sources, the best-fit models reproduce
the residual spectra more closely than the results using
the spline fit (see Figures \ref{fig:spec} and \ref{fig:spec7}),
and $\chi^2/$dof decreases by a factor of $\simali$100.
However, the best-fit model parameters show the consistent
properties of PAHs (\ie large, neutral PAHs in DoAr 21\footnote{%
 Although we categorize the model parameters of DoAr 21
 as less reliable due to its relatively weak PAH features
 and the rough subtraction of the
 9.7$\mum$ silicate absorption feature (Table \ref{tab:para}), 
 it seems that the PAHFIT results consistently indicate the
 existence of large PAHs (\ie the lack of small PAHs)
 in this disk. This supports that the unique environment
 of DoAr 21 results in the efficient destruction of PAHs
 (see Section \ref{sub:dest}).
}
and slightly small, ionized PAHs in Oph IRS48), and the median
values of the top 70 models for each parameter
also agree with each other
(DoAr 21: $\langle\ap\rangle=15.22\pm6.52\Angstrom$
and $\langle\phiion\rangle=0.4\pm0.1$ for PAHFIT,
$\langle\ap\rangle=10.31\pm2.94\Angstrom$
and $\langle\phiion\rangle=0.3\pm0.1$ for spline, Oph IRS48:
$\langle\ap\rangle=5.64\pm1.05\Angstrom$
and $\langle\phiion\rangle=0.9\pm0.1$ for PAHFIT,
$\langle\ap\rangle=7.27\pm1.62\Angstrom$
and $\langle\phiion\rangle=0.7\pm0.1$ for spline).
This indicates that the correlations between the PAH 
parameters and the stellar properties that we found
(Figures \ref{fig:corr_teff}--\ref{fig:corr_peak})
would not be altered significantly by the exact
choice of continuum subtraction.
In addition, we check the effect on the total mass
of PAHs, which is directly affected by the absolute
scale of the residual spectra. We found
$M^{10\AU}_{\rm PAH}=2.91$ and $2.62\times10^{-5}\Mearth$
for DoAr 21 and Oph IRS48, respectively, which increase
by a factor of $\simali$1.1--2.4. This implies that
$M_{\rm PAH}$ derived in Section \ref{sec:res} could
be underestimated slightly (\ie $\simali$10--20\%)
for those with prominent PAH features and 
by a factor of $\simali$2 for those with weak PAH features.

The total mass of PAHs at 10\,AU
($M^{10\AU}_{\rm PAH}$) listed in Table \ref{tab:para}
and its possible underestimation may 
give one an impression that those sources 
with prominent PAH features tend to have 
a lower mass than those with weak PAH features.
This is most likely due to the different starlight intensities 
for different sources: those with strong PAH features
are usually HAeBe stars with relatively high $\Teff$
(\ie high $U$, \eg HD 85667: $\Teff=20900$ K,
MWC 297: $\Teff=24000$ K),
whereas those with weak PAH features are mostly
TTSs with low $\Teff$ (\ie low $U$, \eg EC82:
$\Teff=4060$ K, IC348LRL110: $\Teff=3778$ K).
Even if the PAH features from a source with low $U$
are weak, one may still require a large amount of PAHs 
to account for the observed PAH emission because of
its low $U$ ($M_{\rm PAH}\propto1/U$).
Moreover, the detected PAH features from these
low-$U$ sources might be biased toward unusual cases
emitting stronger PAH emission relative to other sources
with similar $U$ (possibly due to more abundant PAHs)
since we might not be able to detect
those whose PAH feaures are weaker.
Lastly, $M^{10\AU}_{\rm PAH}$ does not represent
the actual total mass of PAHs in PPDs. 
It must be reviewed taking into account 
the spatial distribution of PAHs in each disk, 
which will give a true insight into the relation 
between the stellar properties 
and the PAH mass in PPDs.

\section{Summary}\label{sec:summ}

We have accumulated a sample of 69 PPDs
reported to show PAH features in the literature
and have presented their IR spectra
collated from the literature and data archives.
Our sample includes 14 TTSs and 55 HAeBe stars,
which allows us to examine PAHs in PPDs with a wide
range of stellar parameters. 
We have modeled the PAH emission of 61 PPDs,
excluding eight PPDs of which the IR spectra are not
adequate for model calculations, and the best-fit
models mostly provide an excellent fit to the PAH
emission features of the PPDs. 
In the model calculations, the size distribution of PAHs
characterized by the peak and the width of the log-normal
distribution ($a_0$ and $\sigma$) as well as the ionization
fraction ($\phiion$) are derived coherently for the
large sample of PPDs.
Our principle results are as follows:

1. For the 61 PPDs, the peak of the PAH mass distribution
($\ap$) is calculated from $a_0$ and $\sigma$, which
represents the overall PAH size distributions. 
It is found that
$\ap$ shows a positive correlation with the stellar
effective temperature ($\Teff$), indicating that small
PAHs are dominant in disks around cool stars while
large PAHs are in those associated with hot stars.
This trend is a natural result from the photodissociation 
of PAHs induced by energetic photons, which destroy
small PAHs more rapidly than large PAHs. 
The peak size $\ap$ also shows a correlation with
$\Lstar/\Mstar$, which appears as tight as with $\Teff$.
This is probably due to the dependence of $\Lstar$
on $\Teff$ ($\Lstar\propto\Teff^4$).

2. The peak size $\ap$ does not exhibit any tight correlation with
the stellar age, but the overall trend indicates that
$\ap$ decreases with the stellar age. To maintain
the abundance of small PAHs against complete
destruction by photodissociation, small PAHs
are required to be replenished rapidly. This
might imply that a continuous supply of PAHs 
via the outgassing and collisions of
planetesimals, asteroids, and cometary bodies
is favored as a primary mechanism of the replenishment. 

3. Unlike $\ap$, $\phiion$ shows no significant
correlation with $\Teff$. This lack of correlation is attributed
to the diversity of the spatial distribution of PAHs
in disks from one to another. Since $\phiion$ depends
on $U(r)$ ($U\propto r^{-2}$), the distance of PAHs
from the central star should be taken into account
to properly interpret $\phiion$. 

4. The shift of the peak wavelength of the 7.7$\mum$
PAH feature ($\lambda_{7.7}$) shows a moderate
correlation with $\ap$: as PAH molecules become smaller,
$\lambda_{7.7}$ tends to shift toward longer wavelengths.
This is consistent with the negative correlation between
$\lambda_{7.7}$ and $\Teff$ that has been previously
reported because small PAHs are dominant in disks
around cool stars (\ie a positive correlation between $\ap$ 
and $\Teff$). 

5. Twenty one sources in the sample show the aliphatic feature
at 6.9$\mum$, indicating aliphatic CH sidegroups of 
PAHs in those systems. The aliphatic fractions
($N_{\rm C,\,aliph}/N_{\rm C,\,arom}$) tend to
decrease with increasing $\Teff$ since the aliphatic
component is more susceptible to local physical 
conditions such as $\Teff$ than the aromatic
rings.

Increasing the number of PPDs with spatially resolved
PAH emission obtained with ground-based telescopes
as well as unprecedented results from the up-coming
space mission, {\it James Webb Space Telescope},
will enable us to more precisely quantify the physical
properties of PAHs, in particular the total mass of PAHs,
and to better answer the role of
PAHs in the evolution of PPDs and possible planet formation.

\acknowledgements

We thank X. J. Yang for her kind assistance with Figure 15 and
V. C. Geers for generously providing the VLT data.
We thank X. J. Yang, H. Zhang, and the anonymous referee
for very helpful discussions and suggestion. 
We are supported in part by
NSF AST-1109039, 
NNX13AE63G, 
NSFC\,11173019, 
and the University of Missouri Research Board.

\appendix
\section{Brief description on the individual sources}
\subsection{AB Aur}

AB Aur is one of the well-known transitional disks,
exhibiting a truncated dusty disk at an inner radius of
$\simali$70\,AU (Pi\'etu \etal2005) with a compact
inner disk (Tang \etal2012). 
NIR observations show the presence of spiral arms
in the disk (Fukagawa \etal2004).
The dust disk is highly asymmetric and shows a
horseshoe morphology, which is interpreted
as the birthplace of future planets. Hashimoto \etal(2011)
have estimated an upper limit of 5 $M_J$ for
a possible companion of AB Aur although
planets have not yet been detected in the disk. 

\subsection{AK Sco}

AK Sco A/B (a.k.a. HD 152404A/B) is a binary system
composed of two nearly-identical
F5 stars in an eccentric orbit. The orbital inclination is 
constrained to the range of $65\degr<i<70\degr$
(Alencar \etal2003), but eclipses have not been detected.
Although the IR spectra do not show any significant 
MIR-variability (K\'osp\'al \etal2012), the optical spectra
of AK Sco exhibit emission or absorption features
that show variation in shape (Alencar \etal2003).
This indicates the presence of outflow and infall activites
in this system.
Recent observations with {\it Hubble Space Telescope}
($HST$) reveal a drop of the H$_2$ flux by up to $\simali$10\%
caused by gas infalling (G\'omez de Castro \etal2016).

\subsection{BD+40$\degr$4124}

BD+40$\degr$4124 (a.k.a. V1685 Cyg, MWC 340) is one
of the youngest objects in our sample (\eg Manoj \etal2006;
Liu \etal2011). BD+40$\degr$4124 is located within a small
pre-main-sequence star cluster (\eg Sandell \etal2011),
consisting of two HAeBe stars (BD+40$\degr$4124 and
LkH$\alpha$ 224) and an embedded binary system,
LkH$\alpha$ 225 (=V1318 Cyg).
Sub-millimeter observations reveal that BD+40$\degr$4124
is not associated with any significant disk continuum emission,
instead, the extended emission from the mother cloud dominates
the observed submm emission (Sandell \etal2011). The $Spitzer$
70$\mum$ image shows extended emission features near 
BD+40$\degr$4124, which likely originate from
heated dust from the reflection nebulosity rather than the PPD.
In this context, its MIR spectrum is dominated by PAH emission
features, of which the profiles are similar to those of diffuse ISM,
and this indicates the origin of the PAH emission from the
reflection nebulosity. 

\subsection{BF Ori}

The Herbig Ae star BF Ori is known to have strong photometric
variability (Shevchenko \etal1993), which is similar to that of
UX Ori. MIR spectra also show variability of the 9.7$\mum$
silicate emission feature over a decade (K\'osp\'al \etal2012).
An edge-on circumstellar disk is indicated from an
anti-correlated linear polarization (Grinin \etal1991).
Episodic accretion around BF Ori has been monitored
(de Winter \etal1999), suggesting evaporation of star-grazing
bodies, which is also called the ``$\beta$ Pic phenomenon''
(\eg Grinin \etal1996).
 
\subsection{DoAr 21}

DoAr 21 (a.k.a. Haro 1-6, ROXR1 13, RXJ1626.03-2423,
V2246 Oph) is one of the
brightest X-ray weak-lined TTS (WTTS; Neuh\"auser et al. 1994).
An extended ring structure is detected at NIR H$_2$ emission
(Panic 2009, Ph.D. Thesis), which is at $\simali$73--219\,AU away from the
central star. This is consistent with the estimation of a cavity
size of $r_{\rm cav}=70^{+30}_{-10}\AU$ (van der Marel \etal2016).
Jensen \etal(2009) claimed that the observed
PAH, H$_2$, and continuum emission can be associated with
a small scale PDR rather than a PPD.

\subsection{EC 82}

EC 82 with a spectral type of K7-K8 (Sturm \etal2013;
Rigliaco \etal2015) is associated with the Serpens cluster.
Its VLT/VISIR $N$-band spectrum is dominated by the strong
9.7$\mum$ silicate emission feature (Geers \etal2007b),
and the tentative detection of the 11.2$\mum$ PAH feature
in their spectrum gives a radial spatial extent of $\la0\farcs23$
(\ie $\la95\AU$ at 415\,pc). This object is one of the coolest stars
($\Teff\approx4060 \K$)
that exhibit PAH features in their IR spectra. 
This provides observational support to the theory of 
Li \& Draine (2002) which suggests that the excitation of
PAHs does not require UV photons, instead, visible and
NIR photons are capable of exciting PAHs sufficiently
to emit at the PAH bands (also see Mattioda \etal2005).

\subsection{HD 31648}

The isolated Herbig Ae star HD 31648 (a.k.a. MWC 480) has
a spectral type of A2pshe (Keller \etal2008), located in the
Taurus-Auriga complex. A large circumstellar disk around
HD 31648 has been spatially resolved in CO and (sub-)mm 
continuum (\eg Simon \etal2000; Hamidouche \etal2006;
Pi\'etu \etal2007; Sandell \etal2011), but the disk is not seen
in coronagraphic NIR images obtained with {\it HST}
(Augereau \etal2001). While $^{12}$CO
$J=2-1$ and sub-mm (480 and 850$\mum$) emissions are
extended up to $\simali$8$\arcsec$ (\eg Pi\'etu \etal2007;
Sandell \etal2011), the 1.4\,mm continuum shows
a disk size of only $0\farcs8\times0\farcs7$ (Hamidouche \etal2006).

\subsection{HD 34282}

HD 34282 is a Herbig Ae star (A0V--A3V, \eg Mer\'in \etal2004;
Keller \etal2008), and its distance is rather uncertain. The
$Hipparcos$ parallax gives a distance of $164^{+60}_{-30}\pc$,
which results in an ``anomalous position'' of HD 34282 in
the H-R diagram due to a low luminosity (\ie well below the
expected luminosity of a star of its spectral type, $\Lstar=4.8\Lsun$;
van den Ancker \etal1998). Based on the Keplerian motion inferred
from CO observations (Pi\'etu \etal2003), its distance is newly
constrained as $d=400^{+170}_{-100}\pc$, which yields $\Lstar=
29^{+30}_{-13}\Lsun$ more consistent with the stellar properties
expected for the spectral type of HD 34282 (Mer\'in \etal2004). 
Its disk has been spatially resolved at 1.3\,mm (Pi\'etu \etal2003)
but is unresolved at 480 and 850$\mum$ with SCUBA on the
James Clerk Maxwell Telescope (JCMT; Sandell \etal2011). 
A spatially resolved MIR $Q$-band image obtained with
Gemini north/MICHELLE infers the presence of a large dust
gap in the disk ($r_{\rm cav}=92^{+31}_{-17}\AU$;
Khalafinejad \etal2016). 

\subsection{HD 34700}

HD 34700 (a.k.a. SAO 112630) has a spectral type of
G0IVe (Sch\"utz \etal2009) and is probably in transition
from a TTS to a Vega-like star (\eg Seok \& Li 2015). Its
distance is very uncertain, ranging from $\simali$100
to 430\,AU (\eg Acke \& van den Ancker 2004; Torres 2004;
Seok \& Li 2015), and we adopt 260\,AU following Seok \& Li
(2015). The HD 34700 system composed of a spectroscopic
binary (Torres 2004) and two other faint components is
confirmed with $JHK$-band images and optical
spectroscopy (Sterzik \etal2005).
Although it has not been spatially resolved,
analysis of its SED indicates the dust spatial distribution
of the HD 34700 disk
peaks at $\approx100\AU$ and decreases outward
with a power-law index of $\approx2$ (Seok \& Li 2015).
The PAH parameters derived in this work
($\langle a_0 \rangle=2.5\pm0.5\Angstrom$,
$\langle\sigma\rangle=0.3\pm0.1$, and
$\langle\phiion\rangle=0.2\pm0.1$) are consistent with the
previous results from Seok \& Li (2015), which indicate small PAH
population ($a_0=3.5\Angstrom$ and $\sigma=0.3$)
with a mixture of neutral and ionized PAHs: 
for small PAHs ($a\la5\Angstrom$), $\phiion\simali$0.2
at $r=25\AU$ (see their Figure 3).

\subsection{HD 35187}

The Herbig Ae star HD 35187 (a.k.a. BD+24$\degr$826, SAO 77144)
is a young binary system (HD 35187A: A7V, HD 35187B: A2V;
Dunkin \& Crawford 1998). The separation of two stars is
by $\simali$$1\farcs39$ at a distance of 150\,pc. 
Dunkin \& Crawford (1998) suggested that HD 35187B is
surrounded by a substantial disk of gas and dust whereas
HD 35187A is not. Therefore, the observed IR emission 
including PAH features is likely to originate from 
HD 35187B. HD 35187 is a relatively strong radio source
(\eg Natta \etal2004), which is dominated by free-free
emission and by dust emission at wavelengths 
shorter than 1.3\,mm.
SCUBA submm observations did not resolve
the source (\eg Sheret \etal2004; Sandell \etal2011).
Both stars are located close to the
zero-age main sequence (ZAMS) in the H-R diagram,
and the SED of HD 35187 is similar to those
of evolved HAeBe stars seen by Waelkens \etal(1994).
This implies that this system is in transition between
the HAeBe stars and the Vega-like stars (\eg Dunkin
\& Crawford 1998). 

\subsection{HD 36112}

HD 36112 (a.k.a. MWC 758) has a transition disk around
the Herbig Ae star with a spectral type of A5IVe (\eg van Boekel
\etal2005), which has been extensively observed in
multi-wavelengths (\eg Chapillon \etal2008; Isella
\etal2008, 2010; Grady \etal2013;
Benisty \etal2015; Marino \etal2015). Its disk structure
shows spiral arms, non-axisymmetric dust continuum,
and axisymmetric CO emission, which can be associated
with a massive planet at $\simali$160\,AU (Dong
\etal2015). Dust observations at mm regime infer an inner
cavity with a radius of $\simali$73\,AU possibly due to
a low-mass companion within $\simali$42\,AU (Isella \etal2010;
Andrews \etal2011). NIR observations with the VLTI
(Isella \etal2008) show 
strong emission coming from sub-AU scales, indicating
that the disk can be classified as a pre-transitional disk
(\eg Espaillat \etal2007).

\subsection{HD 36917}

The Herbig Ae star HD 36917 (a.k.a V372 Ori) is a spectroscopic
binary (B9.5+A0.5; Levato \& Abt 1976).
HD 36917 is a member of the young
Orion Nebula Cluster, which implies that its age is younger than
1\,Myr (Manoj \etal2002). High-resolution MIR imaging at 12
and 18$\mum$ with MICHELLE on the Gemini North telescope 
did not resolve its circumstellar disk (Marin\~as \etal2011), which
gives a upper limit of $\la0\farcs25$ for its angular size (\ie
$\la101\AU$ at the distance of $\approx375$\,pc).
 
\subsection{HD 37357}
HD 37357 has a spectral type of A0V and is a member of
the Orion OB1c association (Juh\'asz \etal2010).
Its pulsational variability is discovered from high-precision 
time-series photometry obtained with the Microvariability
and Oscillations of Stars (MOST) and Convection,
Rotation et Transits Plan\'etaires (CoRoT) satellites
(Zwintz \etal2014).

\subsection{HD 37411}

The Herbig Ae star HD 37411 was discovered by
Hu \etal (1989) and is classified as a $\lambda$ Boo star
(Gray \& Corbally 1998), which is metal deficient but
has nearly solar abundances of CNO and S
(Morgan \etal1943). The existence of a disk with an
inclination of 63$\degr$ is
inferred from the SED analysis (Liu \etal2011).
Boersma \etal(2008) reported a strong 11.1$\mum$
feature in HD 37411, which is often detected as a minor
feature accompanying the 11.2$\mum$ PAH feature 
(Hony \etal2001). However, the 11.1$\mum$ feature seen in
HD 37411 is unusually strong relative to the 11.2$\mum$
feature. They claimed the presence of warm crystalline
silicate in the source based on the presence of
the broad 10$\mum$ feature,
which implies that that the emission from forsterite
might account for the 11.1$\mum$ feature. 

\subsection{HD 37806}

HD 37806 (a.k.a. MWC 120) has a spectral type of B9--A2
(Kessler--Silacci \etal2006) and is a probable variable (van 
den Ancker \etal1998). Its circumstellar disk has not been
spatially resolved with MICHELLE on the Gemini North
telescope (\eg Marin\~as \etal2011), which gives an
upper limit of $0\farcs31$ and $0\farcs33$ at 12 and 18$\mum$,
respectively. 

\subsection{HD 38120}

HD 38120 has a spectral type of B9 (\eg Juh\'asz \etal2010),
and its age estimate ranges 
from $\la1.0$\,Myr (\eg Sartori \etal2010; Manoj \etal2010) to 
$\simali$5$^{+0.5}_{-0.5}$\,Myr (\eg Alecian \etal2013)
probably owing to different distance assumption
(\eg $d=422\pm230 \pc$; Donehew \& Brittain 2011). 
CO emission from its disk has been detected (Dent \etal2005),
and the best-fit to the $J=3$--2 CO line profiles with a simple
disk model constrains the properties of the disk,
such as dust mass of $\simali$10$^{-5}\Msun$ 
and an outer radius of $\simali$300\,AU 
at the distance of 420\,pc. 
 
\subsection{HD 58647}

HD 58647 is a young Herbig Be star (B9IV: \eg Juh\'asz \etal2010),
of which the age is $\approx1$\,Myr (Mari\~nas \etal2011)
or even younger (0.4\,Myr: van den Ancker \etal1998,
0.16\,Myr: Montesinos \etal2009). It is located at 543\,pc away
(Mari\~nas \etal2011), but other studies estimate that it can
be closer ($280^{+80}_{-50}$\,pc: van den Ancker \etal1998,
$318^{+65}_{-46}$\,pc: van Leeuwen 2007). 
Using the $K$-band observations of the Keck interferometer,
Monnier \etal(2005) measured the radius of a geometric ring of
the $K$-band continuum emission, which corresponds
$0.82\pm0.13$\,AU at a distance of 280\,pc. 
The $K$-band emission region is considered to arise near
the dust sublimation radius of the accretion disk. 
Some of its hydrogen emission lines (\eg H$\alpha$, Br$\gamma$)
appear to have a double-peak profile. NIR observations with
high angular and high spectral resolution
(\eg Kurosawa \etal2016) indicate that the Br$\gamma$
emission is from a disk wind with a radius of $\sim0.5$\,AU,
which is within the inner radius of the $K$-band
continuum-emitting ring ($0.68$\,AU). 
 
\subsection{HD 72106}
 
HD 72106 is a binary system, consisting of the primary star,
HD 72106A, identified as a magnetic star (Wade \etal2005)
and the secondary star, HD 72106B, identified as a HAeBe
star (Vieira \etal2003). The binary stars are separated by 
$0\farcs805$ (ESA 1997) or $\approx280$\,AU at a distance
of $289^{+204}_{-85}$\,pc (Alecian \etal2013). HD 72106B is the one 
emitting PAH emission as well as other dust emission
(\eg Sch\"utz \etal2005). Sch\"utz \etal(2005) obtained
$N$-band spectra (8--13$\mum$)
with the ESO TIMMI2 camera at La Silla observatory.
It is found that the spectrum of HD 72106 is dominated
by crystalline forsterite and enstatite rather than small,
amorphous silicate grains. Also,
large amorphous silicates and SiO$_2$ are present. 
Such dust species are known to result from dust processing
in circumstellar disks, and only a few pre-main-sequence
stars (\eg HD 100546, HD 179218) have shown them so far.
Its IR spectrum shows a remarkable resemblance with
a combination of spectra of the comets Halley and
Hale-Bopp, which makes this object interesting for searching
for on-going planet formation in its disk. 
While HD 72106A shows strong peculiarities in
chemical abundances (Folsom \etal2008), the large
majority of the chemical elements in HD 72106B is consistent
with solar abundances as commonly found in other
HAeBe stars (\eg Acke \& Waelkens 2004).
Schegerer \etal(2009) performed interferometric observations
in $N$-band with the MIDI at the VLTI and found that
a purely passive disk (\ie no accretion) with an inner and
outer disk radii of $\simali$0.5 and $\simali$40\,AU, respectively.

%{\bf stellar parameters in table 1 seem to be for the primary star
%(see Table 2. in Folsom et al. 2008)}

\subsection{HD 85567}

HD 85567 (a.k.a. V596 Car, Hen 3-331) is a Herbig Be star
(B7-8Ve: Juh\'asz \etal2010) located at $1.5\pm0.5$ kpc
(Verhoeff \etal2012). The existence of a close binary
companion interacting with the circumstellar disk of HD 85567
was proposed (Miroshnichenko \etal2001), and the binary
companion with a separation of $\ga0\farcs5$ is confirmed
later (Baines \etal2006). Vural \etal(2014) performed NIR
interferometric observations to investigate the sub-AU scale
disk structure of HD 85567 with the AMBER instrument at
the VLTI and found that the inner radius of the disk
($R_{\rm in}\simali$$0.67\AU$ with a temperature of
$T_{\rm in}\simali$$2200\K$)
is smaller than that expected from the size-luminosity
relation. This implies that an optically thick
gaseous inner disk exists to shield the stellar radiation
of the central star so that dust can survive closer to
the star than the radius of dust sublimation ($T\simali$$1500\K$).

\subsection{HD 95881}

HD 95881, a Herbig Ae star (A2III/IVe: Acke \&
van den Ancker 2004), is classified as a group II source
following the classification of Meeus \etal(2001). Group II
sources have blue SEDs in the MIR spectral range
(10--60$\mum$), implying a flat or self-shadowed disk geometry.
Although group II sources
usually show weak or no PAH emission (\eg Acke \etal2010),
HD 95881 shows prominent PAH emission as well as
[O I] 6300$\Angstrom$, both of which indicate the presence
of gas in the upper layers of the disk and direct irradiation by
stellar photons. Verhoeff \etal(2010) carried out a
comprehensive study toward the disk of HD 95881
using NIR ($K$-band)
and MIR (10$\mum$) interferometric observations
with the AMBER and MIDI instruments at the VLTI and
$Q$-band imaging and $N$-band spectroscopy with VLT/VISIR.
They found that the PAH emission is more extended
($R\simali$$100\AU$) than the dust continuum (mostly located within
$\la2.5\AU$) and concluded that the inner disk
has a puffed up inner rim containing most of 
the grains emitting in the IR wavebands 
(large grains probably existing deep inside the
disk and detectable only in mm are excluded).
Whereas a mixture of gas and PAHs with a flaring disk geometry
exists in the outer disk, dust grains are depleted due to
coagulation and settled down to the mid-plane. 

%luminosity of 15.4 Lsun (stellar parameter in Tab. 1)

\subsection{HD 97048}

HD 97048 (a.k.a. CU Cha), a Herbig Ae star (A0pshe:
Keller \etal2008), is located at $\simali$$150\pc$ (Alonso-Albi
\etal2009) or $175^{+26}_{-20}\pc$ (van Boekel \etal2005).
This is one of the several PPDs showing prominent nanodiamond
features at 3.43 and 3.53$\mum$ together with the 3.3$\mum$
PAH feature in its NIR spectrum (\eg Habart \etal2004c).
Both PAH and nanodiamond features have been spatially resolved
(\eg Habart \etal2004c, van Boekel \etal2004, Habart \etal2006,
Quanz \etal2012): while the PAH features at 8.6, 11.3 
and 12.7$\mum$ are extended on a scale of 
a few 100$\AU$ (van Boekel \etal2004,
$\sim80$--$370\AU$: Quanz \etal2012), 
the 3.3$\mum$ PAH emission mainly comes from 
a more confined region (Habart \etal2006).
The diamond features are slightly less extended than the
3.3$\mum$ feature (FWHM$\simali$$41\AU$: Habart \etal2006,
$\la15\AU$: Quanz \etal2012), and the adjacent
continuum is even less extended (FWHM$\simali$$23\AU$:
van Boekel \etal2004, or $\simali$$20$--$160\AU$ for the
3.5--18$\mum$ NIR/MIR continuum: Quanz \etal2012).
Recent observations including ALMA, ATCA, and VLT/SPHERE
(Spectro-Polarimetric High-contrast Exoplanet REsearch)
have revealed that the HD 97048 disk
has several ring (and gap) structures (\eg Maaskant \etal2013;
Ginski \etal2016; van der Plas \etal2016; Walsh \etal2016), which can be explained
by on-going low-mass planet formation (\eg $\approx0.7$ $M_{J}$
at 2.5--11\,AU: van der Plas \etal2016).

Using the CRIRES instrument at the VLT, Carmona \etal(2011)
detected H$_2$ 1-0 S(1) line emission at 2.12$\mum$
and found that the emission is extended
at least up to $200\AU$. Line ratios indicate H$_2$ gas at
$T>2000\K$, which implies the observed emission might
be excited by energetic electrons from X-rays.
The intensity of H$_2$ 1-0 S(1) increases by an order
of magnitude from the previous measurement carried out
in 2003 by Bary \etal(2008), which indicates line variability.

\subsection{HD 97300}

HD 97300 (Sp. type: B9V; Keller \etal2008) is one of the two
intermediate-mass stars in Chamaeleon I (the other is HD 97048),
which is one of the closest star-forming regions ($\simali$$150\pc$). 
A largely extended elliptical ring around HD 97300
($\simali$$7500\AU$ or $0.045\times0.03\pc$ in size) has been
found in MIR bands (\eg Siebenmorgen \etal1998; K\'osp\'al \etal2012),
of which the emission is considered to be dominated by PAH emission.
The origin of the ring structure can be explained by a bubble blown
into the surrounding interstellar matter and heated by the star
(K\'osp\'al \etal2012).
No IR-excess at $\la24\mum$ is found, indicating that
HD 97300 is close to a ZAMS (K\'osp\'al \etal2012). 

\subsection{HD 98922}

HD 98922 is a young ($\la0.01$\,Myr, Manoj \etal2006) Herbig Be
star (B9Ve; Juh\'asz \etal2010), and its distance is rather uncertain
(\eg $\simali$$500$--2000\,pc: Alecian \etal2013; Maaskant \etal2014;
Hales \etal2014). 
Like HD 95881, HD 98922 is classified as a group II source
(\ie self-shadowed disk) with the presence of PAH emission
and [O I] line emission (Acke \etal2005; Verhoeff \etal2010).
This implies that the [O I] emission might come from a gaseous
disk inside the dust-sublimation radius (Acke \etal2005).
Recent NIR interferometric observations with VLTI/AMBER
by Caratti o Garatti \etal(2015)
indicate that the Br$\gamma$-emitting region
($\simali$$0.31\pm0.04\AU$) is smaller than the continuum-emitting
region (inner radius of the dust disk: $\simali$$0.7\pm0.2\AU$).
Also, they have spatially resolved the circumstellar disk of HD 98922
with the VLT/SINFONI integral field spectrograph, which is extended
in $K$-band up to $\simali$$140\AU$ in diameter at a distance of
$440^{+60}_{-50}\pc$.
Its MIR spectrum around 10$\mum$ shows a broader and
flatter silicate feature than that typically seen in the ISM
(van Boekel \etal2003), indicating large (1--2$\mum$)
processed grains as well as minor spectral features associated
with crystalline olivines and pyroxenes. 

\subsection{HD 100453}

The Herbig Ae star HD 100453 (spectral type of A9Ve;
Keller \etal2008) is thought to be in transition from
a gas-rich PPD to a gas-depleted debris disk (Collins
\etal2009). 
Its $Spitzer$/IRS spectrum shows a very weak sign
of silicate features at 10 and 20$\mum$. This, together
with SED analysis, indicates the presence of a gap in
the disk (\eg Mari\~nas \etal2011; Maaskant \etal2013;
Khalafinejad \etal2016).
A spatially resolved MIR $Q$-band image obtained
with Gemini north/MICHELLE indicates that the outer
edge of the gap is at 20$^{+3}_{-3}$\,AU and the disk
is extended up to $\simali$200\,AU (Khalafinejad \etal2016). 
 
\subsection{HD 100546}

HD 100546 is one of the nearest Herbig Be stars (Sp. type:
B9Ve, Acke \& van den Ancker 2004; $d\simali$$103\pm6\pc$,
van Boekel \etal2005). It is the first disk in which crystalline
silicates (Hu \etal1989) and PAHs (Malfeit \etal1998) have
been detected.
Its 8.6 and 11.3$\mum$ PAH features
are spatially extended on a few 100$\AU$ scale (van Boekel
\etal2004), and the spatial distribution of the 3.3$\mum$
emission shows a gap in the innermost region ($R\simali$$5$--10$\AU$)
and is extended up to $\simali$$50\AU$ (Habart \etal2006)
or $R\simali$$12\pm3\AU$ (Geers \etal2007b).
Minor features at 3.4 and 3.46$\mum$ are also detected
(Habart \etal2006), which could be attributed to 
aliphatic C-H stretches in methyl or ethyl side-groups
attached to PAHs (\eg Joblin \etal1996; Yang \etal2013, 2016b).
Lisse \etal(2007) compared the MIR spectrum (5--35$\mum$)
of HD 100546 with those of Comet 9P/Tempel 1 and
Comet C/ 1995 O1 (HaleÐBopp), which shows similar 
emission signatures including PAH features as well as
crystalline silicate features.
Similar to HD 97048, H$_2$ 1-0 S(1) emission at 2.12$\mum$
is detected (Carmona \etal2011), which may be extended
at least up to $50\AU$ from the central star.
Based on $HST$/NICMOS2 coronagraphic observations at $1.6\mum$,
Augereau \etal(2001) detected an elliptical structure extended
up to 350--380$\AU$ and inferred that the disk 
is marginally optically thick inside 80$\AU$ and optically thin further out. 
A gas giant planet at $\simali$$50\AU$ from the central star
has been confirmed by direct imaging (Quanz \etal2015),
and multiple planet formation is expected from the HD 100546
system (\eg Currie \etal2015; Pinilla \etal2015; Quanz \etal2015).

\subsection{HD 101412}

HD 101412 (a.k.a. CD-59 3865) is a Herbig Ae star 
(Sp. type of B9.5Ve--A0; Geers \etal2006, Juh\'asz \etal2010). 
The distance to this star is rather uncertain
(\eg $d=600\pm100\pc$, Folsom \etal2012;
$d\simali$160\,pc, Juh\'asz \etal2010;
$d\simali$118\,pc, Maaskant \etal2015).
Geers \etal(2007b) performed $N$-band and $L$-band
observations toward HD 101412 using VISIR and ISAAC
instruments at the VLT, which place an upper limit on the radial spatial
extent of the 3.3, 8.6, and 11.2$\mum$ PAH features to be $\la38$,
$\la26$, and $\la24$\,AU, respectively at a distance of 160\,pc.
In the $N$-band spectrum, a weak and broad feature at 11.2$\mum$
confused with the 11.2$\mum$ PAH feature is confirmed, which 
is possibly due to crystalline silicates (Geers \etal2006).
Fedele \etal(2008) compared the spatial distribution of the gas
traced by [\ion{O}{1}] 6300$\Angstrom$ with that of the dust traced by
the 10$\mum$ emission in the HD 101412 disk,
and found that the disk is strongly flared in gas but
self-shadowed in dust beyond $\simali$2\,AU.
This might imply that the disk was initially flared then
became flat by gas-dust decoupling, grain-growth,
and dust settling. 

\subsection{HD 135344B}

HD 135344B (spectral type of F4Ve; Dunkin \etal1997),
also known as SAO 206462, is located at a distance of 142\,pc.
With an age of $8^{+8}_{-4}$\,Myr (van Boekel \etal2005),
HD 135344B is one of the most studied PPDs in detail.
Its disk structure
has been spatially resolved in multi-wavebands, which
shows dust cavity, spiral structures, asymmetries in the
outer disk (\eg Pontoppidan \etal2008; Brown \etal2009;
Muto \etal2012; P\'erez \etal2014; Stolker \etal2016).
Its remarkable features include the existence of CO 4.7$\mum$
emission inside the dust cavity at 45\,AU (Pontoppidan \etal2008)
and the NIR scattered light extending close to 25\,AU from
the central star (\eg Muto \etal2012). Its NIR-excess
indicates the existence of hot dust with a temperature
of $\simali$1500 K (\eg Coulson \& Walther 1995).
With the above characteristics, HD 135344B is considered
to have a pre-transitional disk. 
Its $Spitzer$/IRS spectra show weak PAH features and
no indication of silicate emission (\eg Geers \etal2006;
Keller \etal2008). The profile of its ``8.0''$\mum$ feature is
a blend of class B and C following the classification of
Peeters \etal(2002). The 8.6$\mum$ PAH feature is marginally
detected (Sloan \etal2005). The 11.2$\mum$ feature
is noticeably shifted to longer wavelengths
and is relatively broader than that typically seen
in the ISM (Sloan \etal2005; Geers \etal2006).
Pech \etal(2002) claimed that the broadening of the
11.2$\mum$ feature could be
due to anharmonicity, which might imply the presence
of very hot PAHs in the innermost part of the HD
135344B disk (Geers \etal2006). 
 
\subsection{HD 139614}
 
HD 139614 (Sp. type: A7Ve; Keller \etal2008) is considered
to be a pre-transitional disk (\eg Espaillat \etal2008), which is
characterized by NIR excess with weak or no silicate feature
at 10$\mum$ (Juh\'asz \etal2010), indicating a gap separation
between inner and outer disks. Based on MIR interferometric
data obtained with VLTI/MIDI combined with SED analysis,
Matter \etal(2014) suggest the presence of an optically thin
inner disk at 0.22--$2.3\AU$, a gap, and an outer disk from
$5.6\AU$. Later, with more comprehensive data,
Matter \etal(2015) found that an interaction between a single
giant planet ($\simali$$3M_J$ at $\simali$$4.5\AU$ from the star)
and the disk can induce the gap structure in warm $\mum$--sized
dust distribution appearing at $\simali$2.5--6$\AU$.

\subsection{HD 141569}

The nearby ($d=116\pm8\pc$; Folsom \etal2012) HAeBe star
HD 141569 (Sp. type: B9.5 Ve; Juh\'asz \etal2010) is well-known
to have a large ($R\simali$400\,AU), optically thin disk with
two ring-like structures: an outer ring at $\simali$325\,AU with 
a width of $\simali$150\,AU and an inner ring at $\simali$200\,AU
(\eg Augereau \etal1999; Weinberger \etal1999;
Mouillet \etal2001). Later, more complex structures including
outer spirals and an additional arc-like feature have been found
(\eg Biller \etal2015). Either an interaction with two M-type
companion stars (\eg Augereau \& Papaloizou 2004) or
planet formation (Wyatt 2005), or both, can account for the
disk structures. Recent observations have further discovered
an inner disk component inside the inner ring
(\eg Currie \etal2016; Konishi \etal2016), but no point-like
source has been detected within the gap between the
inner disk and the inner ring (Konishi \etal2016). 
Li \& Lunine (2003) modeled the IR emission from the
HD 141569 disk using a porous dust model together with
a population of PAH molecules and found that PAHs in the disk are
largely charged and dust grains with a high porosity are preferred.
The PAH parameters derived in this work
($\langle a_0\rangle=3.0\pm1.0\Angstrom$, 
$\langle\sigma\rangle=0.3\pm0.1$, and $\langle\phiion\rangle=0.9\pm0.1$)
are in good agreement with their results
($a_0\approx2.5\Angstrom$ and $\sigma\approx0.3$).

\subsection{HD 142527}

HD 142527 is a Herbig Fe star (Sp. type: F7IIIe;
Acke \& van den Ancker 2004) located in a distance of
$\simali$145\,pc (\eg Acke \& van den Ancker 2004;
Fukagawa \etal2010) or $\simali$200\,pc 
(\eg Alonso-Albi \etal2009; Maaskant \etal2014).
Its MIR (18.8 and 24.5$\mum$) images exhibit an inner disk
($R\simali$5\,AU), an outer disk with bright arc-like emission,
and a wide gap devoid of dust (and gas) between
$\simali$30 and 90\,AU (Fujiwara \etal2006).  
Recent ALMA observations have revealed the detailed disk
structures (\eg Casassus \etal2013, 2015; Fukagawa \etal2013;
Muto \etal2015) and flows of gas through the dust depleted
gap (Casassus \etal2013). It is found that there is a low-mass companion
(HD 142517B) with a stellar mass of $\simali$0.1--0.4$\Msun$
at $\approx12\AU$ from the central star (\eg Biller \etal2012;
Rodigas \etal2014; Lacour \etal2016), which may have played
a role in generating the large gap in the disk of the HD 142527 system.  

\subsection{HD 142666}

HD 142666 (a.k.a. V1026 Sco) is a Herbig Ae star (Sp. type:
A8Ve; Keller \etal2008) located in the Sco OB2-2 association.
Its group II type SED (Meeus \etal2001) and the presence of PAH
emission imply that the HD 142666 disk could be in transition
between a gas-rich flaring disk to a gas-poor self-shadowed disk
(\eg Verhoeff \etal2010). This object is reported to have Br$\gamma$
emission (\eg Eisner \etal2009).
Dent \etal(2005) detected CO emission,
predicting a nearly face-on disk with a radius of $\simali$45\,AU.
Based on the SCUBA 450 and 850$\mum$ images, Sandell \etal(2011)
reported the disk size to be $\la2\farcs2$ 
(\ie $\la330\AU$ at $d\simali$145\,pc).
van Boekel \etal(2003) found that the mass ratio of large (2.0$\mum$)
to small (0.1$\mum$) grains is larger than unity ($\simali$1.54)
and interpreted this as the removal of small grains
from the disk by radiation pressure. Using NIR imaging
polarimetry, Hales \etal(2005) reported a large degree of
polarization ($P\approx1.32\%$ in $J$-band),
which is intrinsic to the system. 

\subsection{HD 144432}

HD 144432 is a Herbig Ae star (Sp. type: A9Ve; Keller \etal2008)
located in the Sco OB2-2 association at a distance of
$145\pm43\pc$ (\eg van Boekel \etal2005) or $d\simali$253\,pc
(Keller \etal2008). Distinctive P Cyg profiles are seen in
its optical spectrum, indicating extensive stellar activity in
the HD 144432 environment (Dunkin \etal1997).
The HD 144432 disk has been spatially resolved in MIR
(Leinert \etal2004), and its size is measured to be $0\farcs014$
corresponding to $2.1\pm0.2\AU$ at $145\pc$.
HD 144432 exhibits strong silicate emission in its IR spectrum
(\eg Keller \etal2008), which is one of the strongest observed
in HAeBe stars (van Boekel \etal2003).

\subsection{HD 145718}

The Herbig Ae star HD 145718 (a.k.a. V718 Sco) has a
spectral type of A8III/IVe (\eg Keller \etal2008) or A5Ve
(Carmona \etal2010). It is likely associated 
with the $\rho$ Ophiuchus cloud.
It is a photometric variable with a small variation
(\eg $\simali$1\% in V), and its binarity is uncertain (\eg
Friedemann \etal1996).
Guimar\~aes \etal(2006) detected red-shifted absorption features
only in the Balmer lines, indicating circumstellar activity with
hydrogen-rich material. Dent \etal(2005) estimated a disk
radius of $\simali$$60\pm30\AU$ from the $J=3$--2 $^{12}$CO emission.
 
\subsection{HD 163296}

HD 163296 (a.k.a. MWC 275) is a young ($\simali$5\,Myr;
\eg van den Ancker \etal1998; Alecian \etal2013) Herbig Ae
star (Sp. type: A1Ve; Juh\'asz \etal2010).
The disk around HD 163296 has been spatially resolved
(\eg Grady \etal2000; Fukagawa \etal2010; Sandell \etal2011;
Garufi \etal2014), which is extended to $\simali$440\,AU
and shows a broken ring structure.
Asymmetric supersonic jets associated with HD 163296
are found, which are roughly perpendicular to the disk plane
(\eg Devine \etal2000; Wassell \etal2006).
Based on its $ISO$ spectra, van den Ancker \etal(2000)
reported that the IR fluxes of HD 163296 are dominated by solid
state features. The detection of a minor feature at 6.85$\mum$
attributed to aliphatic hydrocarbons has been reported
(Bouwman \etal2001), and a nanodiamond feature at 3.53$\mum$
is tentatively detected by Acke \& van den Ancker (2006).
van den Ancker \etal(2000) also pointed out the presence of a population
of very large, mm-sized grains in the HD 163296 disk. 
Recent observations with ALMA and VLA indicate a population
of large grains and pebbles ($\ga 1\cm$) in the inner region
(\ie $\la 50\AU$) of the disk and the presence of small grains
similar to those in the diffuse ISM in the outer (\ie $\ga150\AU$) disk
(Guidi \etal2016). Since the CO snowline at $\simali$90\,AU
has already been identified by ALMA (Qi \etal2015),
the different populations of dust grains along the radial distance
could imply grain growth at the CO snowline and transportation
to the inner regions (Guidi \etal2016). 

\subsection{HD 169142}

HD 169142 (a.k.a. MWC 925, SAO 186777) is a Herbig Ae star 
(A5Ve; Keller \etal2008) showing complex disk structures.
At least two cavities, one ring structure, and an outer disk
have been spatially resolved (\eg Honda \etal2012;
Quanz \etal2013; Osorio \etal2014; Momose \etal2015).
In addition, its SED shows a considerable NIR (2--6$\mum$) excess,
suggesting that the disk is pre-transitional (\ie the presence
of an inner disk; Espaillat \etal2007).
Recent observations revealed a point-like feature within
the inner cavity (Biller \etal2014; Reggiani \etal2014),
implying (multiple) planet formation in the disk. 
Seok \& Li (2016) performed a comprehensive modeling
of its SED as well as the PAH emission features with 
porous dust and astronomical-PAHs and found that
three dust populations and relatively small PAH molecules
($a_0=3.0\Angstrom$ and $\sigma=0.4$) with an ionization
fraction of $\phiion=0.6$ can explain the entire SED and
the observed PAH features. These results are consistent 
with those derived in this work
($\langle a_0\rangle=3.0\pm0.5\Angstrom$, $\langle\sigma
\rangle=0.3\pm0.1$, and $\langle\phiion\rangle=0.5\pm0.2$).

\subsection{HD 179218}

HD 179218 (a.k.a. MWC 614) is a young ($\simali$1\,Myr;
\eg van Boekel \etal2005; Alecian \etal2013) isolated HAeBe star
with a spectral type of B9e (Acke \& van den Ancker 2004).
As its IR spectra are dominated by crystalline forsterite
and enstatite rather than small, amorphous silicate grains
(\eg Bouwman \etal2001; Sch\"utz \etal2005;
van Boekel \etal2005), HD 179218 is known
to have the second largest percentage of crystalline dust
(van Boekel \etal2005). This indicates dust processing in the
circumstellar disk of HD 179218, and the presence of cold
enstatite at $\ga10\AU$ implies that enstatite is mostly
produced in the inner regions 
and is transported outward by radial mixing (van Boekel \etal2005). 
A double-ring-like emission at $\simali$10$\mum$ has been
spatially resolved (\eg Fedele \etal2008), peaking at
$\simali$1 and 20\,AU, respectively.
In addition, a gap at $\simali$10\,AU is reported based on
the MIR interferometry with VLTI/MIDI (Menu \etal2015).
[\ion{O}{1}] 6300$\Angstrom$ line emission tracing the distribution
of the gas in HD 179218 has been observed (\eg Fedele \etal2008;
van der Plas \etal2008), which suggests a flaring disk structure
in HD 179218. The [\ion{O}{1}] line emission is probably
extended to $\simali$65\,AU from the central star
(van der Plas \etal2008) but is mainly produced between
$\simali$1--10\,AU (Fedele \etal2008). 

\subsection{HD 200775}

HD 200775 (a.k.a. MWC 361) is a young, massive spectroscopic
binary ($\simali$0.1\,Myr, $\simali$10$\Msun$ for both components;
\eg Alecian \etal2008). It illuminates the reflection
nebula NGC 7023 which is well-known to show strong PAH emission.
Its distance is commonly adopted to be $\simali$430\,pc (\eg
Acke \& van den Ancker 2004; Alecian \etal2013), 
but more recently, Benisty \etal(2013) claimed 
$d=320\pm51\pc$ based on new radial velocity measurements.
A biconical cavity associated with HD 200775 has been observed
with CO and HI observations (Fuente \etal1998),
which is probably due to an energetic, bipolar outflow
during its earlier evolutionary stage.
Okamoto \etal(2009) found elongated emission extended
in $\simali$1000\,AU scale in MIR, perpendicular
to the cavity wall, which can be explained by an inclined
flared disk. They also reported an amorphous silicate
feature peaking at 9.2$\mum$, which is substantially
shifted from the typical wavelength of 
$\simali$9.5--9.8$\mum$ where the amorphous silicate
features seen in the diffuse ISM and other PPDs typically peak.
This might suggest alternation of grains due to plasma irradiation. 
HD 200775 displays a dipolar magnetic
field geometry with similar characteristics to the magnetic fields
of Ap/Bp stars (\eg Alecian \etal2008).

\subsection{HD 244604}

HD 244604 (a.k.a. V1410 Ori) is a Herbig Ae star (Sp. type:
A0--3; \eg Marin\~as \etal2011).
Depending on the assumed extinction 
($A_V\simali$0.14--0.57\,mag),
its stellar luminosity and age are estimated
to be in the range of $\simali$22--55$\Lsun$ 
and $\simali$1.9--8\,Myr, respectively 
(\eg Marin\~as \etal2011; Alecian \etal2013; 
Fairlamb \etal2015; Kama \etal2015).
We adopt $\Lstar=55\Lsun$ 
and $\tau=2.79$\,Myr from Alecian \etal(2013).
Its IRS spectrum shows strong silicate emission 
with weak PAH features.
The subtraction of the strong 9.7$\mum$ silicate feature
results in a substantial non-PAH residual 
in the 7--12$\mum$ range
(see Figure \ref{fig:spec5}).
Folsom \etal(2012) examined the chemical abundances 
of HD 244604, which show no selective depletion 
or enhancement.

\subsection{HD 250550}

HD 250550 (a.k.a. MWC 789, V1307 Ori)
is a young ($\simali$1\,Myr; 
\eg see Liu \etal2011, Fairlamb \etal2015) 
Herbig Be star with a spectral type of B4--5IIIe 
(\eg Juh\'asz \etal2010).
Its distance is commonly adopted to be $\simali$700\,pc
(\eg Catala \etal1991; Bouret \etal2003; Juh\'asz \etal2010),
but a much shorter distance ($\simali$280\,pc) also has been
listed in the literature (\eg Maaskant \etal2014; Hein Bertelsen
\etal2016). Catala \etal(1991) observed short-term variations
in CaII K line, indicative of rotational modulation seen in AB Aur 
or HD 163296. Using far-UV spectroscopy, Bouret \etal(2003) 
detected \ion{C}{3} and \ion{O}{4} emission lines, 
revealing very high temperature regions ($\approx3\times10^5$ K) 
around the star. This would suggest the existence of 
a circumstellar halo (\eg Leinert \etal2001).
Hein Bertelsen \etal(2016) reported a detection of broad CO line
emission from the HD 250550 disk and the FWHM of the CO
lines increases with $J$, the rotational quantum number,
suggesting a radially extended CO-emitting region.

\subsection{HD 259431}

HD 259431 (a.k.a. V700 Mon, MWC 147) is a Herbig Be
star (Sp. type: B1Ve--B6e; Juh\'asz \etal2010;
Sandell \etal2011; Li \etal2014) associated with the Mon R1
association, and illuminates the reflection nebula NGC 2247.
Its distance is estimated to vary from $\simali$290\,pc
(\eg Juh\'asz \etal2010) to 800\,pc (Kraus \etal2008b).
Depending on the adopted distance, its stellar luminosity
also largely varies (\eg $\approx320\Lsun$: Li \etal2014;
$760^{+500}_{-350}\Lsun$: Menu \etal2015;
1550$\Lsun$: Kraus \etal2008b; 2233$\Lsun$: Alecian \etal2013).
Li \etal(2014) found similarities between HD 259431 and
MWC 1080, suggesting that the extended, filamentary emission
at 11.2 and 18.5$\mum$ surrounding HD 259431 may
arise from an outflow cavity like MWC 1080. The
CO first overtone bandhead emission of HD 259431
has been detected (Ilee \etal2014), which extends from 0.89
to 4.3\,AU at an inclination of $52\degr$. This inclination is
in good agreement with that measured from $K$-band
spectrointerferometry with VLTI/AMBER by Kraus \etal(2008b).
Kraus \etal also spatially resolved NIR (2.2$\mum$) and
MIR (11$\mum$) emission from HD 259431, which are extended
to $\simali$1.3 and 9\,AU, respectively. They conclude that
the NIR emission is dominated by accretion luminosity from
an optically thick inner gaseous disk whereas the MIR
emission is attributed to the outer dust disk.
 
% resolved the PAH emission at $11.2\mum$.

\subsection{HD 281789}
 
HD 281789 (a.k.a. F04101+3103) is a HAeBe star
with a spectral type of A0-1 (\eg Furlan \etal2006; Keller \etal2008).
This object is not well studied in the literature, but it is 
included in a survey of TTSs in the Taurus star-forming regions
(Furlan \etal2006). Following the SED group classification
of Meeus \etal(2001), Keller \etal(2008) classified the HD
281789 disk as a group IIa member (\ie self-shadowed disk).
 
\subsection{IC348 LRL110}

IC348 LRL110 (Sp. type: M0.5, $\Teff\approx3778\K$;
Mer{\'{\i}}n \etal2010) is one of the coldest objects in our sample
that shows clear detection of PAH features (see Figure
\ref{fig:spec5}), supporting that the excitation of PAHs does
not require UV photons (Li \& Draine 2002, Mattioda \etal2005).
This object is associated with the Perseus star-forming region
located at $\simali$250\,pc and is photometrically selected
as a candidate for cold disks (\ie disks with large inner dust holes)
based on the ``Cores to Disks'' (c2d) $Spitzer$ survey data
(Evans \etal2009). Mer{\'{\i}}n \etal(2010) performed follow-up
IRS observations toward IC348 LRL110 and found the presence
of PAH emission as well as 9.7$\mum$ silicate emission but
no hint for a hole inside its disk. 

\subsection{IC348 LRL190}

Similar to IC348 LRL110, IC348 LRL190
(Sp. type: M3.75, $\Teff\approx3306\K$; Mer{\'{\i}}n \etal2010)
is associated with the Perseus star-forming region
($d\simali$250 pc) and is included in the c2d survey.
This object is suggested to have a hole with a radius of
$\simali$5$\pm1\AU$ (Mer{\'{\i}}n \etal2010; or
$R\simali$2--4$\AU$; van der Marel \etal2016).
Its IRS spectrum is rather noisy and dominated by
silicate emission (see Figure \ref{fig:nomod}), so
it is excluded in our model calculation. The presence of
PAH features in this very cold disk (even colder than
IC348 LRL110) is reported in 
Mer{\'{\i}}n \etal(2010, see their Table 8), 
probably based on the broad feature shown at 6.2$\mum$.

\subsection{IRAS 03260+3111}

IRAS 03260+3111 is a young stellar object in the NGC 1333
molecular cloud core (a.k.a. NGC 1333 SVS 3) and has
a binary system (Sp. type: B5+F2; K{\'o}sp{\'a}l \etal2012)
with a wide separation of $\simali$1160\,AU (\ie $3\farcs62$) assuming
a distance of $d=320\pc$ to Perseus (Herbig 1998).
This object shows a prototypical PAH spectrum of class A
defined by Peeters \etal(2002, see Figure~\ref{fig:spec6}
of this paper and Figure~5 of Sloan \etal2005).
Spatial variations in the strength and shape of the
PAH features with increasing distance from the central
star have been noticed (\eg Sloan \etal1999). In particular,
an excess at $\simali$10.8--11.0$\mum$ beside the 11.2$\mum$
PAH feature and a feature seen at $\simali$10$\mum$ disappear
further from the central star (Sloan \etal1999) similar to the trend that
the fraction of PAH cations decreases with increasing distance
(Joblin \etal1996). This indicates these features are associated
with PAH cations. 
Haisch \etal(2006) suggested that the primary is
a Class II based on the MIR photometry and the secondary
is at least a Class II YSO based on their upper limits of the
10$\mum$ fluxes.

\subsection{IRAS 06084--0611}

IRAS 06084--0611 is a star-forming region (also known as
GGD 12--15) embedded in the Monoceros molecular cloud
at a distance of $\simali$1050\,pc (Boersma \etal2009) or
$830\pm50\pc$ (Maaskant \etal2011). Two very bright,
massive stars have been detected (\eg Persi \& Tapia 2003;
Maaskant \etal2011): a cometary H{\sc ii} region VLA1 (IRS4)
and a Herbig Be star VLA4 (IRS2) with a spectral type of B3
(Boersma \etal2009). VLA4 shows the PAH emission features
in its MIR spectrum obtained with the TIMMI2 camera
mounted at the ESO 3.6 m telescope (\eg Boersma \etal2009).
The 11.2$\mum$ PAH emission is spatially resolved
(FWHM$\simali$$2\farcs2$ or 2200\,AU), which is more extended
than the nearby (10$\mum$) continuum (FWHM$\simali$$1\farcs2$
or 1300\,AU). Considering the large extent of the PAH emission,
it is most likely that the surrounding materials around VLA4
dominate the observed PAH emission. 

\subsection{J032903.9+305630}

SSTc2d J032903.9+3056 is a TTS (Sp. type: G7; Mer{\'{\i}}n \etal2010),
which is associated with the Perseus star-forming region.
Its IRS spectrum shows silicate absorption as well as water
absorption features indicating that it is close to an edge-on system
(Mer{\'{\i}}n \etal2010). Although the broad features at 7--9$\mum$
are likely due to PAH emission (see Figure \ref{fig:nomod}),
other PAH features are not clearly detected, so we exclude
this object for model calculation. 

\subsection{J182858.1+001724}

SSTc2d J182858.1+001724 is a TTS (Sp. type: G3; 
Mer{\'{\i}}n \etal2010) and is associated with the Serpens
star-forming region ($d\simali$260\,pc). 
This is one of the coldest objects ($\Teff\simali$5830 K;
Mer{\'{\i}}n \etal2010) in our sample with PAH-dominated spectra
(see Figure \ref{fig:spec6}).
J182858.1+001724 shows strong, extended emission
in the MIPS 70$\mum$ band similar to that of DoAr 21
(Mer{\'{\i}}n \etal2010),
which is likely due to contamination from the surrounding materials.

\subsection{J182907.0+003838}

SSTc2d J182907.0+003838 is a TTS associated with
the Serpens star-forming region (Sp. type: K7; Mer{\'{\i}}n \etal2010). 
Its IRS spectrum is dominated by silicate emission
and shows marginal detection of PAH features
(see Figure \ref{fig:nomod}, and also see Table 8 in Mer{\'{\i}}n \etal2010).
It does not show any hint for a hole in the disk.

\subsection{LkH$\alpha$ 224}

LkH$\alpha$ 224 (a.k.a. V1686 Cyg) is a HAeBe star and belongs 
to a small pre-main-sequence star cluster with BD+40$\degr$4124
and LkH$\alpha$ 225 (\eg van den Ancker \etal2000;
Sandell \etal2011). Its nature reported in the literature
is rather uncertain; its spectral type is reported as B2--3e
(\eg Juh\'asz \etal2010), B5 (\eg Alonso-Albi \etal2009),
A7e (\eg Acke \& van den Ancker 2004), and F9 (\eg Manoj \etal2006).
Similarly, its effective temperature ranges between $\simali$6170
and 15500 K (\eg Acke \& van den Ancker 2004; 
Manoj \etal2006; Alonso-Albi \etal2009).
LkH$\alpha$ 224 has characteristics of UX Ori type stars such
as optical variations with $\ga1$ mag on timescales of days to weeks.
The IR spectra obtained with $ISO$ are examined by van den Ancker
\etal(2000) in detail. IR emission lines detected in the IR spectra can be
explained by a single non-dissociated shock, and
the shock might be produced by a slow outflow from
LkH$\alpha$ 225.

\subsection{LkH$\alpha$ 330}

LkH$\alpha$ 330 (a.k.a. F03426+3214) is a classical TTS
(Sp. type: G3e; Geers \etal2006)
associated with the Perseus molecular clouds.
Brown \etal(2008) carried out  340 GHz (880$\mum$) dust
continuum imaging with the Submillimeter Array (SMA) and
resolved the disk with an inner hole ($r\sim40\AU$) devoid
of dust but gas still remaining. A steep transition between
the hole and the outer disk is found, which could be explained
by the truncation of the outer disk via gravitational
instabilities (\eg Brown \etal2008). 
Andrews \etal(2011) also measured the cavity size with
new SMA data, giving a slightly larger radius ($\simali$68\,AU).
The 1.3\,mm observations with the Combined Array for Research in
Millimeter-wave Astronomy (CARMA) revealed that a lopsided
ring at $\simali$100\,AU in the LkH$\alpha$ 330 disk (Isella \etal2013).
While the 6.2 and 11.2$\mum$ PAH features are clearly shown in the
$Spitzer$/IRS spectrum (see Figure \ref{fig:spec6}), the 7.7 and
8.6$\mum$ features, often coming together with the 6.2$\mum$
feature are only marginally detected. This might be affected
by the disk geometry such as disk thickness (Geers \etal2006),
or more likely, hidden by the strong silicate emission
feature at 9.7$\mum$.

\subsection{MWC 297}

MWC 297 (a.k.a. NZ Ser, PDS 518) 
is a B1.5Ve type Herbig Be star
(\eg Acke \& van den Ancker 2004).
It is one of the closest massive young stars 
($d\simali$250\,pc; \eg Acke \etal2008).
The Br$\gamma$ line and continuum emission
measurements of MWC 297 by Malbet \etal(2007) 
suggest a flat, optically thick accretion disk 
with an outflowing wind around MWC 297. 
The regions emitting the continuum emission 
are much more compact than the dust sublimation 
radius expected for an irradiated dust disk, 
and the line-emitting region is more extended 
than the continuum-emitting region (\eg Kraus \etal2008a).
Using interferometric spectrographs, AMBER and MIDI mounted
on the VLTI, Acke \etal(2008) have spatially resolved the IR emission
including the 9.7$\mum$ silicate feature and found that all these
emissions originate from a very compact region (FWHM$<1.5\AU$)
with no evidence for an inner emission-free gap. To explain
the observed IR emission, Acke \etal(2008) adopted a geometric model
consisting of three Gaussian disks, which can account for the IR
emission but underestimates the observed submm/mm emission.
Verhoeff \etal(2012) have marginally resolved the PAH emission in
the narrow-band peaking at 11.88$\mum$ with VLT/VISIR,
constraining a spatial extension of $\simali$110\,AU in diameter.
The mm observations obtained with the VLA and the IRAM Plateau de
Bure interferometers (Alonso-Albi \etal2009) also showed that
the MWC 297 disk is small, with an outer radius of $\simali$28\,AU.
NIR polarization produced by multiple scattering has been
observed (Pereyra \etal2009), supporting the presence of
an optically thick disk.

\subsection{MWC 865}

MWC 865 (a.k.a. V921 Sco, CD-42$\degr$ 11721, Hen 3-1300)
is a Herbig Be star with a spectral type of B0IVep (\eg Acke
\& van den Ancker 2004), embedded in a small, dark cloud
(Boersma \etal2009) as the most massive member of a small
star cluster (\eg Weidner \etal2010). Its distance is rather uncertain,
ranging from $\simali$160\,pc to 2.2 kpc (\eg Kreplin \etal2011
and references therein), thus other stellar parameters are also
not well-constrained 
(\eg $\Teff\approx14000\pm1000\K$, Borges Fernandes \etal2007;
$\Teff\approx30000\K$, Acke \& van den Ancker 2004). 
While the $ISO$ spectrum is dominated by strong PAH emission,
the IR spectra obtained with ground-based telescopes do not show
any evidence of PAH features but exhibit strong HI lines
such as Pf$\delta$ at 3.297$\mum$ and the Humphreys series
(Acke \& van den Ancker 2006). The discrepancy can be explained
by different aperture sizes, which indicates that the PAH emission
does not originate from a confined region but is rather extended,
possibly from the surrounding nebulosity (\eg Acke \& van den
Ancker 2004; Kraus \etal2008a). Boersma \etal(2009) have
studied the spatial distribution of the MIR emission features 
around MWC 865 and reported that an arc and a patch 
with an extent (FWHM) of $\simali$$1\arcsec$ 
and $\simali$$5\arcsec$ (\ie 400 and 2000\,AU 
at $d\simali$400\,pc) apart from the central star, respectively, 
show strong PAH emission beside one bright peak near the star. 
A tentative detection of the nanodiamond feature 
around 3.53$\mum$ is also reported (Acke \& van den Ancker 2006). 
The extinction along the line of sight toward MWC 865 is
strong (\eg $A_V=5.08$\,mag; Acke \& van den Ancker 2004),
and its SED is dominated by thermal emission of circumstellar dust. 
This, together with the detection of the HI emission lines,
suggests that MWC 865 is most likely surrounded
by an accretion disk.

 \subsection{MWC 1080}
 
MWC 1080 (a.k.a. V628 Cas) is a young ($\la1$\,Myr) 
triple system associated with a small cluster embedded
in the dark cloud LDN 1238 at a distance of 
$\simali2.2$\,kpc (\eg Acke \& van den Ancker 2004; 
Alecian \etal2013). Note that Eisner \etal(2003)
claimed that the distance to MWC 1080 is 
$\simali$$1\pm0.2$\,kpc (Hillenbrand \etal1992).
The primary, MWC 1080A (Sp. type: B0e; 
Acke \& van den Ancker 2004), 
itself is an eclipsing binary (Shevchenko \etal1994), 
and the secondary, MWC 1080B, is also a HAeBe star 
(Leinert \etal1997). CO and CS observations revealed
a biconical outflow cavity with a size of 
$0.3\pc\times0.05$\,pc around MWC 1080, 
which was most likely created by
the bipolar outflow from MWC 1080A (Wang \etal2008).
Using narrow-band MIR imaging at 11.2, 11.6, and 18.5$\mum$,
Li \etal(2014) showed that the filamentary nebulosity of the
MWC 1080 system extends to $\simali$0.15\,pc, which
traces the internal surfaces of the gas cavity.
Strong extended emission at 850$\mum$ surrounding
MWC 1080 is detected (Sandell \etal2011), and a faint
emission peak coincides with MWC 1080A.
With $N$-band (8--13$\mum$) spectra of the MWC 1080 system,
Sakon \etal(2006) recognized a weak feature
around 10.95--11.1$\mum$ (\ie 11.0$\mum$ feature)
attributed to a solo CH out-of-plane
wagging mode of cationic PAHs (Hudgins \& Allamandola 1999).
They found that the ratio of the 11.0$\mum$ feature to the
11.2$\mum$ PAH feature increases toward the central star
and interpreted this as the promotion
of PAHs' ionization to cationic species in the vicinity of the
central star.

\subsection{Oph IRS48}

Oph IRS48 (a.k.a. WLY 2--48), a Herbig Ae star
(Sp. type: A0; \eg Brown \etal2012a) is well-known 
to have a transitional disk, which is characterized 
by a gap in the disk. MIR imaging (18.7$\mum$)
has spatially resolved the dust disk around Oph IRS48,
peaking at $\simali$55\,AU from the central star
with a gap of $\simali$30\,AU in radius (Geers \etal2007a).
Its gas distribution traced by the 4.7$\mum$ CO fundamental 
rovibrational band also confirms 
a hole at $\simali$30\,AU (Brown \etal2012a).
van der Marel \etal(2013) reported a highly asymmetric
structure associated with large, mm-sized grains by ALMA
observations, deviating from the MIR emission tracing
small ($\um$-sized) grains. This is attributed to
a dust trap, which would be the start of core formation
(\ie planet formation) near the central star (\eg van der
Marel \etal2015). The radius of the hole revealed by
mm-sized dust is measured to be $\approx13\AU$,
even inside the gas hole (Brown \etal2012b).
Its IRS spectrum shows strong PAH emission (see Figure
\ref{fig:spec7}), dominated by ionized PAHs ($\phiion=0.8$).
Maaskant \etal(2014) explained that highly ionized PAHs
located in optically thin regions inside the disk cavity overwhelm
the IRS spectrum whereas neutral PAHs from the outer disk
contribute insignificantly. This is consistent with the size
of the resolved PAH-emitting region being smaller than 
the neighboring dust continuum-emitting region 
(Geers \etal2007b).
As Oph IRS48 resides in a region with high extinction
(\eg $A_V\simali$11--12 mag; Brown \etal2012a, Follette \etal2015),
its stellar parameters such as $\Lstar$ and age are not
well-constrained. While Brown \etal(2012a) derived 
$\tau$\,$\simali$15\,Myr and $\Lstar\sim14.3\Lsun$, 
Follette \etal(2015), 
adopting a much grayer extinction curve,
estimated $\tau$\,$\simali$5--8\,Myr
and a much higher $\Lstar$ ($\simali$24--45$\Lsun$).
The latter is more closer to the median age of 
the Ophiuchus members ($\simali$2--5\,Myr).

\subsection{PDS 144N}

PDS 144 (a.k.a. CD-25 11111, IRAS 15462--2551) is a binary
system consisting of two Herbig Ae stars (PDS 144N \& S)
with a separation of $5\farcs5$ on the sky (\eg Torres \etal1995),
which is the first confirmed Herbig Ae--Herbig Ae wide binary
(\eg Hornbeck \etal2012). The distance to PDS 144 is rather
uncertain: while $\simali$1\,kpc is generally adopted
(\eg Perrin \etal2006; Fairlamb \etal2015), Hornbeck \etal(2015)
claimed that its distance is $145\pm2\pc$ considering PDS
144N \& S are members of the Upper Sco association. 
An optically thick edge-on disk around the primary, PDS 144N
(Sp. type: A2IV) has been spatially resolved (Perrin \etal2006),
and a dark lane of the edge-on disk is $0\farcs15$ in height
and $0\farcs8$ in across. Prominent PAH emission is detected
in PDS 144N (\eg Perrin \etal2006; Sch\"utz \etal2009), 
whereas the disk of the secondary, PDS 144S (Sp. type: A5V),
lacks PAH emission.
PDS 144N \& S are found to have intrinsic polarization,
which is aligned with the local magnetic field and the jet
axis (Pereyra \etal2012), and the high NIR polarization of
the PDS 144N disk is in favor of the presence of small
grains similar to those of the diffuse ISM.
Sch\"utz \etal(2009) found that the spectral
profile of the 8.6$\mum$ PAH feature is similar to that of
PAHs in the ISM (\ie PAH class A as determined by
Peeters \etal2002), whereas the peak wavelength of
the 11.2$\mum$ PAH feature is red-shifted compared
to those of class A and B profiles.

\subsection{RR Tau}

RR Tau is a young ($\simali$1\,Myr; \eg Alonso-Albi \etal2009),
highly variable Herbig Ae star (Sp. type: A0IVev; \eg
Juh\'asz \etal2010). It is one of the UX Ori type stars showing 
an anti-correlation between the linear polarization and
the brightness variation like BF Ori or WW Vul
(\eg Grinin \etal1996; Rostopchina \etal1997; Rodgers \etal2002;
Bedell \etal201). Although we adopt
a distance of $\simali$160\,pc and 
a luminosity of $\Lstar=2\Lsun$ from
Acke \& van den Ancker (2004),
they are highly uncertain and sensitive 
to the assumed extinction.
Larger distances 
($d\simali$600--2100\,pc, Blondel \& Djie 2006,
Alonso-Albi \etal2009, Montesinos \etal2009) 
and much higher luminosity
(\eg $\Lstar=150\Lsun$ at $d=800\pc$, Alonso-Albi \etal2009)
also have been reported in the literature.
Using mm interferometric observations, Boissier \etal(2011)
detected a disk around RR Tau with a size of $\simali$720\,AU
(assuming $d=800\pc$), which is much larger than 
the outer disk radius of $\simali$80\,AU 
derived from fitting the SED. 
This is possibly due to a complex geometry 
in the outer disk such as spiral arms or a secondary ring.

\subsection{RXJ1615.3-3255}

RXJ1615.3-3255 is a TTS (Sp. type: K4; Mer{\'{\i}}n \etal2010)
associated with the Ophiuchus star-forming region at
$\simali$120\,pc (or 185\,pc; van der Marel \etal2016).
Its IRS spectrum as well as SED indicates
the existence of a hole in its disk ($R\approx2\pm1\AU$;
Mer{\'{\i}}n \etal2010), and millimeter imaging suggets
a slightly large gap ($R=10^{+10}_{-2}\AU$: van der Marel \etal2016,
or $\simali30\AU$: Andrews \etal2011).
Its IRS spectrum dominated by silicate emission is noisy 
and shows a marginal detection of the 11.2$\mum$ PAH feature
possibly contaminated by the 11.2$\mum$ crystalline
forsterite feature (see Figure \ref{fig:nomod}). 

\subsection{SR 21N}

SR 21N (a.k.a. Elias 2-30) is a young ($\simali$1--3\,Myr;
\eg Prato \etal2003; Brown \etal2009) TTS binary system 
located at the $\rho$ Ophiuchus cloud. 
The primary and secondary, 
separated by $\simali$$6\farcs4$,
have spectral types of G2.5 and M4, respectively
(\eg Geers \etal2006).
The disk around SR 21N is known to be transitional;
its cavity ($R\approx33$--$36\AU$) has been spatially resolved
(\eg Brown \etal2009; Andrews \etal2011; P\'erez \etal2014),
and the outer disk is extended to $\approx140\AU$ as revealed
by the SMA 340 GHz (880$\mum$) imaging.
Inside the cavity, SR 21N shows detectable warm CO
emission arising from the inner regions of the disk
($R\approx7\AU$; \eg Pontoppidan \etal2008, P\'erez \etal2014),
which is relatively further away from the central star
compared to other transitional disks.
Recent ALMA observations reveal the existence of
a bright asymmetry in the south of the dust disk (P\'erez \etal2014).
The $H$-band scattered light emission shows smooth
distribution extending to $\simali$80\,AU without apparent asymmetry
(Follette \etal2013), which differs from that of submm imaging
(\eg P\'erez \etal2014). This indicates the different spatial distributions
of small and large (mm-sized) grains, which could result from
a local gas-density enhancement (\eg via disk vortices).
Geers \etal(2006) pointed out that the 11.2$\mum$ feature
detected in the IRS spectrum is broader and red-shifted
similar to those of HD 135344B and LkH$\alpha$ 330.

\subsection{SU Aur}

SU Aur (a.k.a. HD 282624) is probably a classical TTS
(\eg Furlan \etal2006) with a spectral type of G1--2III (\eg
Calvet \etal2004; Jeffers \etal2014) and strong
photometric variability (DeWarf \etal2003).
X-ray emission from SU Aur has been observed (\eg
Franciosini \etal2007), of which the origin is most likely
associated with magnetic activity rather than accretion
like other young late-type stars with hot corona and flares.
The inner radius of its disk is derived to be $\approx$0.14\,AU
from the central star with a high inclination of $\simali$62$\degr$
(\eg Akeson \etal2002). Direct imaging of the SU Aur disk
indicates an outer disk radius of $\simali$500\,AU (\eg
Chakraborty \& Ge 2004; Jeffers \etal2014).
Jeffers \etal(2014) suggested the presence of 
very small grains in the surface layers of the disk with strong
turbulence indicating enhanced dust mixing in the disk.
Using $H$-band observations, de Leon \etal(2015) revealed
a pair of asymmetric tail structures, reaching to
a few hundred\,AU from the disk, which might be explained by tidal
interaction with an unseen brown dwarf.   
The presence of weak PAH emission unlike other late-type
TTSs indicates that SU Aur has a flared disk (\eg Keller
\etal2008; Jeffers \etal2014). The PAH profiles at 7--9$\mum$ belong to
a ``Class C'' PAH spectrum (\eg Peers \etal2002).
The ``7.7$\mum$'' feature peaks at $\approx$\,8.2$\mum$
and the 8.6$\mum$ feature is only marginal 
(\eg Keller \etal2008).

\subsection{T Cha}

T Cha is a nearby TTS
(Sp. type: K0; \eg Murphy \etal2013), 
of which the distance was previously 
considered to be $\simali$66\,pc 
(\eg van den Ancker \etal1998)
and is now commonly accepted 
to be $\simali108$\,pc (\eg Torres \etal2008)
with the recent confirmation of its membership 
in the $\epsilon$ Cha association (\eg Murphy \etal2013).
It is a wide binary (T Cha AB) 
with a separation of 0.2\,pc (Kastner \etal2012).
Geers \etal(2007b) detected weak 3.3 and 11.2$\mum$ features
in ISAAC $L$-band and VISIR $N$-band spectra, 
both of which are unresolved placing an upper limit 
of $\la0\farcs34$ and $\la0\farcs19$ 
($\approx37$ and 21\,AU at 108\,pc)
on the spatial extents, respectively. 
Olofsson \etal(2013) suggested that PAH molecules
cannot be located at the inner edge of 
the outer disk ($\approx12\AU$)
and might have been transported outwards 
by radiation pressure or stellar wind.  
Similar to HD 135344, 
without evident crystalline silicate emission,
the 11.2$\mum$ feature is broad 
and red-shifted with respect to 
that of the diffuse ISM and is possibly 
indicative of the presence of very hot PAHs 
in the innermost part of the disk (Geers \etal2006).
By modeling the SED, Brown \etal(2007) inferred 
a disk structure with an inner radius of 0.08\,AU 
and an inner hole at 0.2--15\,AU 
(assuming a distance of 66\,pc).
Within this inner hole, a planet candidate has been found
($R\approx6.7\AU$; Hu\'elamo \etal2011). Also,
Olofsson \etal(2011) spatially resolved the innermost dusty
disk to be extremely narrow and close to the star 
($\simali$0.13--0.17\,AU),
and the outer dust disk extends up to $\simali$80\,AU
whereas the gas disk is more extended with a radius of
$\simali$230\,AU (Hu\'elamo \etal2015).

\subsection{TY CrA}

TY CrA (a.k.a. CD-37$\degr$ 13024) is a long-known tertiary
system with an eclipsing binary (all within 1.5\,AU to each other;
\eg Casey \etal1998).
Recent observations revealed the fourth star in the system,
with a separation of $\simali$$0\farcs3$ or 40\,AU from TY CrA,
making the TY CrA system quadruple (\eg Chauvin \etal2003). 
TY CrA is embedded in the reflection nebula NGC 6726/6727
located near the R Corona Australis star-forming region.
In the IR images, TY CrA is surrounded by a nebulosity
including a bar-like structure (\eg Boersma \etal2009).
The observed $ISO$ spectrum (with a large aperture) is
most likely dominated by the nebulosity (\eg the bar) rather
than TY CrA itself (\eg Siebenmorgen \etal2000; Geers \etal2007b).
The 3.3$\mum$ PAH emission has been spatially resolved
with VLT/ISAAC ($\simali$54\,AU; Geers \etal2007b), which
confirms the presence of PAH emission 
in the TY CrA disk.
Boersma \etal(2009) extracted spatially-resolved silicate
and PAH spectra and found that the amorphous silicate
emission peaking at $\simali$9.8$\mum$ 
is seen up to $\simali$130\,AU,
whereas the PAH emission is more extended 
up to $\simali$390\,AU.
In addition to the major PAH features, a prominent 11.0$\mum$
feature (\eg Roche \etal1991) as well as a weak feature at
$\simali$12.0$\mum$ (\eg Boersma \etal2009) are detected.
The spatial distributions of the 8.6, 11.0, and 11.2$\mum$ features
show differences: the 11.0$\mum$ emission is most confined
while the 11.2$\mum$ emission is most extended. This
can be explained in terms of different band carriers with different
ionization fraction of PAHs (\ie 8.6$\mum$: charged PAHs, 
11.0$\mum$: the out-of-plane bending mode in cationic PAHs,
and 11.2$\mum$: the out-of-plane bending mode in neutral PAHs;
\eg Hony \etal2001, Bauschlicher \etal2008).

\subsection{UX Tau}

UX Tau (a.k.a. HD 285846) is a pre-main sequence triple
system consisting of one classical TTS (UX Tau A) and
two WTTSs (UX Tau B and C: Sp. type of M2 and M5,
respectively; Furlan \etal2006).
The primary, UX Tau A (Sp. type: K2--G8; \eg Furlan \etal2006;
Wahhaj \etal2010; Pinilla \etal2014), is the only component
which shows evidence for a circumstellar disk. The disk is known
to be pre-transitional, characterized by an inner optically
thick disk ($R\la0.21\AU$; \eg Espaillat \etal2010) and a gap
(\eg Espaillat \etal2007). Andrews \etal(2011) presented
a spatially resolved cavity with a radius of $\simali$25\,AU
at 880$\mum$, also confirmed by 3\,mm (100\,GHz)
observations (Pinilla \etal2014).
The $H$-band polarimetric observations show a strongly
polarized disk around UX Tau A which extends to $\simali$120\,AU
(Tanii \etal2012).
The observed PAH features in the $Spitzer$/IRS spectra are
likely to be dominated by UX Tau A (Furlan \etal2006),
one of the coolest TTSs ($\Teff\simali$5520 K)
with detection of PAH emission. Also, the IRS spectra show
a very weak 10$\mum$ silicate feature, 
a strong 20$\mum$ feature and some minor ones 
at $\ga20\mum$ due to crystalline silicates 
(Furlan \etal2006; Espaillat \etal2010).
This indicates a lack of small silicate grains,
implying grain growth by collisional coagulation 
in the disk (\eg Tanii \etal2012).

\subsection{V590 Mon}

V590 Mon (a.k.a. LkH$\alpha$ 25) is a HAeBe star
with a spectral type of B8ep+sh (Juh\'asz \etal2010)
or A0 (Habart \etal2004a). 
The distance to this object is widely adopted 
to be $\simali$800\,pc (\eg Liu \etal2011), 
but a recent survey with the
X-Shooter spectrograph mounted at the VLT 
suggests a distance of $\simali$$1722^{+171}_{-160}\pc$
(Fairlamb \etal2015).
Current estimates of its luminosity show large discrepancy
ranging from $\simali$9$\Lsun$ (Habart \etal2004a) 
to $\simali$$295\Lsun$ (Liu \etal2011). 
Habart \etal(2004a) placed an upper limit 
of $\simali$1200\,AU for the spatial extent 
of the PAH emission.
A tentative detection of the nanodiamond feature 
at 3.53$\mum$ is reported (Acke \& van den Ancker 2006).

\subsection{V892 Tau}

V892 Tau (a.k.a. Elias 3-1) is a HAeBe star with a spectral type
of B8--A6 (\eg Furlan \etal2006; Alonso-Albi \etal2009;
van der Marel \etal2016) located in the Taurus-Aurigae
star-forming region and suffering severe extinction 
($A_V\simali$4.05--11\,mag; \eg Keller \etal2008; 
Monnier \etal2008; Liu \etal2011; van der Marel \etal2016). 
An effective temperature of $\Teff=8000\K$ 
from Alonso-Albi \etal(2009) is adopted in this paper, 
but somewhat higher $\Teff$ 
(\eg $\simali$11000\,K) has also been reported in the literature
(\eg Keller \etal2008; van der Marel \etal2016).
Depending on the assumed $A_V$, its $\Lstar$ ranges from
$\simali$21$\Lsun$ up to a few hundred $\Lsun$ 
(\eg Monnier \etal2008; Alonso-Albi \etal2009; 
Menu \etal2015; van der Marel \etal2016).
V892 Tau is associated with a faint reflection nebula 
most brightly seen at 24$\mum$ (Mooley \etal2013).
High-resolution IR observations showed that V892 Tau is
a close binary with a separation of $\simali$7.7\,AU 
and nearly equal brightness 
(\eg Smith \etal2005; Monnier \etal2008).
It also has a faint T Tauri stellar companion,
separated by $\simali$$4\farcs1$ (Smith \etal2005).
Monnier \etal(2008) have spatially resolved the V892 Tau
disk in MIR showing an asymmetric circumbinary disk inclined
at $\simali$60$\degr$ with an inner hole diameter of 35\,AU.
Using interferometric observations at 1.3 and 2.7\,mm,
Hamidouche (2010) resolved its disk with a radius of
$\simali$100\,AU and inferred grain growth in the disk 
(up to cm size) based
on a low opacity ($\beta=1.1$), which is also supported
by submm observations (Sandell \etal2011).

\subsection{VV Ser}

VV Ser is a Herbig Ae star 
(Sp. type: A0Vevp, Juh\'asz \etal2010; 
or B7, Cauley \etal2016). 
The distance to VV Ser is unknown, 
ranging from $\simali$230 to $614^{+99}_{-88}\pc$
(\eg Alecian \etal2013; Montesinos \etal2009).
This object shows the characteristics of UX Ori type stars
(\eg Herbst \& Shevchenko 1999) and evidence of an optically
thin disk from NIR polarimetric measurements (Pereyra \etal2009).
Habart \etal(2004a) estimated the PAH emission extension
to be $\approx$410\,AU at a distance of 440\,pc. 
$Spitzer$ images at 5.6--70$\mum$ show a very extended
($\ga4\arcmin$), bright nebulosity around VV Ser, which
is most likely due to transiently-heated very small grains
and PAH molecules (Pontoppidan \etal2007). 
In addition, a wedge-shaped dark bend extending across the
nebulosity appears, attributed to a large
shadow cast by the small ($\simali$50\,AU) central disk.
This is consistent with an almost edge-on geometry of the
disk around VV Ser, as revealed by NIR interferometry (\eg Eisner
\etal2003) and NIR polarization (Pereyra \etal2009).
NIR \ion{H}{1} emission lines with variable line profiles are
detected (Garcia Lopez \etal2016), which can be explained
in terms of contributions of an extended wind such as a
bipolar outflow in a complex inner disk region. 
 
\subsection{VX Cas}

VX Cas is a Herbig Ae star (Sp. type: A1Ve+sh; Acke \& van
den Ancker 2004) located at $d=620\pm60\pc$ (Alecian
\etal2013) or $\simali$760\,pc (Dent \etal2005).
VX Cas is considered to have a geometrically flat disk
(\ie group II object; Meeus \etal2001).
Broad and weak [\ion{O}{1}] 6300$\Angstrom$ emission line
is detected (Acke \etal2005), which shows a double-peaked profile. 
Similar to other group II objects with the detection of the [\ion{O}{1}]
emission line (\eg HD 101412), this emission might originate
from a rotating gaseous disk within the dust-sublimation radius
(Acke \etal2005). Oudmaijer \etal(2001) found a slight
anti-correlation between photometry and polarization of VX Cas
and classified VX Cas as a UX Ori type variable.
Folsom \etal(2012) identified suspiciously weak metal
lines in the optical spectrum of VX Cas, which might imply
$\lambda$ Boo peculiarities.

\subsection{WL\,16}

WL\,16 is a Herbig Ae star (Sp. type: B8--A7; Geers \etal2007b)
embedded in the $\rho$ Ophiuchus molecular cloud
at a distance of $d\simali$125\,pc (Ressler \& Barsony 2003).
This object is well-known to have one of the most extended
disks in MIR, of which the MIR emission is largely dominated
by PAH features (\eg Moor \etal1998; Ressler \& Barsony 2003). 
Ressler \& Barsony (2003) obtained diffraction-limited images
of the WL\,16 disk from 7.9 to 24.5$\mum$ and revealed that 
the disk size is $\simali$900\,AU in diameter. 
Also, they confirmed the existence of ionized PAHs in the
central regions and reported the discovery of a population
of larger ($\ge50$--80 C atom) or more hydrogenated PAHs
in the periphery of the disk. Recently, Zhang \etal(2016)
presented MIR (8--25$\mum$) polarimetric images and
spectra of WL\,16 obtained with the CanariCam at the Gran
Telescopio Canarias. A polarization fraction of
$\simali$2\% is detected, which mainly arises from aligned
elongated dust grains in the foreground. 

\subsection{Wray 15-1484}

Wray 15-1484 (a.k.a. Hen 3-1191) has a spectral type of B0:[e]
(Acke \& van den Ancker 2004), and its nature is rather uncertain.  
It was classified as a young compact proto-planetary nebula
(Le Bertre \etal1989), but later it was considered to be a
Herbig Be star (\eg De Winter \etal1994). Its distance was
estimated to be $\approx750\pc$ (Acke \& van den Ancker 2004) 
although it is not well constrained (Lachaume \etal2007).
Wray 15-1484 shows a bipolar structure in $R$-band
(Le Bertre \etal1989), and an axis of NIR polarization
is nearly parallel to the bipolar axis (Pereyra \etal2009),
implying scattering in an optically thin disk.
Lachaume \etal(2007) performed $N$-band (8--13$\mum$)
interferometric observations with VLTI/MIDI and found that
the Wray 15-1484 system can be explained in terms of a disk with
an unusually high mass accretion (or excretion), a large central gap,
and a binary made of two IR sources.

\subsection{WW Vul}

WW Vul (a.k.a HD 344361) is a Herbig Ae star (Sp. type:
A2 IVe; Juh\'asz \etal2010) located at 
$d=700^{+260}_{-150}\pc$ (Alecian \etal2013) or
$d\approx440$--550\,pc (Alonso-Albi \etal2009; Juh\'asz \etal2010).
The extended PAH emission is reported with a radius of
$\simali$88\,AU adopting a distance of 700\,pc (Habart \etal2004a).
WW Vul is considered as a UX Ori type star (\eg Grinin \etal1996), which is
characterized by high-amplitude variability in its light curve,
Algol-like minima, an increase of polarization when its
brightness decreases and so on (\eg Mora \etal2004).
WW Vul exhibits ``$\beta$ Pic phenomenon'' like BF Ori
(\eg Grinin \etal1996). It shows transient absorption
features in metallic lines without counterparts in hydrogen
lines, suggesting the presence of circumstellar gas clouds
with enhanced metallicity around WW Vul (Mora \etal2004).
Pereyra \etal(2009) found a polarization reversal between
optical and NIR data for WW Vul.

%%%%%%%%%%%%%%% References %%%%%%%%%%%%%%%%%
%


\begin{thebibliography}{30}
%%
            
\bibitem[]{}Acke, B., \& van den Ancker, M.~E.\ 
             2004, A\&A, 426, 151 

\bibitem[Acke \& Waelkens(2004)]{2004A&A...427.1009A} Acke, B., \& Waelkens, C.\ 2004, \aap, 427, 1009 

\bibitem[Acke et al.(2005)]{2005A&A...436..209A} Acke, B., van den Ancker, M.~E., \& Dullemond, C.~P.\ 2005, \aap, 436, 209 

\bibitem[Acke \& van den Ancker(2006)]{2006A&A...457..171A} Acke, B., \& van den Ancker, M.~E.\ 2006, \aap, 457, 171 

\bibitem[]{}Acke, B., Bouwman, J., 
            Juh{\'a}sz, A., et al.\ 2010, ApJ, 718, 558              

\bibitem[Akeson et al.(2002)]{2002ApJ...566.1124A} Akeson, R.~L., Ciardi, D.~R., van Belle, G.~T., \& Creech-Eakman, M.~J.\ 2002, \apj, 566, 1124 

\bibitem[Alecian et al.(2008)]{2008MNRAS.385..391A} Alecian, E., Catala, C., Wade, G.~A., et al.\ 2008, \mnras, 385, 391 

\bibitem[Alecian et al.(2013)]{2013MNRAS.429.1001A} Alecian, E., Wade, G.~A., Catala, C., et al.\ 2013, \mnras, 429, 1001

\bibitem[Alencar et al.(2003)]{2003A&A...409.1037A} Alencar, S.~H.~P., Melo, C.~H.~F., Dullemond, C.~P., et al.\ 2003, \aap, 409, 1037 

\bibitem[]{}Allamandola, L. J., Tielens, A. G. G. M.,
            \& Barker, J.R.\ 1985, ApJ, 290, L25
%
\bibitem[Alonso-Albi et al.(2009)]{2009A&A...497..117A} Alonso-Albi, T., Fuente, A., Bachiller, R., et al.\ 2009, \aap, 497, 117 
%
\bibitem[]{}Andersson, M.~P., Blomquist, J., \& Uvdal, P.\ 2005,
           J. Chem. Phys., 123, 224714

\bibitem[Andrews et al.(2011)]{2011ApJ...732...42A} Andrews, S.~M., Wilner, D.~J., Espaillat, C., et al.\ 2011, \apj, 732, 42 

\bibitem[Augereau et al.(1999)]{1999A&A...350L..51A} Augereau, J.~C., Lagrange, A.~M., Mouillet, D., \& M{\'e}nard, F.\ 1999, \aap, 350, L51 

\bibitem[Augereau et al.(2001)]{2001A&A...365...78A} Augereau, J.~C., Lagrange, A.~M., Mouillet, D., \& M{\'e}nard, F.\ 2001, \aap, 365, 78 

\bibitem[Augereau \& Papaloizou(2004)]{2004A&A...414.1153A} Augereau, J.~C., \& Papaloizou, J.~C.~B.\ 2004, \aap, 414, 1153 

\bibitem[Baines et al.(2006)]{2006MNRAS.367..737B} Baines, D., Oudmaijer, R.~D., Porter, J.~M., \& Pozzo, M.\ 2006, \mnras, 367, 737 

\bibitem[]{}Bakes, E. L. O., \& Tielens, A. G. G. M.\ 
            1994, ApJ, 427, 822
\bibitem[]{}Barker, J.R., Allamandola, L.J., \& Tielens, A.G.G.M.\
            1987, ApJ, 315, L61

\bibitem[Bary et al.(2008)]{2008ApJ...678.1088B} Bary, J.~S., Weintraub, D.~A., Shukla, S.~J., Leisenring, J.~M., \& Kastner, J.~H.\ 2008, \apj, 678, 1088-1098 

\bibitem[Bauschlicher et al.(2008)]{2008ApJ...678..316B} Bauschlicher, C.~W., Jr., Peeters, E., \& Allamandola, L.~J.\ 2008, \apj, 678, 316-327 

\bibitem[Bauschlicher et al.(2010)]{2010ApJS..189..341B} Bauschlicher, C.~W., Jr., Boersma, C., Ricca, A., et al.\ 2010, \apjs, 189, 341-351 

\bibitem[Benisty et al.(2015)]{2015A&A...578L...6B} Benisty, M., Juhasz, A., Boccaletti, A., et al.\ 2015, \aap, 578, L6 

\bibitem[Benisty et al.(2013)]{2013A&A...555A.113B} Benisty, M., Perraut, K., Mourard, D., et al.\ 2013, \aap, 555, A113 

%\bibitem[]{}Bernstein, M.~P., Elsila, J.~E., Dworkin, J.~P., et al.\ 
%		2002, ApJ, 576, 1115 
%
\bibitem[]{}Bernstein, M.P., Sandford, S.A., \& Allamandola, L.J.\
            1996, ApJ, 472, L127
%
\bibitem[Biller et al.(2012)]{2012ApJ...753L..38B} Biller, B.~A.,, Lacour, S., Juh{\'a}sz, A., et al.\ 2012, \apjl, 753, L38
 
\bibitem[]{} Biller, B.~A., Males, J., Rodigas, T., et al.\ 
		2014, ApJL, 792, L22 

\bibitem[Biller et al.(2015)]{2015MNRAS.450.4446B} Biller, B.~A., Liu, M.~C., Rice, K., et al.\ 2015, \mnras, 450, 4446 

\bibitem[Blondel \& Djie(2006)]{2006A&A...456.1045B} Blondel, P.~F.~C., \& Djie, H.~R.~E.~T.~A.\ 2006, \aap, 456, 1045 

\bibitem[Boersma et al.(2008)]{2008A&A...484..241B} Boersma, C., Bouwman, J., Lahuis, F., et al.\ 2008, \aap, 484, 241 

\bibitem[Boersma et al.(2009)]{2009A&A...502..175B} Boersma, C., Peeters, E., Mart{\'{\i}}n-Hern{\'a}ndez, N.~L., et al.\ 2009, \aap, 502, 175 

\bibitem[Boissier et al.(2011)]{2011A&A...531A..50B} Boissier, J., Alonso-Albi, T., Fuente, A., et al.\ 2011, \aap, 531, A50

\bibitem[Borges Fernandes et al.(2007)]{2007MNRAS.377.1343B} Borges Fernandes, M., Kraus, M., Lorenz Martins, S., \& de Ara{\'u}jo, F.~X.\ 2007, \mnras, 377, 1343 

\bibitem[]{}Borowski, P.\ 2012,
           %An Evaluation of Scaling Factors
           %for Multiparameter Scaling Procedures
           %Based on DFT Force Fields.
           J. Phys. Chem. A, 116, 3866

\bibitem[Bouret et al.(2003)]{2003A&A...410..175B} Bouret, J.-C., Martin, C., Deleuil, M., Simon, T., \& Catala, C.\ 2003, \aap, 410, 175 
           
\bibitem[Bouwman et al.(2001)]{2001A&A...375..950B} Bouwman, J., Meeus, G., de Koter, A., et al.\ 2001, \aap, 375, 950 

\bibitem[]{}Brooke, T.~Y., Tokunaga, A.~T., \& Strom, S.~E.\ 
           1993, AJ, 106, 656 
           

\bibitem[Brown et al.(2007)]{2007ApJ...664L.107B} Brown, J.~M., Blake, G.~A., Dullemond, C.~P., et al.\ 2007, \apjl, 664, L107 

\bibitem[Brown et al.(2008)]{2008ApJ...675L.109B} Brown, J.~M., Blake, G.~A., Qi, C., Dullemond, C.~P., \& Wilner, D.~J.\ 2008, \apjl, 675, L109 

\bibitem[Brown et al.(2009)]{2009ApJ...704..496B} Brown, J.~M., Blake, G.~A., Qi, C., et al.\ 2009, \apj, 704, 496 

\bibitem[Brown et al.(2012)]{2012ApJ...744..116B} Brown, J.~M., Herczeg, G.~J., Pontoppidan, K.~M., \& van Dishoeck, E.~F.\ 2012a, \apj, 744, 116 

\bibitem[Brown et al.(2012)]{2012ApJ...758L..30B} Brown, J.~M., Rosenfeld, K.~A., Andrews, S.~M., Wilner, D.~J., \& van Dishoeck, E.~F.\ 2012b, \apjl, 758, L30 


\bibitem[Calvet et al.(2004)]{2004AJ....128.1294C} Calvet, N., Muzerolle, J., Brice{\~n}o, C., et al.\ 2004, \aj, 128, 1294 

\bibitem[Caratti o Garatti et al.(2015)]{2015A&A...582A..44C} Caratti o Garatti, A., Tambovtseva, L.~V., Garcia Lopez, R., et al.\ 2015, \aap, 582, A44 

\bibitem[Carmona et al.(2010)]{2010A&A...517A..67C} Carmona, A., van den Ancker, M.~E., Audard, M., et al.\ 2010, \aap, 517, A67 

\bibitem[Carmona et al.(2011)]{2011A&A...533A..39C} Carmona, A., van der Plas, G., van den Ancker, M.~E., et al.\ 2011, \aap, 533, A39 

\bibitem[Casassus et al.(2013)]{2013Natur.493..191C} Casassus, S., van der Plas, G., M, S.~P., et al.\ 2013, \nat, 493, 191 

\bibitem[Casassus et al.(2015)]{2015ApJ...812..126C} Casassus, S., Wright, C.~M., Marino, S., et al.\ 2015, \apj, 812, 126 

\bibitem[Casey et al.(1998)]{1998AJ....115.1617C} Casey, B.~W., Mathieu, R.~D., Vaz, L.~P.~R., Andersen, J., \& Suntzeff, N.~B.\ 1998, \aj, 115, 1617 

\bibitem[Catala et al.(1991)]{1991A&A...244..166C} Catala, C., Czarny, J., Felenbok, P., Talavera, A., \& The, P.~S.\ 1991, \aap, 244, 166 

\bibitem[Cauley \& Johns-Krull(2016)]{2016ApJ...825..147C} Cauley, P.~W., \& Johns-Krull, C.~M.\ 2016, \apj, 825, 147 

\bibitem[Chakraborty \& Ge(2004)]{2004AJ....127.2898C} Chakraborty, A., \& Ge, J.\ 2004, \aj, 127, 2898 

\bibitem[Chapillon et al.(2008)]{2008A&A...488..565C} Chapillon, E., Guilloteau, S., Dutrey, A., \& Pi{\'e}tu, V.\ 2008, \aap, 488, 565 

\bibitem[Chauvin et al.(2003)]{2003A&A...406L..51C} Chauvin, G., Lagrange, A.-M., Beust, H., et al.\ 2003, \aap, 406, L51 

\bibitem[]{}Chen, C.H., Sargent, B.A.,
            Bohac, C., et al.\ 2006, ApJS, 166, 351 

\bibitem[]{}Chiang, E., \& Youdin, A.~N.\ 2010,
		 Annual Review of Earth and Planetary Sciences, 38, 493 
%
%
\bibitem[]{}Chiar, J.~E., Tielens, A.~G.~G.~M.,
		Whittet, D.~C.~B., et al.\ 2000, \apj, 537, 749 
%
\bibitem[Collins et al.(2009)]{2009ApJ...697..557C} Collins, K.~A., Grady, C.~A., Hamaguchi, K., et al.\ 2009, \apj, 697, 557 

\bibitem[Connelley et al.(2008)]{2008AJ....135.2496C} Connelley, M.~S., Reipurth, B., \& Tokunaga, A.~T.\ 2008, \aj, 135, 2496 

\bibitem[Coulson \& Walther(1995)]{1995MNRAS.274..977C} Coulson, I.~M., \& Walther, D.~M.\ 1995, \mnras, 274, 977 

\bibitem[Currie et al.(2015)]{2015ApJ...814L..27C} Currie, T., Cloutier, R., Brittain, S., et al.\ 2015, \apjl, 814, L27 

\bibitem[Currie et al.(2016)]{2016ApJ...819L..26C} Currie, T., Grady, C.~A., Cloutier, R., et al.\ 2016, \apjl, 819, L26 

\bibitem[de Leon et al.(2015)]{2015ApJ...806L..10D} de Leon, J., Takami, M., Karr, J.~L., et al.\ 2015, \apjl, 806, L10 

\bibitem[Dent et al.(2005)]{2005MNRAS.359..663D} Dent, W.~R.~F., Greaves, J.~S., \& Coulson, I.~M.\ 2005, \mnras, 359, 663 

\bibitem[Devine et al.(2000)]{2000ApJ...542L.115D} Devine, D., Grady, C.~A., Kimble, R.~A., et al.\ 2000, \apjl, 542, L115 

\bibitem[DeWarf et al.(2003)]{2003ApJ...590..357D} DeWarf, L.~E., Sepinsky, J.~F., Guinan, E.~F., Ribas, I., \& Nadalin, I.\ 2003, \apj, 590, 357 

\bibitem[de Winter et al.(1999)]{1999A&A...343..137D} de Winter, D., Grady, C.~A., van den Ancker, M.~E., P{\'e}rez, M.~R., \& Eiroa, C.\ 1999, \aap, 343, 137 

\bibitem[de Winter et al.(1994)]{1994ASPC...62..413D} de Winter, D., The, P.~S., \& Perez, M.~R.\ 1994, The Nature and Evolutionary Status of Herbig Ae/Be Stars, 62, 413 

\bibitem[Donehew \& Brittain(2011)]{2011AJ....141...46D} Donehew, B., \& Brittain, S.\ 2011, \aj, 141, 46 

\bibitem[Dong et al.(2015)]{2015ApJ...809L...5D} Dong, R., Zhu, Z., Rafikov, R.~R., \& Stone, J.~M.\ 2015, \apjl, 809, L5 

\bibitem[]{}Draine, B.T., \& Li, A.\
            2001, ApJ, 551, 807
\bibitem[]{}Draine, B.T., \& Li, A.\
            2007, ApJ, 657, 810

\bibitem[Dunkin et al.(1997)]{1997MNRAS.290..165D} Dunkin, S.~K., Barlow, M.~J.,
\& Ryan, S.~G.\ 1997, \mnras, 290, 165 

\bibitem[Dunkin \& Crawford(1998)]{1998MNRAS.298..275D} Dunkin, S.~K., \& Crawford, I.~A.\ 1998, \mnras, 298, 275 

\bibitem[Eisner et al.(2003)]{2003ApJ...588..360E} Eisner, J.~A., Lane, B.~F., Akeson, R.~L., Hillenbrand, L.~A., \& Sargent, A.~I.\ 2003, \apj, 588, 360 

\bibitem[Eisner et al.(2009)]{2009ApJ...692..309E} Eisner, J.~A., Graham, J.~R., Akeson, R.~L., \& Najita, J.\ 2009, \apj, 692, 309 

\bibitem[Espaillat et al.(2007)]{2007ApJ...670L.135E} Espaillat, C., Calvet, N., D'Alessio, P., et al.\ 2007, \apjl, 670, L135 

\bibitem[Espaillat et al.(2010)]{2010ApJ...717..441E} Espaillat, C., D'Alessio, P., Hern{\'a}ndez, J., et al.\ 2010, \apj, 717, 441 

\bibitem[Evans et al.(2009)]{2009ApJS..181..321E} Evans, N.~J., II, Dunham, M.~M., J{\o}rgensen, J.~K., et al.\ 2009, \apjs, 181, 321-350 

%\bibitem[Evans et al.(2003)]{2003PASP..115..965E} Evans, N.~J., II, Allen, L.~E., Blake, G.~A., et al.\ 2003, \pasp, 115, 965 

\bibitem[Fairlamb et al.(2015)]{2015MNRAS.453..976F} Fairlamb, J.~R., Oudmaijer, R.~D., Mendigut{\'{\i}}a, I., Ilee, J.~D., \& van den Ancker, M.~E.\ 2015, \mnras, 453, 976

\bibitem[Fedele et al.(2008)]{2008A&A...491..809F} Fedele, D., van den Ancker, M.~E., Acke, B., et al.\ 2008, \aap, 491, 809 

\bibitem[Follette et al.(2013)]{2013ApJ...767...10F} Follette, K.~B., Tamura, M., Hashimoto, J., et al.\ 2013, \apj, 767, 10 

\bibitem[Follette et al.(2015)]{2015ApJ...798..132F} Follette, K.~B., Grady, C.~A., Swearingen, J.~R., et al.\ 2015, \apj, 798, 132 

\bibitem[Folsom et al.(2008)]{2008MNRAS.391..901F} Folsom, C.~P., Wade, G.~A., Kochukhov, O., et al.\ 2008, \mnras, 391, 901 

\bibitem[Folsom et al.(2012)]{2012MNRAS.422.2072F} Folsom, C.~P., Bagnulo, S., Wade, G.~A., et al.\ 2012, \mnras, 422, 2072 

\bibitem[Franciosini et al.(2007)]{2007A&A...471..951F} Franciosini, E., Scelsi, L., Pallavicini, R., \& Audard, M.\ 2007, \aap, 471, 951 

\bibitem[Friedemann et al.(1996)]{1996A&AS..117..205F} Friedemann, C., Guertler, J., \& Loewe, M.\ 1996, \aaps, 117, 205 

\bibitem[Fuente et al.(1998)]{1998A&A...339..575F} Fuente, A., Martin-Pintado, J., Rodriguez-Franco, A., \& Moriarty-Schieven, G.~D.\ 1998, \aap, 339, 575 

\bibitem[Fujiwara et al.(2006)]{2006ApJ...644L.133F} Fujiwara, H., Honda, M., Kataza, H., et al.\ 2006, \apjl, 644, L133 

\bibitem[Fukagawa et al.(2004)]{2004ApJ...605L..53F} Fukagawa, M., Hayashi, M., Tamura, M., et al.\ 2004, \apjl, 605, L53 

\bibitem[Fukagawa et al.(2010)]{2010PASJ...62..347F} Fukagawa, M., Tamura, M., Itoh, Y., et al.\ 2010, \pasj, 62, 347 

\bibitem[Fukagawa et al.(2013)]{2013PASJ...65L..14F} Fukagawa, M., Tsukagoshi, T., Momose, M., et al.\ 2013, \pasj, 65, L14 

\bibitem[]{}Furlan, E., Hartmann, L., Calvet, N., et al.\ 
            2006, ApJS, 165, 568 

\bibitem[Garcia Lopez et al.(2016)]{2016MNRAS.456..156G} Garcia Lopez, R., Kurosawa, R., Caratti o Garatti, A., et al.\ 2016, \mnras, 456, 156 

\bibitem[Garufi et al.(2014)]{2014A&A...568A..40G} Garufi, A., Quanz, S.~P., Schmid, H.~M., et al.\ 2014, \aap, 568, A40 

\bibitem[]{}Geers, V.~C., Augereau, J.-C., Pontoppidan,
		 K.~M., et al.\ 2006, A\&A, 459, 545 

\bibitem[Geers et al.(2007)]{2007A&A...469L..35G} Geers, V.~C., Pontoppidan, K.~M., van Dishoeck, E.~F., et al.\ 2007a, \aap, 469, L35 

\bibitem[Geers et al.(2007)]{2007A&A...476..279G} Geers, V.~C., van Dishoeck, E.~F., Visser, R., et al.\ 2007b, \aap, 476, 279 
% 

\bibitem[Ginski et al.(2016)]{2016A&A...595A.112G} Ginski, C., Stolker, T., Pinilla, P., et al.\ 2016, \aap, 595, A112 

\bibitem[G{\'o}mez de Castro et al.(2016)]{2016ApJ...818L..17G} G{\'o}mez de Castro, A.~I., Loyd, R.~O.~P., France, K., Sytov, A., \& Bisikalo, D.\ 2016, \apjl, 818, L17 

\bibitem[Grady et al.(2000)]{2000ApJ...544..895G} Grady, C.~A., Devine, D., Woodgate, B., et al.\ 2000, \apj, 544, 895 

\bibitem[Grady et al.(2013)]{2013ApJ...762...48G} Grady, C.~A., Muto, T., Hashimoto, J., et al.\ 2013, \apj, 762, 48 

\bibitem[Gray \& Corbally(1998)]{1998AJ....116.2530G} Gray, R.~O., \& Corbally, C.~J.\ 1998, \aj, 116, 2530 

\bibitem[Grinin et al.(1991)]{1991Ap&SS.186..283G} Grinin, V.~P., Kiselev, N.~N., Chernova, G.~P., Minikulov, N.~K., \& Voshchinnikov, N.~V.\ 1991, \apss, 186, 283 

\bibitem[Grinin et al.(1996)]{1996A&A...309..474G} Grinin, V.~P., Kozlova, O.~V., The, P.~S., \& Rostopchina, A.~N.\ 1996, \aap, 309, 474 

\bibitem[Guidi et al.(2016)]{2016A&A...588A.112G} Guidi, G., Tazzari, M., Testi, L., et al.\ 2016, \aap, 588, A112 

\bibitem[Guimar{\~a}es et al.(2006)]{2006A&A...457..581G} Guimar{\~a}es, M.~M., Alencar, S.~H.~P., Corradi, W.~J.~B., \& Vieira, S.~L.~A.\ 2006, \aap, 457, 581 

\bibitem[]{}Habart, E., Natta, A., \& Kr{\"u}gel, E.\ 
	    2004a, A\&A, 427, 179 
      
\bibitem[Habart et al.(2004)]{2004A&A...414..531H} Habart, E., Boulanger, F., Verstraete, L., Walmsley, C.~M., \& Pineau des For{\^e}ts, G.\ 2004b, \aap, 414, 531
          
\bibitem[Habart et al.(2004)]{2004ApJ...614L.129H} Habart, E., Testi, L., Natta, A., \& Carbillet, M.\ 2004c, \apjl, 614, L129 

\bibitem[]{}Habart, E., Natta, A., Testi, L., \& Carbillet, M.\
            2006, A\&A, 449, 1067        

\bibitem[Haisch et al.(2006)]{2006AJ....132.2675H} Haisch, K.~E., Jr., Barsony, M., Ressler, M.~E., \& Greene, T.~P.\ 2006, \aj, 132, 2675 

\bibitem[Hales et al.(2006)]{2006MNRAS.365.1348H} Hales, A.~S., Gledhill, T.~M., Barlow, M.~J., \& Lowe, K.~T.~E.\ 2006, \mnras, 365, 1348 

\bibitem[Hales et al.(2014)]{2014AJ....148...47H} Hales, A.~S., De Gregorio-Monsalvo, I., Montesinos, B., et al.\ 2014, \aj, 148, 47 

\bibitem[Hamidouche et al.(2006)]{2006ApJ...651..321H} Hamidouche, M., Looney, L.~W., \& Mundy, L.~G.\ 2006, \apj, 651, 321 

\bibitem[Hamidouche(2010)]{2010ApJ...722..204H} Hamidouche, M.\ 2010, \apj, 722, 204 

\bibitem[Harvey et al.(1984)]{1984ApJ...278..156H} Harvey, P.~M., Wilking, B.~A., \& Joy, M.\ 1984, \apj, 278, 156 

\bibitem[Hashimoto et al.(2011)]{2011ApJ...729L..17H} Hashimoto, J., Tamura, M., Muto, T., et al.\ 2011, \apjl, 729, L17 

\bibitem[Hein Bertelsen et al.(2016)]{2016A&A...590A..98H} Hein Bertelsen, R.~P., Kamp, I., van der Plas, G., et al.\ 2016, \aap, 590, A98 

\bibitem[Herbig(1998)]{1998ApJ...497..736H} Herbig, G.~H.\ 1998, \apj, 497, 736 

\bibitem[Herbst \& Shevchenko(1999)]{1999AJ....118.1043H} Herbst, W., \& Shevchenko, V.~S.\ 1999, \aj, 118, 1043 

\bibitem[Hillenbrand et al.(1992)]{1992ApJ...397..613H} Hillenbrand, L.~A., Strom, S.~E., Vrba, F.~J., \& Keene, J.\ 1992, \apj, 397, 613 

\bibitem[]{}Honda, M., Maaskant, K., 
	Okamoto, Y.~K., et al.\ 
	2012, ApJ, 752, 143 

\bibitem[Hony et al.(2001)]{2001A&A...370.1030H} Hony, S., Van Kerckhoven, C., Peeters, E., et al.\ 2001, \aap, 370, 1030 

\bibitem[Hornbeck et al.(2012)]{2012ApJ...744...54H} Hornbeck, J.~B., Grady, C.~A., Perrin, M.~D., et al.\ 2012, \apj, 744, 54 

\bibitem[Hu et al.(1989)]{1989A&A...208..213H} Hu, J.~Y., The, P.~S., \& de Winter, D.\ 1989, \aap, 208, 213 

\bibitem[Hubrig et al.(2009)]{2009A&A...502..283H} Hubrig, S., Stelzer, B., Sch{\"o}ller, M., et al.\ 2009, \aap, 502, 283 

\bibitem[Hudgins \& Allamandola(1999)]{1999ApJ...516L..41H} Hudgins, D.~M., \& Allamandola, L.~J.\ 1999, \apjl, 516, L41 

\bibitem[]{}Hudgins, D. M., Bauschlicher, C. W., Allamandola, L. J., \& Fetzer, J. C.
	2000, J. Phys. Chem., 104, 3655

\bibitem[Hu{\'e}lamo et al.(2011)]{2011A&A...528L...7H} Hu{\'e}lamo, N., Lacour, S., Tuthill, P., et al.\ 2011, \aap, 528, L7 

\bibitem[Hu{\'e}lamo et al.(2015)]{2015A&A...575L...5H} Hu{\'e}lamo, N., de Gregorio-Monsalvo, I., Macias, E., et al.\ 2015, \aap, 575, L5 

\bibitem[Ilee et al.(2014)]{2014MNRAS.445.3723I} Ilee, J.~D., Fairlamb, J., Oudmaijer, R.~D., et al.\ 2014, \mnras, 445, 3723 

\bibitem[Isella et al.(2008)]{2008A&A...483L..13I} Isella, A., Tatulli, E., Natta, A., \& Testi, L.\ 2008, \aap, 483, L13 

\bibitem[Isella et al.(2010)]{2010ApJ...725.1735I} Isella, A., Natta, A., Wilner, D., Carpenter, J.~M., \& Testi, L.\ 2010, \apj, 725, 1735 

\bibitem[Isella et al.(2013)]{2013ApJ...775...30I} Isella, A., P{\'e}rez, L.~M., Carpenter, J.~M., et al.\ 2013, \apj, 775, 30 

%\bibitem[]{}Jayawardhana, R., Fisher, R.~S., 
%		Telesco, C.~M., et al.\ 
%		2001, AJ, 122, 2047 

\bibitem[Jeffers et al.(2014)]{2014A&A...561A..23J} Jeffers, S.~V., Min, M., Canovas, H., Rodenhuis, M., \& Keller, C.~U.\ 2014, \aap, 561, A23 

\bibitem[Jensen et al.(2009)]{2009ApJ...703..252J} Jensen, E.~L.~N., Cohen, D.~H., \& Gagn{\'e}, M.\ 2009, \apj, 703, 252 

\bibitem[]{}Joblin, C., Tielens, A.G.G.M., Allamandola, L.J.,
           \& Geballe, T.R.\ 1996, ApJ, 458, 610

\bibitem[Jonkheid et al.(2004)]{2004A&A...428..511J} Jonkheid, B., Faas, F.~G.~A., van Zadelhoff, G.-J., \& van Dishoeck, E.~F.\ 2004, \aap, 428, 511 

\bibitem[Juh{\'a}sz et al.(2010)]{2010ApJ...721..431J} Juh{\'a}sz, A., Bouwman, J., Henning, T., et al.\ 2010, \apj, 721, 431 

\bibitem[Kama et al.(2015)]{2015A&A...582L..10K} Kama, M., Folsom, C.~P., \& Pinilla, P.\ 2015, \aap, 582, L10 

\bibitem[Kamp \& Dullemond(2004)]{2004ApJ...615..991K} Kamp, I., \& Dullemond, C.~P.\ 2004, \apj, 615, 991 

\bibitem[Kastner et al.(2012)]{2012ApJ...747L..23K} Kastner, J.~H., Thompson, E.~A., Montez, R., et al.\ 2012, \apjl, 747, L23 

\bibitem[]{}Keller, L.~D., Sloan, G.~C., Forrest, W.~J., et al.\ 
		2008, ApJ, 684, 411 

\bibitem[Kessler-Silacci et al.(2006)]{2006ApJ...639..275K} Kessler-Silacci, J., Augereau, J.-C., Dullemond, C.~P., et al.\ 2006, \apj, 639, 275 

\bibitem[Khalafinejad et al.(2016)]{2016A&A...587A..62K} Khalafinejad, S., Maaskant, K.~M., Mari{\~n}as, N., \& Tielens, A.~G.~G.~M.\ 2016, \aap, 587, A62 

\bibitem[Konishi et al.(2016)]{2016ApJ...818L..23K} Konishi, M., Grady, C.~A., Schneider, G., et al.\ 2016, \apjl, 818, L23 

\bibitem[K{\'o}sp{\'a}l et al.(2012)]{2012ApJS..201...11K} K{\'o}sp{\'a}l, {\'A}., {\'A}brah{\'a}m, P., Acosta-Pulido, J.~A., et al.\ 2012, \apjs, 201, 11 
 
\bibitem[Kraus et al.(2008)]{2008A&A...489.1157K} Kraus, S., Hofmann, K.-H., Benisty, M., et al.\ 2008a, \aap, 489, 1157 

\bibitem[Kraus et al.(2008)]{2008ApJ...676..490K} Kraus, S., Preibisch, T., \& Ohnaka, K.\ 2008b, \apj, 676, 490-508 

\bibitem[Kreplin et al.(2012)]{2012A&A...537A.103K} Kreplin, A., Kraus, S., Hofmann, K.-H., et al.\ 2012, \aap, 537, A103 

\bibitem[Kurosawa et al.(2016)]{2016MNRAS.457.2236K} Kurosawa, R., Kreplin, A., Weigelt, G., et al.\ 2016, \mnras, 457, 2236 
 
\bibitem[]{}Kurucz, R.~L.\ 1979, ApJS, 40, 1

\bibitem[Lachaume et al.(2007)]{2007A&A...469..587L} Lachaume, R., Preibisch, T., Driebe, T., \& Weigelt, G.\ 2007, \aap, 469, 587 

\bibitem[Lacour et al.(2016)]{2016A&A...590A..90L} Lacour, S., Biller, B., Cheetham, A., et al.\ 2016, \aap, 590, A90 

\bibitem[]{}Langhoff, S. R. 1996, J. Phys. Chem., 100, 2819

\bibitem[Le Bertre et al.(1989)]{1989A&A...225..417L} Le Bertre, T., Heydari-Malayeri, M., Epchtein, N., Gouiffes, C., \& Perrier, C.\ 1989, \aap, 225, 417 

\bibitem[]{}L\'{e}ger, A., \& Puget, J. 
	    1984, A\&A, 137, L5

\bibitem[Leinert et al.(1997)]{1997A&A...318..472L} Leinert, C., Richichi, A., \& Haas, M.\ 1997, \aap, 318, 472 

\bibitem[Leinert et al.(2001)]{2001A&A...375..927L} Leinert, C., Haas, M., {\'A}brah{\'a}m, P., \& Richichi, A.\ 2001, \aap, 375, 927 

\bibitem[Leinert et al.(2004)]{2004A&A...423..537L} Leinert, C., van Boekel, R., Waters, L.~B.~F.~M., et al.\ 2004, \aap, 423, 537 

\bibitem[Levato \& Abt(1976)]{1976PASP...88..712L} Levato, H., \& Abt, H.~A.\ 1976, \pasp, 88, 712 

\bibitem[]{}Li, A. 2004, 
	in Astrophysics of Dust (ASP Conf. Ser. 309), 
	ed. A. N. Witt, G. C. Clayton, \& B. T. Draine 
	(San Francisco, CA: ASP), 417

\bibitem[]{}Li, A.\ 2009, 
            in Small Bodies in Planetary Sciences 
            (Lecture Notes in Physics vol.\,758), 
            ed. I. Mann, A. Nakamura, \& T. Mukai, 
            Springer, Chapter 6, 167
%\bibitem[]{}Li, A.\ 2009, 
%	    in Deep Impact as a World Observatory Event: 
%	    Synergies in Space, Time, and Wavelength, 
%	    ed. H.~U. K{\"a}ufl \& C. Sterken, 161 
\bibitem[]{}Li, A., \& Draine, B.T.\ 2001a, 
            ApJ, 550, 213
\bibitem[]{}Li, A., \& Draine, B.T.\ 2001b, 
            ApJ, 554, 778
\bibitem[]{}Li, A., \& Draine, B.T.\ 2002, 
            ApJ, 572, 232
\bibitem[]{}Li, A., \& Draine, B.T.\ 2012, 
            ApJ, 760, L35
%%\bibitem[]{}Li, A., \& Greenberg, J.M.\
%%            1997, A\&A, 323, 566
%\bibitem[]{}Li, A., \& Greenberg, J.M.\ 
%            1998, A\&A, 331, 291
%\bibitem[]{}Li, A., \& Lunine, J. I.\
%            2003a, ApJ, 590, 368
\bibitem[]{}Li, A., \& Lunine, J. I.\
            2003, ApJ, 594, 987

\bibitem[]{}Li, D., Mari{\~n}as, N., \& Telesco, C.~M.\ 2014, \apj, 796, 74 

\bibitem[Lisse et al.(2007)]{2007Icar..187...69L} Lisse, C.~M., Kraemer, K.~E., Nuth, J.~A., Li, A., \& Joswiak, D.\ 2007, \icarus, 187, 69 

\bibitem[Liu et al.(2011)]{2011ApJ...734...22L} Liu, T., Zhang, H., Wu, Y., Qin, S.-L., \& Miller, M.\ 2011, \apj, 734, 22 

\bibitem[Maaskant et al.(2011)]{2011A&A...531A..27M} Maaskant, K.~M., Bik, A., Waters, L.~B.~F.~M., et al.\ 2011, \aap, 531, A27 

\bibitem[]{}Maaskant, K.~M., Honda, M., 
		Waters, L.~B.~F.~M., et al.\ 
		2013, A\&A, 555, A64 

\bibitem[]{}Maaskant, K.~M., Min, M., Waters, L.~B.~F.~M.,
		\& Tielens, A.~G.~G.~M.\ 
		2014, A\&A, 563, A78 

\bibitem[Maaskant et al.(2015)]{2015A&A...574A.140M} Maaskant, K.~M., de Vries, B.~L., Min, M., et al.\ 2015, \aap, 574, A140 

\bibitem[Magazzu et al.(1991)]{1991A&A...249..149M} Magazzu, A., Martin, E.~L., \& Rebolo, R.\ 1991, \aap, 249, 149 

\bibitem[Malbet et al.(2007)]{2007A&A...464...43M} Malbet, F., Benisty, M., de Wit, W.-J., et al.\ 2007, \aap, 464, 43 

\bibitem[Malfait et al.(1998)]{1998A&A...331..211M} Malfait, K., Bogaert, E., \& Waelkens, C.\ 1998, \aap, 331, 211 

\bibitem[Mann et al.(2006)]{2006A&ARv..13..159M} Mann, I., K{\"o}hler, M., Kimura, H., Cechowski, A., \& Minato, T.\ 2006, \aapr, 13, 159 

\bibitem[Mann et al.(2007)]{2007P&SS...55.1000M} Mann, I., Murad, E., \& Czechowski, A.\ 2007, \planss, 55, 1000 

\bibitem[Manoj et al.(2002)]{2002MNRAS.334..419M} Manoj, P., Maheswar, G., \& Bhatt, H.~C.\ 2002, \mnras, 334, 419 

\bibitem[Manoj et al.(2006)]{2006ApJ...653..657M} Manoj, P., Bhatt, H.~C., Maheswar, G., \& Muneer, S.\ 2006, \apj, 653, 657 

\bibitem[]{}Mari{\~n}as, N., Telesco, C.~M., 
		Fisher, R.~S., \& Packham, C.\ 
		2011, ApJ, 737, 57 
		
\bibitem[Marino et al.(2015)]{2015ApJ...813...76M} Marino, S., Casassus, S., Perez, S., et al.\ 2015, \apj, 813, 76 
          
\bibitem[]{}Mathis, J.~S., Mezger, P.~G., \& Panagia, N.\ 
	 	1983, A\&A, 128, 212   

\bibitem[Matter et al.(2014)]{2014A&A...561A..26M} Matter, A., Labadie, L., Kreplin, A., et al.\ 2014, \aap, 561, A26 

\bibitem[Matter et al.(2016)]{2016A&A...586A..11M} Matter, A., Labadie, L., Augereau, J.~C., et al.\ 2016, \aap, 586, A11 

\bibitem[Mattioda et al.(2005)]{2005ApJ...629.1183M} Mattioda, A.~L., Allamandola, L.~J., \& Hudgins, D.~M.\ 2005, \apj, 629, 1183 
		
\bibitem[]{}Meeus, G., Waters, L.~B.~F.~M., Bouwman, J., et al.\ 
	   2001, A\&A, 365, 476 	   

\bibitem[Menu et al.(2015)]{2015A&A...581A.107M} Menu, J., van Boekel, R., Henning, T., et al.\ 2015, \aap, 581, A107 

\bibitem[Mer{\'{\i}}n et al.(2004)]{2004A&A...419..301M} Mer{\'{\i}}n, B., Montesinos, B., Eiroa, C., et al.\ 2004, \aap, 419, 301 

\bibitem[Mer{\'{\i}}n et al.(2010)]{2010ApJ...718.1200M} Mer{\'{\i}}n, B., Brown, J.~M., Oliveira, I., et al.\ 2010, \apj, 718, 1200-1223 

\bibitem[]{}Merrick, J.P., Moran, D., \& Radom, L.\ 2007,
           %An Evaluation of Harmonic Vibrational
           %Frequency Scale Factors.
            J. Phys. Chem. A, 111, 11683
%	
\bibitem[Miroshnichenko et al.(2001)]{2001A&A...371..600M} Miroshnichenko, A.~S., Levato, H., Bjorkman, K.~S., \& Grosso, M.\ 2001, \aap, 371, 600 

\bibitem[]{}Momose, M., Morita, A., Fukagawa, M., et al.\ 
		2015, PASJ, 233 
		
\bibitem[Monnier et al.(2005)]{2005ApJ...624..832M} Monnier, J.~D., Millan-Gabet, R., Billmeier, R., et al.\ 2005, \apj, 624, 832 

\bibitem[Monnier et al.(2008)]{2008ApJ...681L..97M} Monnier, J.~D., Tannirkulam, A., Tuthill, P.~G., et al.\ 2008, \apjl, 681, L97 

\bibitem[Montesinos et al.(2009)]{2009A&A...495..901M} Montesinos, B., Eiroa, C., Mora, A., \& Mer{\'{\i}}n, B.\ 2009, \aap, 495, 901 

\bibitem[Mooley et al.(2013)]{2013ApJ...771..110M} Mooley, K., Hillenbrand, L., Rebull, L., Padgett, D., \& Knapp, G.\ 2013, \apj, 771, 110 

\bibitem[Moore et al.(1998)]{1998MNRAS.299.1209M} Moore, T.~J.~T., Emerson, J.~P., Skinner, C.~J., et al.\ 1998, \mnras, 299, 1209 

\bibitem[Mora et al.(2004)]{2004A&A...419..225M} Mora, A., Eiroa, C., Natta, A., et al.\ 2004, \aap, 419, 225 

\bibitem[]{}Morgan, W. W., Keenan, P. C., \& Kellman, E. (ed.) 1943, in An Atlas of Stellar Spectra, with an Outline of Spectral Classification (Chicago, IL: Univ. Chicago Press)

\bibitem[Mouillet et al.(2001)]{2001A&A...372L..61M} Mouillet, D., Lagrange, A.~M., Augereau, J.~C., \& M{\'e}nard, F.\ 2001, \aap, 372, L61 

\bibitem[Murphy et al.(2013)]{2013MNRAS.435.1325M} Murphy, S.~J., Lawson, W.~A., \& Bessell, M.~S.\ 2013, \mnras, 435, 1325 

\bibitem[Muto et al.(2012)]{2012ApJ...748L..22M} Muto, T., Grady, C.~A., Hashimoto, J., et al.\ 2012, \apjl, 748, L22 

\bibitem[Muto et al.(2015)]{2015PASJ...67..122M} Muto, T., Tsukagoshi, T., Momose, M., et al.\ 2015, \pasj, 67, 122 

\bibitem[Natta et al.(2004)]{2004A&A...416..179N} Natta, A., Testi, L., Neri, R., Shepherd, D.~S., \& Wilner, D.~J.\ 2004, \aap, 416, 179 

\bibitem[Neuhaeuser et al.(1995)]{1995A&A...297..391N} Neuhaeuser, R., Sterzik, M.~F., Schmitt, J.~H.~M.~M., Wichmann, R., \& Krautter, J.\ 1995, \aap, 297, 391 

\bibitem[Okamoto et al.(2009)]{2009ApJ...706..665O} Okamoto, Y.~K., Kataza, H., Honda, M., et al.\ 2009, \apj, 706, 665 

\bibitem[Olofsson et al.(2011)]{2011A&A...528L...6O} Olofsson, J., Benisty, M., Augereau, J.-C., et al.\ 2011, \aap, 528, L6 

\bibitem[Olofsson et al.(2013)]{2013A&A...552A...4O} Olofsson, J., Benisty, M., Le Bouquin, J.-B., et al.\ 2013, \aap, 552, A4 

\bibitem[]{}Osorio, M., Anglada, G., 
		Carrasco-Gonz{\'a}lez, C., et al.\ 
		2014, ApJL, 791, L36 

\bibitem[Oudmaijer et al.(2001)]{2001A&A...379..564O} Oudmaijer, R.~D., Palacios, J., Eiroa, C., et al.\ 2001, \aap, 379, 564 
	
\bibitem[Panic(2009)]{2009PhDT.......210P} Panic, O.\ 2009, Ph.D.~Thesis,  

\bibitem[Pech et al.(2002)]{2002A&A...388..639P} Pech, C., Joblin, C., \& Boissel, P.\ 2002, \aap, 388, 639 

\bibitem[]{}Peeters, E., Hony, S., Van Kerckhoven, C., 
            et al.\ 2002, A\&A, 390, 1089 

\bibitem[Pendleton \& Allamandola(2002)]{2002ApJS..138...75P} Pendleton, Y.~J., \& Allamandola, L.~J.\ 2002, \apjs, 138, 75 

\bibitem[Pereyra et al.(2009)]{2009A&A...501..595P} Pereyra, A., Girart, J.~M., Magalh{\~a}es, A.~M., Rodrigues, C.~V., \& de Ara{\'u}jo, F.~X.\ 2009, \aap, 501, 595 

\bibitem[Pereyra et al.(2012)]{2012A&A...538A..59P} Pereyra, A., Rodrigues, C.~V., \& Magalh{\~a}es, A.~M.\ 2012, \aap, 538, A59 

\bibitem[P{\'e}rez et al.(2014)]{2014ApJ...783L..13P} P{\'e}rez, L.~M., Isella, A., Carpenter, J.~M., \& Chandler, C.~J.\ 2014, \apjl, 783, L13 

\bibitem[Perrin et al.(2006)]{2006ApJ...645.1272P} Perrin, M.~D., Duch{\^e}ne, G., Kalas, P., \& Graham, J.~R.\ 2006, \apj, 645, 1272 

\bibitem[Persi \& Tapia(2003)]{2003A&A...406..149P} Persi, P., \& Tapia, M.\ 2003, \aap, 406, 149 

\bibitem[Pi{\'e}tu et al.(2003)]{2003A&A...398..565P} Pi{\'e}tu, V., Dutrey, A., \& Kahane, C.\ 2003, \aap, 398, 565 

\bibitem[Pi{\'e}tu et al.(2005)]{2005A&A...443..945P} Pi{\'e}tu, V., Guilloteau, S., \& Dutrey, A.\ 2005, \aap, 443, 945 

\bibitem[Pi{\'e}tu et al.(2007)]{2007A&A...467..163P} Pi{\'e}tu, V., Dutrey, A., \& Guilloteau, S.\ 2007, \aap, 467, 163 

\bibitem[Pinilla et al.(2014)]{2014A&A...564A..51P} Pinilla, P., Benisty, M., Birnstiel, T., et al.\ 2014, \aap, 564, A51 

\bibitem[Pinilla et al.(2015)]{2015A&A...580A.105P} Pinilla, P., Birnstiel, T., \& Walsh, C.\ 2015, \aap, 580, A105 

\bibitem[Pontoppidan et al.(2007)]{2007ApJ...656..980P} Pontoppidan, K.~M., Dullemond, C.~P., Blake, G.~A., et al.\ 2007, \apj, 656, 980 

\bibitem[Pontoppidan et al.(2008)]{2008ApJ...684.1323P} Pontoppidan, K.~M., Blake, G.~A., van Dishoeck, E.~F., et al.\ 2008, \apj, 684, 1323-1329 

\bibitem[Prato et al.(2003)]{2003ApJ...584..853P} Prato, L., Greene, T.~P., \& Simon, M.\ 2003, \apj, 584, 853 

\bibitem[Qi et al.(2015)]{2015ApJ...813..128Q} Qi, C., {\"O}berg, K.~I., Andrews, S.~M., et al.\ 2015, \apj, 813, 128 

\bibitem[Quanz et al.(2012)]{2012A&A...538A..92Q} Quanz, S.~P., Birkmann, S.~M., Apai, D., Wolf, S., \& Henning, T.\ 2012, \aap, 538, A92 

\bibitem[]{}Quanz, S.~P., Avenhaus, H., Buenzli, E., et al.\ 
		2013, ApJL, 766, L2 

\bibitem[Quanz et al.(2015)]{2015ApJ...807...64Q} Quanz, S.~P., Amara, A., Meyer, 
M.~R., et al.\ 2015, \apj, 807, 64 

\bibitem[]{}Reggiani, M., Quanz, 
		S.~P., Meyer, M.~R., et al.\ 
		2014, ApJL, 792, L23 
		
\bibitem[Ressler \& Barsony(2003)]{2003ApJ...584..832R} Ressler, M.~E., \& Barsony, M.\ 2003, \apj, 584, 832 

\bibitem[Ricca et al.(2012)]{2012ApJ...754...75R} Ricca, A., Bauschlicher, C.~W., Jr., Boersma, C., Tielens, A.~G.~G.~M., \& Allamandola, L.~J.\ 2012, \apj, 754, 75 

\bibitem[Rigliaco et al.(2015)]{2015ApJ...801...31R} Rigliaco, E., Pascucci, I., Duchene, G., et al.\ 2015, \apj, 801, 31 

\bibitem[Roche et al.(1991)]{1991MNRAS.252..282R} Roche, P.~F., Aitken, D.~K., \& Smith, C.~H.\ 1991, \mnras, 252, 282 

\bibitem[Rodigas et al.(2014)]{2014ApJ...791L..37R} Rodigas, T.~J., Follette, K.~B., Weinberger, A., Close, L., \& Hines, D.~C.\ 2014, \apjl, 791, L37 

\bibitem[Rostopchina et al.(1997)]{1997A&A...327..145R} Rostopchina, A.~N., Grinin, V.~P., Okazaki, A., et al.\ 1997, \aap, 327, 145 

\bibitem[Sakon et al.(2006)]{2006aogs....7..143S} Sakon, I., Onaka, T., Okamoto, Y.~K., et al.\ 2006, Advances in Geosciences, Volume 7: Planetary Science (PS), 7, 143 

\bibitem[Sandell et al.(2011)]{2011ApJ...727...26S} Sandell, G., Weintraub, D.~A., \& Hamidouche, M.\ 2011, \apj, 727, 26 

\bibitem[Sch{\"u}tz et al.(2005)]{2005A&A...431..165S} Sch{\"u}tz, O., Meeus, G., \& Sterzik, M.~F.\ 2005, \aap, 431, 165 

\bibitem[Sch{\"u}tz et al.(2009)]{2009A&A...507..261S} Sch{\"u}tz, O., Meeus, G., Sterzik, M.~F., \& Peeters, E.\ 2009, \aap, 507, 261 

\bibitem[Schegerer et al.(2009)]{2009A&A...502..367S} Schegerer, A.~A., Wolf, S., Hummel, C.~A., Quanz, S.~P., \& Richichi, A.\ 2009, \aap, 502, 367 

\bibitem[]{}Seok, J.~Y., \& Li, A.\ 2015, ApJ, 809, 22 

\bibitem[Seok \& Li(2016)]{2016ApJ...818....2S} Seok, J.~Y., \& Li, A.\ 2016, \apj, 818, 2 

\bibitem[Sheret et al.(2004)]{2004MNRAS.348.1282S} Sheret, I., Dent, W.~R.~F., \& Wyatt, M.~C.\ 2004, \mnras, 348, 1282 

\bibitem[Shevchenko et al.(1993)]{1993Ap&SS.202..137S} Shevchenko, V.~S., Grankin, K.~N., Ibragimov, M.~A., Melnikov, S.~Y., \& Yakubov, S.~D.\ 1993, \apss, 202, 137 

\bibitem[Siebenmorgen et al.(1998)]{1998A&A...339..134S} Siebenmorgen, R., Natta, A., Kruegel, E., \& Prusti, T.\ 1998, \aap, 339, 134 

\bibitem[]{}Siebenmorgen, R., Prusti, T., Natta, A., 
            \& M\"uller, T.~G.\ 2000, A\&A, 361, 258 
%

\bibitem[Simon et al.(2000)]{2000ApJ...545.1034S} Simon, M., Dutrey, A., \& Guilloteau, S.\ 2000, \apj, 545, 1034 

\bibitem[Sloan et al.(1999)]{1999ApJ...513L..65S} Sloan, G.~C., Hayward, T.~L., Allamandola, L.~J., et al.\ 1999, \apjl, 513, L65 

\bibitem[Sloan et al.(2003)]{2003ApJS..147..379S} Sloan, G.~C., Kraemer, K.~E., Price, S.~D., \& Shipman, R.~F.\ 2003, \apjs, 147, 379       
%
\bibitem[]{}Sloan, G.~C., Keller, L.~D., Forrest, W.~J., 
            et al.\ 2005, ApJ, 632, 956       

\bibitem[]{}Sloan, G.~C., Jura, M., 
		Duley, W.~W., et al.\ 2007, ApJ, 664, 1144 
		
\bibitem[Smith et al.(2007)]{2007ApJ...656..770S} Smith, J.~D.~T., Draine, B.~T., Dale, D.~A., et al.\ 2007, \apj, 656, 770 
        
\bibitem[Smith et al.(2005)]{2005A&A...431..307S} Smith, K.~W., Balega, Y.~Y., Duschl, W.~J., et al.\ 2005, \aap, 431, 307 

\bibitem[]{}Smith, T. L., Clayton, G. C., \& Valencic, L.\
            2004, AJ, 128, 357
            
\bibitem[Sterzik et al.(2005)]{2005A&A...434..671S} Sterzik, M.~F., Melo, C.~H.~F., Tokovinin, A.~A., \& van der Bliek, N.\ 2005, \aap, 434, 671 

\bibitem[Stolker et al.(2016)]{2016A&A...595A.113S} Stolker, T., Dominik, C., Avenhaus, H., et al.\ 2016, \aap, 595, A113 

\bibitem[Sturm et al.(2013)]{2013A&A...553A...5S} Sturm, B., Bouwman, J., Henning, T., et al.\ 2013, \aap, 553, A5 

\bibitem[Szczepanski et al.(1995)]{1995CPL...232..221S} Szczepanski, J., Wehlburg, C., \& Vala, M.\ 1995, Chemical Physics Letters, 232, 221 

\bibitem[Tang et al.(2012)]{2012A&A...547A..84T} Tang, Y.-W., Guilloteau, S., Pi{\'e}tu, V., et al.\ 2012, \aap, 547, A84 

\bibitem[Tanii et al.(2012)]{2012PASJ...64..124T} Tanii, R., Itoh, Y., Kudo, T., et al.\ 2012, \pasj, 64, 124 

\bibitem[]{}Tielens, A. G. G. M.\ 2008, 
            ARA\&A, 46, 289
         
\bibitem[Torres et al.(1995)]{1995AJ....109.2146T} Torres, C.~A.~O., Quast, G., de La Reza, R., Gregorio-Hetem, J., \& Lepine, J.~R.~D.\ 1995, \aj, 109, 2146 
   
\bibitem[]{}Torres, C. A. O., Quast, G. R., Melo, C. H. F., \& Sterzik, M. F. 2008, 
Young Nearby Loose Associations, ed. B. Reipurth, 757

\bibitem[Torres(2004)]{2004AJ....127.1187T} Torres, G.\ 2004, \aj, 127, 1187 

\bibitem[van Boekel et al.(2003)]{2003A&A...400L..21V} van Boekel, R., Waters, L.~B.~F.~M., Dominik, C., et al.\ 2003, \aap, 400, L21 

\bibitem[]{}van Boekel, R., Waters, L.~B.~F.~M., 
            Dominik, C., et al.\ 2004, A\&A, 418, 177    

\bibitem[van Boekel et al.(2005)]{2005A&A...437..189V} van Boekel, R., Min, M., Waters, L.~B.~F.~M., et al.\ 2005, \aap, 437, 189 

\bibitem[van den Ancker et al.(1998)]{1998A&A...330..145V} van den Ancker, M.~E., de Winter, D., \& Tjin A Djie, H.~R.~E.\ 1998, \aap, 330, 145 

\bibitem[van den Ancker et al.(2000)]{2000A&A...357..325V} van den Ancker, M.~E., Bouwman, J., Wesselius, P.~R., et al.\ 2000, \aap, 357, 325 

\bibitem[van der Marel et al.(2013)]{2013Sci...340.1199V} van der Marel, N., van Dishoeck, E.~F., Bruderer, S., et al.\ 2013, Science, 340, 1199 

\bibitem[van der Marel et al.(2015)]{2015ApJ...810L...7V} van der Marel, N., Pinilla, P., Tobin, J., et al.\ 2015, \apjl, 810, L7 

\bibitem[van der Marel et al.(2016)]{2016A&A...592A.126V} van der Marel, N., Verhaar, B.~W., van Terwisga, S., et al.\ 2016, \aap, 592, A126 

\bibitem[van der Plas et al.(2008)]{2008A&A...485..487V} van der Plas, G., van den Ancker, M.~E., Fedele, D., et al.\ 2008, \aap, 485, 487 

\bibitem[van der Plas et al.(2016)]{2016arXiv160902488V} van der Plas, G., Wright, C.~M., M{\'e}nard, F., et al.\ 2016, arXiv:1609.02488 

\bibitem[van Diedenhoven et al.(2004)]{2004ApJ...611..928V} van Diedenhoven, B., Peeters, E., Van Kerckhoven, C., et al.\ 2004, \apj, 611, 928 

\bibitem[van Leeuwen(2007)]{2007ASSL..350.....V} van Leeuwen, F.\ 2007, Astrophysics and Space Science Library, vol. 350,  
Hipparcos, the New Reduction of the Raw Data. Springer, Heidelberg

\bibitem[Verhoeff et al.(2010)]{2010A&A...516A..48V} Verhoeff, A.~P., Min, M., Acke, B., et al.\ 2010, \aap, 516, A48 

\bibitem[Verhoeff et al.(2012)]{2012A&A...538A.101V} Verhoeff, A.~P., Waters, L.~B.~F.~M., van den Ancker, M.~E., et al.\ 2012, \aap, 538, A101 

\bibitem[Vieira et al.(2003)]{2003AJ....126.2971V} Vieira, S.~L.~A., Corradi, W.~J.~B., Alencar, S.~H.~P., et al.\ 2003, \aj, 126, 2971 

\bibitem[Vural et al.(2014)]{2014A&A...569A..25V} Vural, J., Kraus, S., Kreplin, A., et al.\ 2014, \aap, 569, A25 

\bibitem[Wade et al.(2005)]{2005A&A...442L..31W} Wade, G.~A., Drouin, D., Bagnulo, S., et al.\ 2005, \aap, 442, L31 

\bibitem[Waelkens et al.(1994)]{1994ASPC...62..405W} Waelkens, C., Bogaert, E., \& Waters, L.~B.~F.~M.\ 1994, The Nature and Evolutionary Status of Herbig Ae/Be Stars, 62, 405 

\bibitem[Wahhaj et al.(2010)]{2010ApJ...724..835W} Wahhaj, Z., Cieza, L., Koerner, D.~W., et al.\ 2010, \apj, 724, 835-854 

\bibitem[Walsh et al.(2016)]{2016ApJ...831..200W} Walsh, C., Juh{\'a}sz, A., Meeus, G., et al.\ 2016, \apj, 831, 200 

\bibitem[Wang et al.(2008)]{2008ApJ...673..315W} Wang, S., Looney, L.~W., Brandner, W., \& Close, L.~M.\ 2008, \apj, 673, 315-330 

\bibitem[Wassell et al.(2006)]{2006ApJ...650..985W} Wassell, E.~J., Grady, C.~A., Woodgate, B., Kimble, R.~A., \& Bruhweiler, F.~C.\ 2006, \apj, 650, 985 

\bibitem[Weidner et al.(2010)]{2010MNRAS.401..275W} Weidner, C., Kroupa, P., \& Bonnell, I.~A.~D.\ 2010, \mnras, 401, 275 

\bibitem[Weinberger et al.(1999)]{1999ApJ...525L..53W} Weinberger, A.~J., Becklin, E.~E., Schneider, G., et al.\ 1999, \apjl, 525, L53 

\bibitem[]{}Weingartner, J. C., \& Draine, B. T.\
            2001, ApJS, 134, 263

\bibitem[Wyatt(2005)]{2005A&A...440..937W} Wyatt, M.~C.\ 2005, \aap, 440, 937 
                        
\bibitem[Yang et al.(2013)]{2013ApJ...776..110Y} Yang, X.~J., Glaser, R., Li, A., \& Zhong, J.~X.\ 2013, \apj, 776, 110 

\bibitem[Yang et al.(2016)]{2016MNRAS.462.1551Y} Yang, X.~J., Glaser, R., Li, A., \& Zhong, J.~X.\ 2016a, \mnras, 462, 1551 

\bibitem[]{} Yang, X.~J., Li, A., Glaser, R., \& Zhong, J.~X.\ 2016b, \apj, 825, 22

\bibitem[Zhang et al.(2016)]{2016arXiv161007906Z} Zhang, H., Telesco, C.~M., Pantin, E., et al.\ 2016, arXiv:1610.07906 

\bibitem[Zwintz et al.(2014)]{2014Sci...345..550Z} Zwintz, K., Fossati, L., Ryabchikova, T., et al.\ 2014, Science, 345, 550 

\end{thebibliography}
\end{document}